\documentclass{aa}  

\usepackage{graphicx}
\usepackage{txfonts}
\usepackage{lipsum}
\usepackage{subcaption}

\usepackage{lscape}

\usepackage{placeins}

\usepackage{orcidlink}
\usepackage{xspace}
\usepackage{natbib}
\usepackage{xcolor}
\usepackage{etoolbox}
\usepackage{bm}
\usepackage{longtable}

\makeatletter
\pretocmd{\NAT@citex}{\color{blue}}{}{}
\makeatother

\newcommand{\XRISM}{\textit{XRISM}\xspace}
\newcommand{\XMM}{\textit{XMM-Newton}\xspace}
\newcommand{\chandra}{\textit{Chandra}\xspace}
\newcommand{\suzaku}{\textit{Suzaku}\xspace}
\newcommand{\nustar}{\textit{NuSTAR}\xspace}
\newcommand{\hitomi}{\textit{Hitomi}\xspace}

\hypersetup{
    colorlinks=true,
    linkcolor=blue,
    citecolor=blue,
    filecolor=blue,
    urlcolor=blue
}

\begin{document}
   
  \title{Chemical composition and enrichment of the Centaurus cluster core seen by \XRISM/Resolve}

   \author{F. Mernier\inst{1,2,3,4}\orcidlink{0000-0002-7031-4772}
        \and K. Fukushima\inst{5}\orcidlink{0000-0001-8055-7113}
        \and A. Simionescu\inst{6}\orcidlink{0000-0002-9714-3862}
        \and M. Kondo\inst{7}\orcidlink{0009-0005-5685-1562}
        \and A. Majumder\inst{8,6}\orcidlink{0000-0002-3525-7186}
        \and T. Pl\v{s}ek\inst{9}\orcidlink{0000-0001-6411-3651}
        \and N. Werner\inst{9}\orcidlink{0000-0003-0392-0120}
        \and Y. Fujita\inst{10}\orcidlink{0000-0003-0058-9719}
        \and K. Sato\inst{11}\orcidlink{0000-0001-5774-1633}
        \and K. Matsushita\inst{12}
        \and M. Loewenstein\inst{2,3,4}\orcidlink{0000-0002-1661-4029}
        \and R. Mushotzky\inst{2}\orcidlink{0000-0002-7962-5446}
        \and J.-P. Breuer \inst{13}\orcidlink{0000-0001-6131-4802}
        \and R. Fujimoto\inst{5}\orcidlink{0000-0002-2374-7073}
        \and Y. Fukazawa\inst{13}
        \and I. Hatsukade\inst{14}\orcidlink{0000-0003-3518-3049}
        \and K. Nakazawa\inst{15}\orcidlink{0000-0003-2930-350X}
        \and M. Urata\inst{13}
        \and N. Yamasaki\inst{5}\orcidlink{0000-0003-4885-5537}
        }

   \institute{IRAP, CNRS, Université de Toulouse, CNES, UT3-UPS, Toulouse, France \\
             \email{francois.mernier@irap.omp.eu}
            \and Department of Astronomy, University of Maryland, 4296 Stadium Dr., College Park, MD 20742-2421, USA 
            \and NASA/Goddard Space Flight Center, Greenbelt, MD 20771, USA
            \and Center for Research and Exploration in Space Science and Technology, NASA/GSFC (CRESST II), Greenbelt, MD 20771, USA
            \and Institute of Space and Astronautical Science (ISAS), Japan Aerospace Exploration Agency (JAXA), 3-1-1 Yoshinodai, Chuo-ku, Sagamihara, Kanagawa 252-5210, Japan
            \and SRON Netherlands Institute for Space Research, Niels Bohrweg 4, 2333 CA Leiden, the Netherlands
            \and Department of Physics, Saitama University, 255 Shimo-Okubo, Sakura, Saitama, Saitama 338-8570, Japan
            \and Waterloo Centre for Astrophysics, Department of Physics and Astronomy, 200 University Avenue West, Waterloo, N2L 3G1, Canada
            \and Department of Theoretical Physics and Astrophysics, Faculty of Science, Masaryk University, Kotlářská 2, Brno, 611 37, Czech Republic
            \and Department of Physics, Tokyo Metropolitan University, 1-1 Minami-Osawa, Hachioji, Tokyo 192-0397, Japan
            \and Department of Astrophysics and Atmospheric Sciences, Kyoto Sangyo University, Kamigamo-motoyama, Kita-ku, Kyoto, Kyoto 603-8555, Japan
            \and Faculty of Physics, Tokyo University of Science, 1-3 Kagurazaka, Shinjuku-ku, Tokyo 162-8601, Japan
            \and Department of Physics, Hiroshima University, 1-3-1 Kagamiyama, Higashi-Hiroshima, Hiroshima 739-8526, Japan
            \and Faculty of Engineering, University of Miyazaki, 1-1 Gakuen-kibanadai-nishi, Miyazaki, Miyazaki 889-2192, Japan
            \and Department of Physics, Nagoya University, Furo-cho, Chikusa-ku. Nagoya, Aichi 464-8602, Japan
            }

   \date{Received September 26, 2025 / Accepted December 9, 2025}

  \abstract
   {Hot, X-ray emitting atmospheres pervading galaxy clusters (and groups) are rich in metals, which have been synthesised and released by asymptotic giant branch (AGB) stars, core-collapse supernovae (SNcc), and Type Ia supernovae (SNIa) over cosmic history. This makes the intracluster medium (ICM) an ideal astrophysical system to constrain its chemical composition, and hence ultimately understand metal production and enrichment on megaparsec scales.}
   {In this work, we take advantage of the unprecedented $\sim$5~eV resolution offered by the Resolve instrument on board the \XRISM observatory to measure the chemical composition of the core of the bright, nearby, and metal-rich Centaurus cluster with unprecedented accuracy. We use these measurements to provide constraints on the stellar populations having enriched the cluster core.} 
   {Through a deep (287 ks) Resolve full-array spectral analysis of Centaurus, we derived the Fe abundance and its relative Si/Fe, S/Fe, Ar/Fe, Ca/Fe, Cr/Fe, Mn/Fe, and Ni/Fe ratios. We completed this high-resolution view with N/Fe, O/Fe, Ne/Fe, and Mg/Fe ratios obtained with \XMM/RGS archival data. This abundance pattern was then fitted with various combinations of AGBs, SNcc and SNIa nucleosynthesis yields with the aim of constraining their explosion and/or progenitor models.}
   {Similarly to the core of Perseus (from previous \hitomi/SXS results), we find that nine out of our 11 measured abundance ratios are formally consistent with the chemical composition of our Solar System (within uncertainties of the latter). However, the (super-solar) N/Fe and (half-solar) Mg/Fe ratios significantly differ from Perseus and/or other systems, and thus they provide tension with the picture of a fully solar composition ubiquitous to all systems. In addition, possible uncertainties in O/Fe and Ne/Fe with atomic codes highlight the need for studying more systems at high spectral resolution to assess (or rule out) the universality of the ICM composition in clusters' cool cores. Combinations of (AGB+)SNcc+SNIa yield models can reproduce our observed X/Fe ratios in all cases. However, whether two distinct populations of SNIa are needed depends on the weight of our RGS measurements. We also briefly discuss the possibility of a multi-metallicity gas phase in this respect.}
   {}


   \maketitle

\section{Introduction}\label{sec:intro}

Whereas primordial nucleosynthesis fixed the chemical composition of the early Universe almost exclusively to H and He ($\sim$ 75\% and 25\%, respectively; \citealt{planck2020}), X-ray spectroscopy has revealed that the hot intracluster medium (ICM) pervading galaxy clusters and groups is rich in metals \citep{mitchell1976,serlemitsos1977}. This major finding showcases a tight interplay between stellar activity and end products on the one hand, and the largest gravitationally bound structures of the Universe on the other hand. Specifically, among the 13 elements detected in the ICM so far, (i) C and N are largely produced and released by asymptotic giant branch (AGB) stars; (ii) O, Ne, and Mg are synthesised in massive stars and core-collapse supernovae (SNcc); (iii) Cr, Mn, Fe, and Ni mostly originate from Type Ia supernovae (SNIa); and (iv) intermediate elements (Si, S, Ar, and Ca) are produced by both SNcc and SNIa in comparable amounts (for a review, see \citealt{nomoto2013}). 

Although the big picture of stars and supernovae as sources of metals is well established, many open questions remain. First, the nature of the SNIa progenitors is still unclear (for reviews, see \citealt{maoz2014,ruiter2025}). In the single-degenerate scenario, the explosive C burning of a carbon-oxygen (CO) white dwarf (WD) is triggered, often near its Chandrasekhar mass, following steady accretion from a main sequence companion. The majority of single-degenerate models predict a burning flame that either propagates subsonically through the entire explosion (deflagration), or transitions from subsonic to supersonic velocities during the process (delayed detonation). In the double-degenerate scenario, the SNIa explosion results from an interaction -- often a merger -- between two WDs that have not necessarily reached their Chandrasekhar mass, likely triggering a pure detonation (although the latter is not excluded in a number of single-degenerate models as well). A better understanding of SNIa explosion mechanisms is thus vital to constrain the nature of their progenitors. In this respect, interestingly, nucleosynthesis yields predicted by SNIa models are sensitive to the physics of the explosion. Second, the initial mass function (IMF) and initial metallicity of massive stars responsible for the enrichment on galactic and extragalactic scales are both largely unknown. In a similar fashion, AGB and SNcc yields depend on the initial mass and metallicity of the modelled stellar progenitor, offering us a concrete way to address these questions. 

Due to its immense (yet gravitationally closed) volume and its relatively simple X-ray emission process in collisional ionisation equilibrium, the ICM is arguably the most suitable astrophysical system in which elemental abundances, as a fossil imprint of the end product of billions of AGBs, SNcc, and SNIa, can be robustly constrained. Nowadays measured routinely at moderate resolution ($\sim$ 120~eV) with \chandra/ACIS, \XMM/EPIC, and \suzaku/XIS, as well as at dispersive high resolution with \XMM/RGS, X-ray spectroscopy of galaxy clusters (and groups) provides us with invaluable information on the nature of stellar populations having enriched the majority of today's Universe, in addition to the history of metal injection and mixing at megaparsec scales (for reviews, see \citealt{biffi2018,mernier2018c}). In addition, measuring accurate abundances in the ICM can also help to better understand the cosmic star formation rate as well as to constrain the fraction of metals that have escaped from their galactic potential well.

While a large number of studies have focused on the overall metallicity of the ICM -- as traced by Fe whose K-shell emission line at $\sim$6.7~keV is easily identified even at moderate resolution -- less effort has been dedicated to measuring abundances of individual elements (X) and their ratios relative to iron (X/Fe). The challenge mostly resides in the spectral resolution offered by X-ray instruments so far, through which individual lines are not well resolved. For this reason, individual abundance studies required not only deep exposures but also extra caution on subtle systematic effects that could increase our uncertainty on such measurements (e.g. background treatment, instrumental calibration, and modelling biases left unnoticed by the fitting statistics) and, eventually, could trigger a (too) extreme reliance on current spectral models. Though valuable, even at higher energy resolution, the RGS gratings on board \XMM provide somewhat limited information due to their instrumental nature (which broadens line profiles in extended sources) and limited bandwidth (which hampers our access to the continuum and several key elements). 

Despite these limitations, studies performed on individual systems \citep[e.g.][]{tamura2001,finoguenov2002,matsushita2003,werner2006,deplaa2006,matsushita2007,simionescu2009,loewenstein2012,mernier2015} and cluster (and/or group) samples \citep[e.g.][]{mushotzky1996,finoguenov2000,tamura2004,baumgartner2005,deplaa2007,sato2007,mernier2016a,mernier2018} have converged towards an overall trend of X/Fe ratios being rather close to their proto-solar\footnote{For convenience, in the following we will use the wording `solar' instead of `proto-solar'.} reference values, although with a few deviations from the latter (e.g. Ca/Fe and Ni/Fe). In 2016, the flight of \hitomi and its Soft X-ray Spectrometer (SXS) instrument allowed non-dispersive high-resolution X-ray spectroscopy for the very first time and paved the way towards a transformative era in our quest of constraining the ICM chemical composition. Despite its unexpectedly short lifetime, \hitomi observed the core of the Perseus cluster and revealed that the abundance pattern of its hot atmosphere is fully consistent with solar in {all} investigated ratios from Si/Fe to Ni/Fe \citep{hitomi2017}. These results were further confirmed and extended to the O/Fe, Ne/Fe, and Mg/Fe ratios through an additional deep \XMM/RGS analysis \citep{simionescu2019}. A first-order comparison of all these ratios to SNcc and SNIa yields led to the conclusion that: (i) unlike earlier trends, SNIa deflagration models are still competitive to explain the ICM enrichment; (ii) a mix of near-Chandrasekhar (near-$M_\mathrm{Ch}$) {and} sub-Chandrasekhar (sub-$M_\mathrm{Ch}$) SNIa models may be necessary to reproduce the Fe-peak ratios satisfactorily; and (iii) no combination of SNcc+SNIa models could successfully reproduce all ratios at once -- with the S/Ar ratio being the most challenging -- suggesting a need for improvement among SNcc nucleosynthesis calculations.

Besides the above considerations, the remarkably near-solar values of the X/Fe ratios in the core of the Perseus ICM were not necessarily expected, as the chemical composition of our own Solar System is not representative of stellar populations in (early-type) cluster galaxies. Instead, the latter exhibit stellar populations with enhanced $\alpha$/Fe ratios \citep[e.g.]{conroy2014} as a plausible result from `downsizing' -- i.e. a rapid, early, and intense episode of star formation at the origin of the assembly of massive elliptical galaxies \citep{thomas2010}. The invaluable heritage of \hitomi on ICM metal abundances thus naturally leads to a fundamental question: are the solar X/Fe ratios of Perseus a coincidence or do they reflect a universal chemical composition in cluster (cool) cores? Beyond its importance for understanding enrichment mechanisms in and around the clusters' brightest central galaxy (BCG), the answer to this question would also shed light on the history of the central ICM enrichment. Seen at large scales, the majority of the ICM (mostly residing in cluster outskirts) is thought to have been enriched at early stages of -- or even before -- cluster assembly, i.e. at redshifts of at least $z \sim 2$--3, when cosmic star formation and supermassive black hole activities were at their peak \citep{madau2014,hickox2018}. This early- (or pre-) enrichment scenario has been indirectly corroborated so far by converging observational results -- including flat radial metallicity profiles in cluster outskirts \citep[e.g.][]{fujita2008,werner2013,urban2017,ghizzardi2021,sarkar2022} and redshift studies consistent with no redshift evolution at $z \lesssim 1$ \citep[e.g.][]{mcdonald2016,mantz2017,liu2020} -- and it is further supported by simulations \citep{biffi2017,biffi2018b}. In cluster cool cores, however, the large scatter of central metallicities \citep[e.g.][]{ghizzardi2021} and the correlation between central Fe gas and BGC stellar masses \citep[e.g.][]{degrandi2004} raises the possibility of later gas-phase enrichment episodes from the stellar population of central galaxies. On paper, such a late central enrichment contribution could be revealed by a diversity of central abundance patterns in cool-core clusters (provided that its relative weight in shaping the abundance pattern is comparable to that of the pre-enriched ICM falling into the core). 

In this context, the recent launch of \XRISM is perfectly suited to extend and complete the transformative results on ICM abundances delivered by \hitomi. Its Resolve instrument (a microcalorimeter very similar to SXS) enables a resolution of $<$5~eV within the 1.7--12~keV band (as is currently in closed gate valve configuration), which constitutes a unique opportunity to measure the hot gas abundance pattern in the core of clusters other than Perseus. In this regard, the Centaurus cluster (Abell\,3526) constitutes an ideal target. Not only is this cluster very nearby ($z = 0.01003$) and does it host a bright cool core onto its BCG (NGC\,4696), but its core also peaks at an ideal temperature of $kT \sim$ 2--2.5~keV, ensuring a wealth of emission lines to be detectable above the continuum. Last but certainly not least, the core of Centaurus is famously known for being unusually metal-rich ($\sim$ 2 solar), with central abundances about twice as high as in most other cool-core clusters \citep[e.g.][]{sanders2008,sanders2016,lakhchaura2019,fukushima2022}, making its emission lines even more prominent than one could otherwise expect.

\XRISM observed Centaurus during its Performance Verification (PV) phase as one of the first targets to follow the calibration phase. A first study presented its Resolve spectrum as well as a map of its (bulk and turbulent) gas velocity measurements \citep{xrism2025}. This present work utilises the same deep dataset to focus on the elemental abundances integrated over the full Resolve array. The goal of this paper is twofold. First, we aim to accurately derive the abundance pattern in the core of the Centaurus ICM from high-resolution spectroscopy (Si/Fe, S/Fe, Ar/Fe, Ca/Fe, Cr/Fe, Mn/Fe, and Ni/Fe) and discuss it in light of comparable measurements in Perseus (\hitomi/SXS) and other clusters (moderate spectral resolution) from the literature. Inspired by the Perseus results presented in \citet{simionescu2019}, we chose to include an additional analysis from \XMM/RGS archival observations as well in order to derive constraints on N/Fe, O/Fe, Ne/Fe,  and Mg/Fe and obtain a full picture of the chemical composition of the Centaurus core. Second, we compared our derived abundance ratios with a series of (historical and recent) stellar nucleosynthesis yields from the literature, with the aim of bringing further constraints on SNcc (and AGB) enrichment and SNIa explosion models. A forthcoming paper \textcolor{blue}{(Fukushima et al., to be subm.)} focuses on a sub-array analysis, with the goal of mapping the spatial abundance distribution within the cluster core and discussing the existence of a currently unexplained central abundance drop, seen as moderate spectral resolution in Centaurus \citep{sanders2016,lakhchaura2019,fukushima2022} and other systems as well \citep[e.g.][]{panagoulia2015,mernier2017}. 

Throughout this work we assume a standard cosmology with $H_0 = 70$~km/s/Mpc, $\Omega_m = 0.30$, and $\Omega_\Lambda = 0.70$. Our abundances are expressed in solar units of \citet{lodders2009}. Unless stated otherwise, the quoted error bars are shown at 1$\sigma$ and correspond to a 68\% confidence interval. Section~\ref{sec:methods} details the data reduction used for the main part of this analysis (mostly \XRISM/Resolve and \XMM/RGS) as well as our adopted spectral fitting strategies. Our full-array absolute abundances and X/Fe ratios are presented in Sect.~\ref{sec:results} and further discussed in Sect.~\ref{sec:discussion}. These measurements are then compared with (and discussed in light of) AGB, SNcc, and/or SNIa yield models in Sect.~\ref{sec:yields}. Section~\ref{sec:conclusions} concludes our study.

\section{Methods}\label{sec:methods}

The \XRISM PV data of Centaurus were obtained between 2023/12/28 and 2024/01/03 for an effective exposure of 287 ks\footnote{The total exposure, including Earth occultation, is 592~ks.}. At the time of the observation, the closed position of the gate valve in front of Resolve made the instrument insensitive to the soft X-ray band (below $\lesssim 2$~keV). In practice, this means that Resolve was not able to access elements lighter than Si. To investigate the abundance of $\alpha$ elements such as N, O, Ne, and Mg in the Centaurus cluster, we rely on the grating instrument RGS on board \XMM. Although RGS is slitless and its dispersive spectroscopy broadens the line profiles in the case of extended sources, this instrument remains able to resolve important emission lines at lower energies (e.g. \ion{O}{viii}). The combination RGS+Resolve offers thus high-resolution spectroscopy through the entire 0.3--10~keV band, which is invaluable for our science case. Figure~\ref{fig:images} shows the core of the Centaurus cluster as well as the regions covered by the above-mentioned two instruments. Table~\ref{table:obslog} summarises the Resolve and RGS observations analysed in this work.

\subsection{Data reduction}\label{sec:methods:data}

\subsubsection{\XRISM/Resolve}\label{sec:methods:data:resolve}

 The reduction of the Resolve data is done with the \XRISM Build8 data reduction software released on June 20, 2024 -- which is effectively identical to the publicly available package included in \texttt{HEASOFT} v6.34 (released on August 22, 2024). The software uses an early post-flight version of the \XRISM CalDB that does not differ from the current versions in any of the relevant files used in pipeline processing. The procedure follows a recipe similar as in \citet{xrism2025}, the main steps of which are summarised below. After applying the pre-pipeline task \texttt{xapipeline}, hence retrieving a clean event list, we perform an additional screening criterion based on a recommended RISE\_TIME expression\footnote{\href{https://heasarc.gsfc.nasa.gov/docs/xrism/analysis/ttwof/index.html}{https://heasarc.gsfc.nasa.gov/docs/xrism/analysis/ttwof/index.html}}. 
 
\begin{table}[t!]
\caption{Summary of the observations of the Centaurus core analysed spectroscopically in this paper.}
\label{table:obslog}
\centering
\begin{tabular}{l | c c}
\hline\hline
 & \XRISM/Resolve & \XMM/RGS \\
\hline
ObsID                   & 000138000 & 0406200101 \\
Obs. starting date      & 2023-12-28 & 2006-07-24 \\
Total duration (ks)     & 592.1 & 125.4 \\
Net exposure (ks)       & 287.4 & 104.0 \\

\hline
\end{tabular}
\end{table}

After inspecting visually the light curve and an image of the Resolve counts on the detector, we use the \texttt{HEASOFT} tool \texttt{xselect} to extract a spectrum covering 34 out of the 36 pixels of the array. Pixel 12 is the calibration pixel outside the instrument field of view and thus is not used here. Although pixel 27 was operational at the time of the observation, it encountered unexpected gain excursions very soon after launch. As a precaution and to remain in line with other PV data analyses, we decide to exclude this pixel as well. This represent less than 2\% of the total counts over the full array.

The redistribution matrix file (RMF) is generated via the task \texttt{rslmkrmf} and is set to be extra-large (\texttt{whichrmf=X} in the expression) in order to reproduce the entire line spread function as well as to account for the electron loss continuum which becomes non-negligible below $E \lesssim 3$~keV. We ensure to use the up-to-date RMF parameter CalDB file updated based on in-flight calibration (\texttt{xa\_rsl\_rmfparam\_20190101v006.fits}; see also \citealt{leutenegger2025}).  

Last but not least, we extract the ancillary response file (ARF) via the task \texttt{xaarfgen}. A key subroutine of this task, \texttt{xrtraytrace}, uses a sophisticated ray-tracing tool to mimic the point spread function (PSF) of the telescope at various energies and, ultimately, to simulate the mixing of photons coming from different sky regions onto different positions on the detector \citep{yaqoob2018}. Although it is crucial to model accurately this `spatial-spectral mixing' (SSM) when investigating physical parameters over several spatial bins \citep[][; Fukushima et al. to be subm.]{hitomi2018,xrism2026}, the full-array approach of this work assumes by definition negligible spatial variation of abundance ratios within the core of Centaurus \citep[see also][]{sanders2016,lakhchaura2019}, thus rendering the need for SSM unnecessary in our case. In order to remain as realistic as possible, however, we choose to extract our ARF based on a \textit{Chandra}/ACIS flux image of the source (in the 2--8~keV energy band; Sect.~\ref{sec:methods:data:chandra}) to capture its extended emission within the Resolve array. We have verified that, except the emission measure, our parameters of interests (temperature, abundances, etc.) are not altered by the choice of this image-based ARF over a more simplistic point-source-based ARF.

\begin{figure*}[t!]
\centering
\includegraphics[width=0.8\hsize]{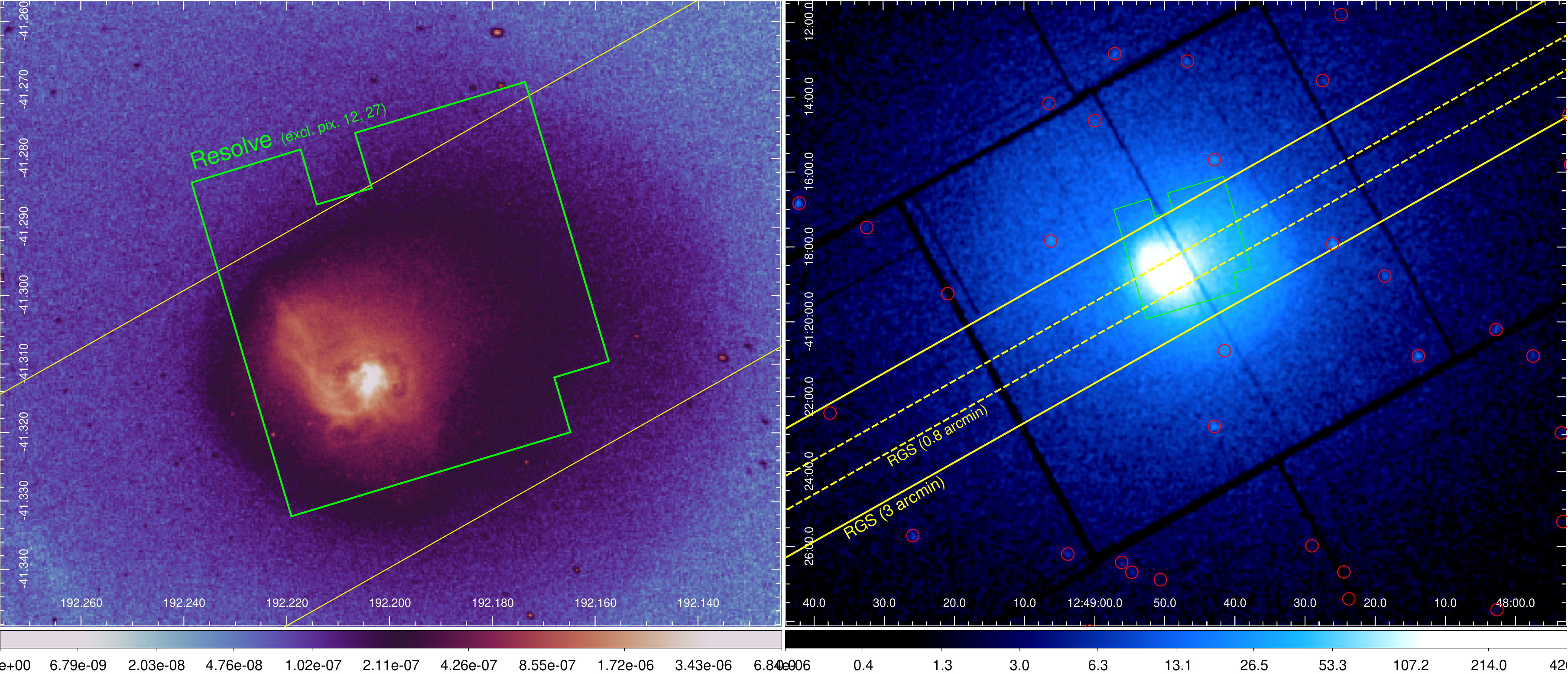}
      \caption{Core of the Centaurus cluster and its \XRISM/Resolve and \XMM/RGS extraction regions. The full-array Resolve FoV (excluding pixels 12 and 27; Sect.~\ref{sec:methods:data:resolve}) is shown in solid green contours while the cross-dispersion apertures of RGS (0.8$'$ and 3$'$) chosen for this work are shown in yellow (dashed and solid contours, respectively). \textit{Left:} Exposure-corrected 0.3-7~keV \chandra/ACIS fluxed image showing the soft X-ray features  of NGC\,4696 (plume, cavities). \textit{Right:} \XMM/EPIC MOS\,2 counts image showing the gas extent at larger scale. Detected point sources (red circles) fall mostly outside of our regions and are thus negligible.}
         \label{fig:images}
\end{figure*}

\subsubsection{\XMM/RGS (and EPIC)}\label{sec:methods:data:rgs}

Although two \XMM observations are centred on NGC\,4696, and are thus suitable for RGS analysis, we chose to focus on the deepest one (125.4~ks raw exposure; Table~\ref{table:obslog}) and discard the other ($<$50~ks). Doing so, we avoid possible systematics due to calibration and/or different roll angles while keeping good statistics.

The \XMM extraction procedure was done using XMM SAS (Science Analysis Software) v21.0. Since the RGS\,1 and RGS\,2 instruments share the same telescope as MOS\,1 and MOS\,2 respectively, we start by extracting EPIC MOS data using the pipeline task \texttt{emproc} with a twofold aim. First, we derive light curves in the 10--12~keV band, to estimate the good time intervals (GTI) devoid of solar proton flares. Via the task \texttt{espfilt}, we extract count rate histograms of these light cuves (in bins of 10~s), which we fitted with a Gaussian function of free mean ($\mu_\mathrm{lc}$) and standard deviation ($\sigma_\mathrm{lc}$). We then choose to discard all events whose count rate exceeds $\mu_\mathrm{lc} + 2\,\sigma_\mathrm{lc}$ \citep[\textcolor{black}{a.k.a. the 2$\sigma$ clipping method; see e.g.}][]{deluca2004,snowden2008}. We repeat the task for the 0.3--10~keV band in order to capture flares at lower energy as well \citep{deluca2004}. Second, we use \texttt{evselect} to extract MOS images (after filtering the MOS events with the appropriate GTIs), which will be necessary for appropriate fitting of the RGS spectra (Sect.~\ref{sec:methods:spectra}). We also verify that no point sources were detected within the Resolve array region of our EPIC image (Fig.~\ref{fig:images} right).

As a next step, we extract the RGS\,1 and RGS\,2 data using the pipeline task \texttt{rgsproc}, taking care of applying the respective MOS\,1 and MOS\,2 GTIs accordingly. The same procedure is used to extract the first and second orders spectra, along with their corresponding RMFs and ARFs. Although -- unlike EPIC -- the grating nature of the RGS instrument does not allow to extract spatial regions of our choice, the \texttt{xpsfincl} parameter in \texttt{rgsproc} allows to select events within a specified fraction (in percent) of the telescope's point spread function (PSF). To quantify possible systematics and compare our measurements with previous work, we choose to extract RGS spectra within apertures of $\sim$0.8$'$ (\texttt{xpsfincl=90}) and $\sim$3.0$'$ (\texttt{xpsfincl=95}). The latter is of particular interest since its extracted region has good overlap with the Resolve full-array pointing (Fig.~\ref{fig:images}).

For comparison purposes with our high-resolution analysis (Sect.~\ref{sec:discussion:abunpattern}), it is also useful to extract EPIC (MOS\,1, MOS\,2, and pn) spectra that follow the Resolve full-array coverage on the sky. Starting from the SAS pipeline \texttt{epproc} task, we thus reduce the pn data in the same way as described above for MOS. We then generate the appropriate spectra, RMFs, and ARFs in the three instruments using the tasks \texttt{evselect}, \texttt{rmfgen}, and \texttt{arfgen}, respectively.

\subsubsection{\chandra/ACIS}\label{sec:methods:data:chandra}

The \chandra/ACIS data were extracted for imaging purpose only. For comprehensive spectral analysis of the Centaurus core with ACIS specifically, we refer the reader to recent work in the literature \citep[e.g.][]{sanders2016,lakhchaura2019,fukushima2022}. We used \texttt{CIAO} v4.16 to download and reprocess (via the task \texttt{chandra\_repro}) the data of the 15 ACIS-S observations pointing to -- or within $\le$4$'$ from -- the centre of NGC\,4696 (ObsID: 504, 1560, 4190, 4191, 4954, 4955, 5310, 16223, 16224, 16225, 16534, 16607, 16608, 16609, 16610)\footnote{\href{https://doi.org/10.25574/cdc.500}{https://doi.org/10.25574/cdc.500}}. After individual cleaning from solar flares using the \texttt{deflare} task, the observations were merged and imaged as a mosaic in various energy bands using \texttt{merge\_obs}. The image used as input for the Resolve ARF generation in IMAGE mode is a 2--8~keV image following the position and roll-angle of the Resolve array and cropped outside of its 3.05$' \times$~3.05$'$ limits. For display purpose, the ACIS image shown in Fig.~\ref{fig:images} left covers soft energies as well (0.3--7~keV) and exhibits all remarkable features known in the Centaurus core (see also \citealt{sanders2016}).

\subsection{Spectral fitting strategies}\label{sec:methods:spectra}

\subsubsection{XRISM/Resolve}\label{sec:methods:spectra:resolve}

Since ICM abundance measurements directly depend on the equivalent width of their X-ray emission lines, they may be subject to a number of systematic effects, essentially coming from two sources of uncertainties: (i) the assumed atomic codes and (ii) the (multi-)temperature structure of the ICM. Two widely used atomic codes, which we use in this work, are SPEXACT (v3.08.01)\footnote{\href{https://spex-xray.github.io/spex-help/index.html}{https://spex-xray.github.io/spex-help/index.html} \citep{spex2024}} and AtomDB (v3.0.9)\footnote{\href{http://www.atomdb.org/}{http://www.atomdb.org/} \citep{smith2001}}, available in the SPEX and Xspec fitting packages, respectively\footnote{Since the time of the analysis, AtomDB has been updated to v3.1.3. Using 1T modelling, we have verified that the impact of these updates to our results are negligible.}. These two packages offer a range of possibilities for modelling redshifted\footnote{While the redshift is included in Xspec collisional plasma models directly, it needs to be modelled separately in SPEX through the multiplicative component \texttt{red}.}, absorbed\footnote{The (multiplicative) absorption models we use from SPEX and Xspec are, respectively, \texttt{TBabs} and \texttt{hot}, the latter of which is used with a temperature of $5 \times 10^{-4}$.} single- or multi-temperature collisional plasmas\footnote{Even though the Centaurus ICM is known to be multi-temperature \citep[e.g.][]{fukushima2022}, understanding the impact of measured abundances with the choice of the temperature modelling at high spectral resolution motivates us to include single-temperature models as well in our analysis.}. We assume a Galactic hydrogen column density $n_H$ of $1.22 \times 10^{21}$~atoms/cm$^2$ (accounting for both atomic and molecular hydrogen; \citealt{willingale2013}) with solar abundances \citep{lodders2009}. The models that are used in this paper are described below. \\

From SPEX:
\begin{itemize}
        \item 1T$_\mathrm{SPEX}$ is a single-temperature model (\texttt{red*hot*cie}) with the emission measure (EM), temperature ($kT_1$), redshift, velocity dispersion ($v_\mathrm{rms,1}$) and Si, S, Ar, Ca, Cr, Mn, Fe, and Ni abundances left free;
        \item 2T$_\mathrm{SPEX}$ is a two-temperature model (\texttt{red*hot*(cie+cie)}) with the two emission measures (EM$_1$, EM$_2$), temperatures ($kT_1$, $kT_2$), redshifts, and velocity dispersions ($v_\mathrm{rms,1}$, $v_\mathrm{rms,2}$) left free and independent from each other. To avoid overfitting, the abundances of the two components are tied with each other;
        \item lin-gdem$_\mathrm{SPEX}$ is a multi-temperature model with differential emission measure shaped as a Gaussian distribution (\texttt{red*hot*cie} with the temperature width parameter $\sigma_{\mathrm{lin,}T} > 0$ left free). The temperature parameter, $kT_\mathrm{mean}$ represents the mean temperature of the distribution and is left free, as the other parameters mentioned for the 1T$_\mathrm{SPEX}$ model;
        \item log-gdem$_\mathrm{SPEX}$ is a multi-temperature model with differential emission measure shaped as a log-normal distribution (\texttt{red*hot*cie} with $\sigma_{\mathrm{log,}T}$ left free). A parameter in the \texttt{cie} component allows to switch easily between lin-gdem$_\mathrm{SPEX}$ and this model. We free the parameters similarly as in the lin-gdem$_\mathrm{SPEX}$ model.
\end{itemize}

From Xspec:
\begin{itemize}
        \item 1T$_\mathrm{Xspec}$ is a single-temperature model (\texttt{TBabs*bvvapec}) with the normalisation (norm$_1$), temperature ($kT_1$), redshift, velocity dispersion ($v_\mathrm{rms,1}$) and Si, S, Ar, Ca, Cr, Mn, Fe, and Ni abundances left free. It is the Xspec equivalent of the 1T$_\mathrm{SPEX}$ model;
        \item 2T$_\mathrm{Xspec}$ is a two-temperature model (\texttt{TBabs*(bvvapec} \texttt{+bvvapec)}) and is the Xspec equivalent of the 2T$_\mathrm{SPEX}$ model;
        \item lin-gdem$_\mathrm{Xspec}$ is a multi-temperature model with differential emission measure shaped as a Gaussian distribution (\texttt{TBabs*bvvgadem}) and is the Xspec equivalent of the lin-gdem$_\mathrm{SPEX}$ model;
        \item wdem$_\mathrm{Xspec}$ is a multi-temperature model with differential emission measure shaped as a power law (\texttt{TBabs*bvvwdem}). It follows $dY/dT \propto kT^{1/\alpha_T}$ when $\beta_T kT_\mathrm{W} < kT < kT_\mathrm{W}$ and $dY/dT = 0$ otherwise. The temperature parameter, $kT_\mathrm{W}$ represents the maximum temperature of the distribution and is left free, as well as the inverse slope parameter $\alpha_T$. The temperature ratio $\beta_T = 0.1$ is fixed to its default value. The other parameters follow the 1T$_\mathrm{Xspec}$ prescription.
\end{itemize}

For reasons that are highlighted further in Sect.~\ref{sec:results}, we choose 2T$_\mathrm{SPEX}$ as our baseline model, both in our Resolve and our RGS analyses. All these models are first fitted within 1.9--10~keV and exclude the \ion{Fe}{xxv} resonance (w) line (Fig.~\ref{fig:spectra_app}) to avoid biases due to possible resonant scattering. We refer this approach further as our `broad-band fit'. 

Following common practice in the literature \citep[e.g.][]{mernier2016a,lakhchaura2019,fukushima2022}, we then refit the spectrum locally to ensure that our derived abundances are robust towards possible imperfections in the ARFs (which, if fitted over a broad band, might lead to an incorrect estimate of the local continuum and, in turn, of the equivalent width of the lines). This `narrow-band fit' approach is achieved by fixing all parameters to their broad-band best-fit values, leaving free the EM (or norm)\footnote{In the case of our 2T models, EM$_2$ (norm$_2$) is tied to EM$_1$ (norm$_1$) with a fixed ratio obtained from the best broad-band fit.}, $z$, $v_\mathrm{rms}$, and the elemental abundance of the considered line. Unlike previous work on CCD data, the exquisite spectral resolution offered by Resolve allows now to do this exercise on lines of separate ions. The local bands used around H-like lines (\ion{Si}{xiv}, \ion{S}{xvi}, \ion{Ar}{xviii}, \ion{Ca}{xx}, \ion{Fe}{xxvi}) and He-like lines (\ion{S}{xv}, \ion{Ar}{xvii}, \ion{Ca}{xix}, \ion{Cr}{xxiii}, \ion{Mn}{xxiv}, \ion{Fe}{xxv}, \ion{Ni}{xxvii}) are shown in Fig.~\ref{fig:spectra_app}. 

Last but not least, the Resolve non-X-ray background (NXB) is modelled following the standard \XRISM data analysis recommendations. The model is based on a series of Resolve observations stacked since the launch of the mission and consists of a power law, 17 fluorescent lines of well known energies and three scaling factors which are left free in our fits. The model is then folded by a dedicated diagonal RMF to account of the particle nature of the NXB. We note that this model is available for analysis with Xspec only. Because currently the routine used to convert OGIP files into SPEX formats (\texttt{trafo}) has issues with loading standard and diagonal RMFs at once, we simulate a mock spectrum of the best-fit Xspec NXB component (with an arbitrarily high exposure, e.g. 1~Ms, to maximise the statistics), which we then subtract from our total full-array spectrum in SPEX. Because the NXB is at least one order of magnitude smaller than the photon count rate in the 2--7~keV band, this differing method is not expected to affect our estimates. More generally, the impact of the (relatively low and stable) Resolve NXB to spectral analysis results is known to be at most marginal \citep[e.g.][]{xrism2025_A2029}, especially for a bright source like Centaurus. We also verify that the spectral contributions from low-mass X-ray binaries and from the central AGN, both modelled with a power-law component, are negligible for this source.

\subsubsection{The \texttt{clus} model (XRISM/Resolve)}\label{sec:methods:spectra:clus}

Along with all other spectral models, we also measure abundances using the \texttt{clus} model in SPEX \citep{stofanova2025}\footnote{\href{https://spex-xray.github.io/spex-help/models/clus.html}{https://spex-xray.github.io/spex-help/models/clus.html}}. The \texttt{clus} model simulates a spherically symmetric model of a cluster given any radial 3D density, temperature and abundance profile. Once provided with these profiles, SPEX calculates a \texttt{cie} model in each 3D shell and projects all emission in the plane of the sky. As a next step, observational quantities are calculated as spectral components between any projected radii by adding up the contribution from each 3D shell. The \texttt{clus} model spectrum can then be redshifted and absorbed through the galactic neutral hydrogen column. We use all available \textit{Chandra} observations in the data archive to derive the radial density, temperature and abundance profiles from the 2--8~keV spectrum (see \citealt{majumder2025} for more details). This energy range discards any emission from an excess cooler gas component at the centre with potentially a different metallicity (see \citealt{xrism2025,majumder2025} and Sect.~\ref{sec:discussion}). The spectrum is then calculated over the entire Resolve field of view. Once the \texttt{clus} model is set up using \textit{Chandra} profiles, it is possible to fit the Resolve spectrum to independently obtain velocity dispersion, redshift and the relative abundances of elements like Si, S, Ar, Ca, Cr, Mn, and Ni with respect to Fe. The \texttt{clus} model parameters are detailed in Sect.~\ref{sec:app:clus}.

\subsubsection{XMM-Newton/RGS}\label{sec:methods:spectra:rgs}

Because the instrument is slitless and gratings are dispersive by nature, the spectral information delivered by RGS convolves with the spatial extent of the observed source. While this is not an issue for point-like sources, the effect on extended sources such as Centaurus is essentially an instrumental broadening $\Delta \lambda$ of the lines as 
\begin{equation}
        \Delta \lambda~(\AA)= \frac{0.138}{m} \Delta \theta~(\mathrm{arcmin}),
\end{equation}
where $m$ is the spectral order and $\Delta \theta$ is the offset angle of the source. This effect needs to be accounted for during the spectral fitting. In SPEX, this is done via the \texttt{rgsvprof} task, which takes as input an image of the source and recreates a surface brightness profile projected along the dispersion direction and which can then be convolved with any SPEX model through the \texttt{lpro} component. For instrument consistency, we use an EPIC MOS\,1 image of our unique \XMM ObsID (Sect.~\ref{sec:methods:data:rgs}). Through the RGS analysis, our baseline model 2T$_\mathrm{SPEX}$ thus becomes \texttt{red*hot*lpro*(cie+cie)} where the $\Delta \lambda$ and offset parameters, though close to their expected values (1.0 and 0, respectively) are left free to account for residual effects. The redshift is fixed to that of NGC\,4696 ($z = 0.0094$) while the velocity dispersion is set to be consistent with that of the cooler Resolve component of our 2T$_\mathrm{SPEX}$ model (Table~\ref{table:results_Resolve}). 

In order to test our RGS abundance measurements to various systematic effects we first perform four experiments, each following one combination of the two prescriptions below:
\begin{enumerate}
        \item The cross-dispersion aperture is set to either 0.8$'$ or 3$'$ (Fig.~\ref{fig:images}) -- the latter being roughly comparable to the Resolve array;
        \item The first order spectrum is fitted either on its own or simultaneously with the second order spectra.
\end{enumerate}

Since abundances in the RGS band can also be sensitive to the choice of $n_\mathrm{H}$ and of the atomic codes, we add two extra experiments: one with the $n_\mathrm{H}$ parameter left free and the other using Xspec instead of SPEX (still within the 2T modelling approach). In the latter case, the instrumental broadening effect is accounted for via the \texttt{rgsxsrc} model in Xspec, in which we inject the same input EPIC MOS\,1 image as in the SPEX \texttt{rgsvprof} convolution tool (see above).

In all our fits, the RGS\,1 and RGS\,2 spectra are kept separated and fitted simultaneously. Because it allows a better control of parameters and systematics, this approach is preferred over that of combining the two spectra (and responses) in one -- usually done via the SAS task \texttt{rgscombine} as sometimes mentioned in the literature. The comparison between our simultaneous fitting approach and \texttt{rgscombine} is shown in Fig.~\ref{fig:rgscombine_app} (Appendix~\ref{sec:app:rgscombine}).

Last but not least, the background is obtained via the SAS task \texttt{rgsbkgmodel} from a series of previous blank sky observations and is subtracted to our spectra before fitting. We show in a companion paper that modelling the background directly in our RGS fit provides self-consistent abundance ratios \textcolor{blue}{(Kondo et al. to be subm)}.

\section{Results}\label{sec:results}

\subsection{Resolve abundances}\label{sec:results:resolve}

\begin{figure*}[t!]
\centering
\includegraphics[width=0.375\hsize, trim={0cm 0.2cm 0cm 0.3cm},clip]{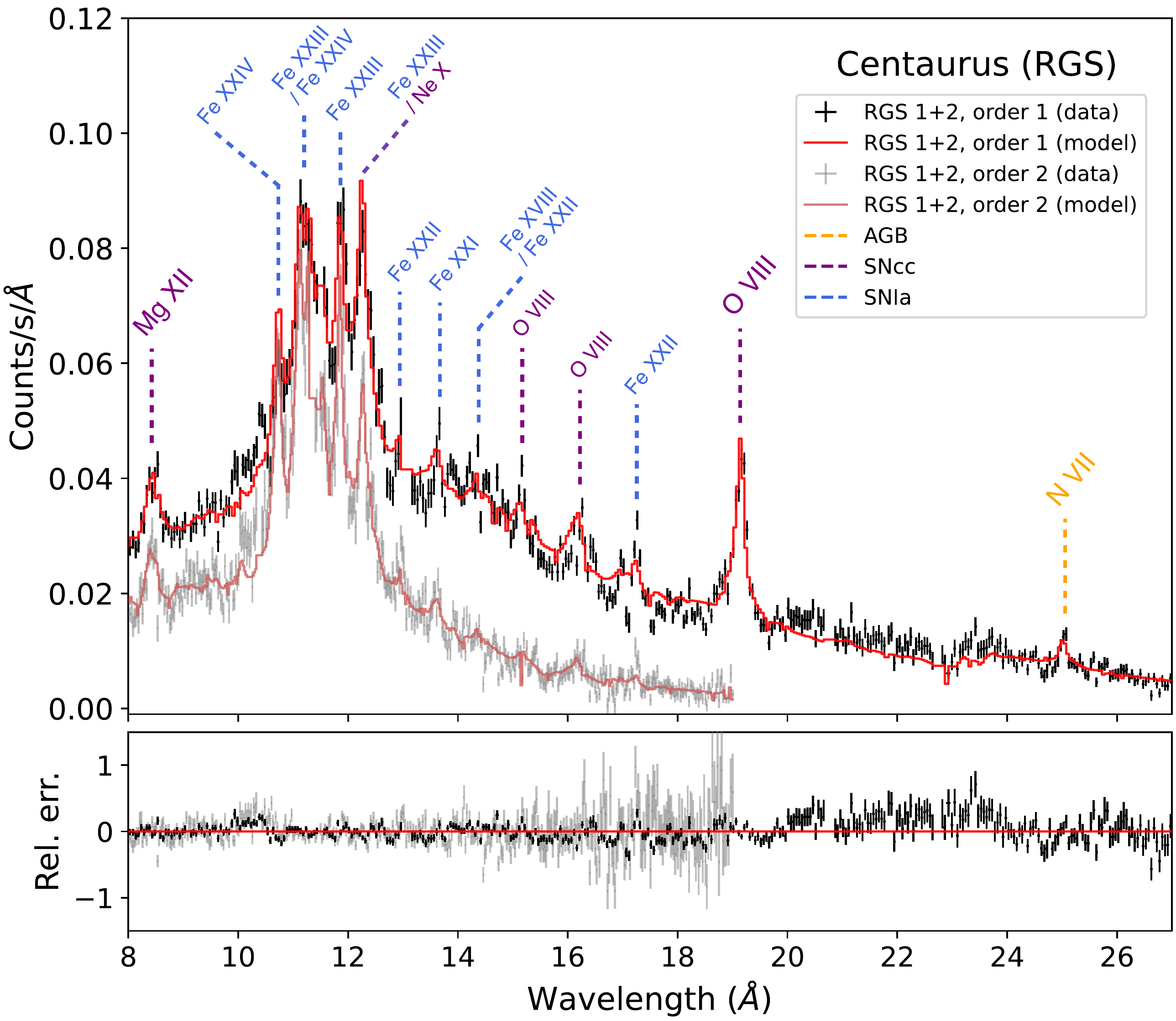}
\includegraphics[width=0.6190\hsize, trim={0cm 0.9cm 0cm 0.9cm},clip]{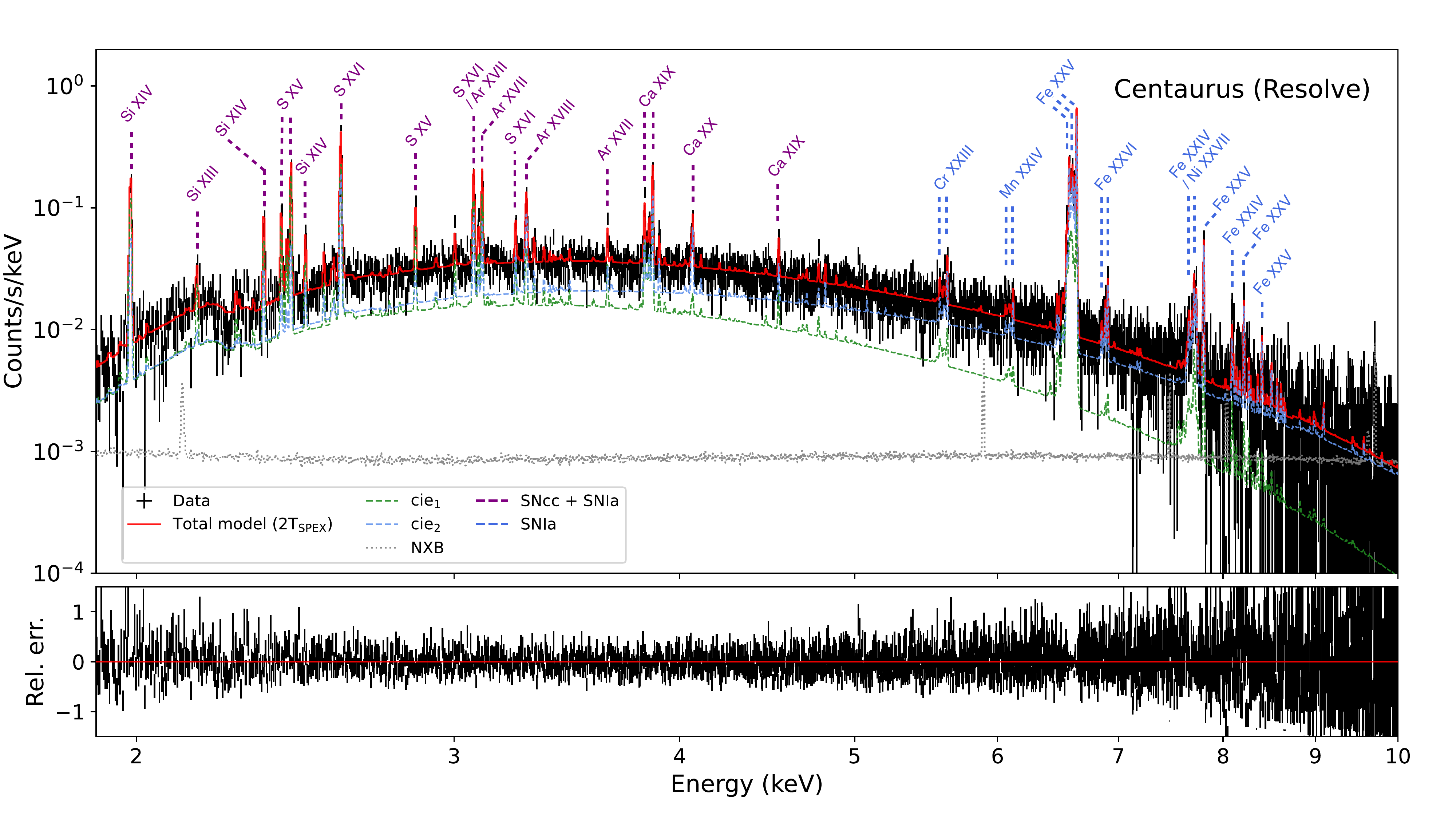}
      \caption{\textit{Left:} \XMM/RGS first and second order spectra extracted with a 3$'$ cross-dispersion aperture (see Fig.~\ref{fig:images}). The RGS\,1 and RGS\,2 data are combined for clarity. Each order is fitted separately with a 2T$_\mathrm{SPEX}$ model. Channels with significantly lower counts (originating from gaps in the detector plane) are accounted for in the fits but ignored in the plot for clarity. \textit{Right:} \XRISM/Resolve full-array spectrum fitted with a 2T$_\mathrm{SPEX}$ model (Sect.~\ref{sec:methods:spectra:resolve}) with the two thermal components and the NXB shown separately. A zoomed-in version of this plot is displayed in Fig.~\ref{fig:spectra_app}. We also show the most prominent emission lines, colour-coded by their astrophysical origin: AGB (left panel only), SNcc and SNIa (both panels).
          }
         \label{fig:spectra}
\end{figure*}

Our RGS (see Sect.~\ref{sec:results:rgs}) and Resolve (this section) spectra are shown along with our best-fit baseline models in, respectively, the left and right panels of Figure~\ref{fig:spectra}. A zoomed-in version of the Resolve spectrum and best-fit model, available in Fig.~\ref{fig:narrowfits_app} (Appendix~\ref{sec:app:zoomed}), shows the plethora of H-like and He-like emission lines resolved at $\sim$5~eV energy resolution. 

As discussed in Sect.~\ref{sec:methods:spectra:resolve}, in addition to our baseline model (2T$_\mathrm{SPEX}$) other modelling strategies are also considered. Table~\ref{table:results_Resolve} lists the best-fit values and 1$\sigma$ errors obtained through our full-array analysis considering every case. In both SPEXACT and AtomDB we find similar C-stat/d.o.f. estimates for different multi-temperature approaches, indicating that none is strongly preferred by the data. On the other hand, the single-temperature fits provide systematically higher C-stat/d.o.f. Given these considerations, the choice of 2T$_\mathrm{SPEX}$ as baseline model ensures a simple, yet physically and statistically realistic modelling approach. For detailed work on the temperature structure of Centaurus at high spectral resolution, see \textcolor{blue}{Kondo et al. (to be subm)}. The bulk velocities ($v_\mathrm{bulk}$) and velocity broadening values ($v_\mathrm{rms}$) are of high interest for the community and we note that they are consistent with the dedicated sub-array study of \citet[][]{xrism2025}, in which these values are discussed in detail. The temperatures are comparable to those reported at similar radial distance in previous work \citep[e.g.][]{matsushita2007,gatuzz2023} and we note that the 2T modelling approaches give lower and upper temperature ranges of, respectively, 1.3--1.7~keV and 2.8--3.1~keV. These temperatures will be of interest when compared with our RGS best-fit values (Sect.~\ref{sec:results:rgs}).

\begin{figure}[h!]
\centering
\includegraphics[width=0.49\textwidth, trim={1.0cm 0cm 1.5cm 0cm},clip]{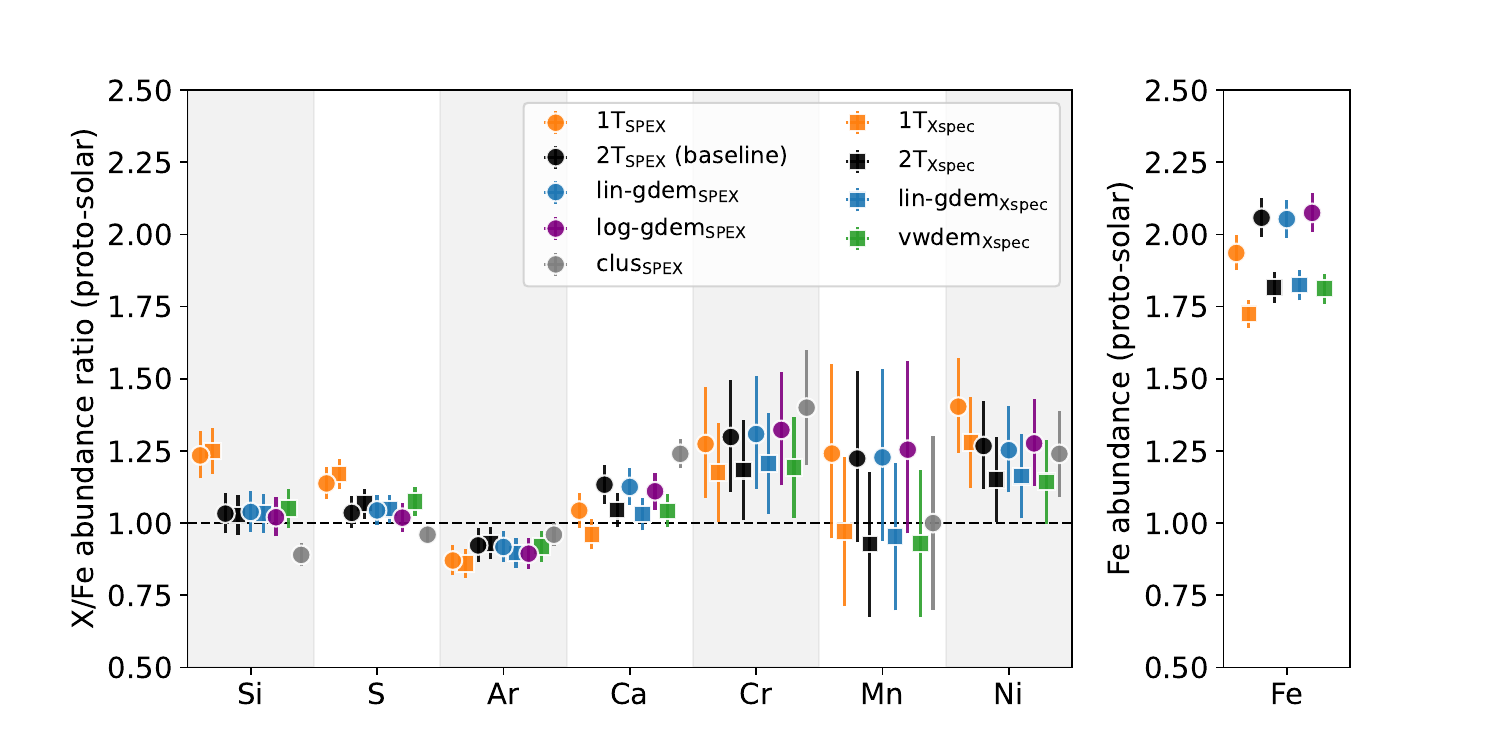} 

      \caption{Best-fit parameters of X/Fe abundance ratios (left panel) and absolute Fe abundance (right panel) from our Resolve full-array analysis. As explained in Sect.~\ref{sec:methods:data:resolve}, our fits are performed assuming different (multi-) temperature modelling and atomic databases. Our baseline model is 2T$_\mathrm{SPEX}$ (black circles).
      }
         \label{fig:broadfits}
\end{figure}

Table~\ref{table:results_Resolve} also lists our best-fit absolute Fe abundance and associated X/Fe ratios. These numbers are plotted in Fig.~\ref{fig:broadfits} and show remarkable consistency among the different spectral codes and our multi-temperature modelling approaches. The 1T models, however, exhibit slight but significant deviations in Fe, Si/Fe, S/Fe, and to some extent Ar/Fe and Ca/Fe. The larger deviations towards lighter elements are likely attributed to the low effective area near 2--3~keV, where the fit allows larger deviations from the data. We interpret this as the need for a cooler gas component which cannot be accounted for in the 1T cases. This incorrect modelling of the temperature structure converts the line equivalent widths into incorrect abundances. This illustrates the importance of modelling temperatures correctly even in high-resolution spectra to avoid biased ICM abundance measurements. The \texttt{clus} model also gives us interesting clues about the multi-temperature state of the gas. As seen from the results listed in Table.~\ref{table:results_clus} (see also Fig.~\ref{fig:broadfits}), we note an excellent consistency of their X/Fe ratios with those from multi-temperature classical (SPEX and Xspec) models above. One interesting tension, however, is on Si/Fe. Because the \texttt{clus} model is multi-temperature by projection only (i.e. all local 3D shells are assumed to be single-temperature) and because the Si lines tend to trace cooler gas, this tension may be seen as a hint for an intrinsic, \textit{in situ} cooler gas component. This is in line with the reduced C-stat value seen to be slightly lower in the 2T$_\mathrm{SPEX}$ model than in the \texttt{clus} model.

\begin{table*}[h!]
\tiny
\caption{Best-fit values and 1$\sigma$ errors of the Resolve pointing on the Centaurus core.}
\label{table:results_Resolve}
\centering
\begin{tabular}{l | c c c c | c c c c}
\hline\hline
 & \multicolumn{4}{c|}{SPEXACT v3.08.01} & \multicolumn{4}{c}{AtomDB v3.0.9} \\
 & 1T$_\mathrm{SPEX}$ & 2T$_\mathrm{SPEX}^{(*)}$ & lin-gdem$_\mathrm{SPEX}$ & log-gdem$_\mathrm{SPEX}$ & 1T$_\mathrm{Xspec}$ & 2T$_\mathrm{Xspec}$ & lin-gdem$_\mathrm{Xspec}$ & wdem$_\mathrm{Xspec}$ \\         
\hline

EM$_1$ / norm$_1$               &       $4.52 \pm 0.08$ &       $2.5_{-0.5}^{+0.3}$     &       $4.99 \pm 0.14$       &       $4.8 \pm 0.09$  &       $2.06 \pm 0.03$ &       $0.89 \pm 0.14$       &       $2.27 \pm 0.05$ &       $2.33 \pm 0.06$ \\
EM$_2$ / norm$_2$               &       $-$     &       $2.2_{-0.3}^{+0.5}$     &       $-$     &       $-$     &       $-$     &       $1.29 \pm 0.15$       &       $-$     &       $-$     \\
$kT_1$ / $kT_\mathrm{mean}$     / $kT_W$ &      $2.333 \pm 0.018$       &       $1.64_{-0.13}^{+0.07}$  &       $2.18 \pm 0.04$       &       $2.15 \pm 0.03$ &       $2.397 \pm 0.017$       &       $1.49 \pm 0.11$       &       $2.24 \pm 0.04$ &       $3.32\pm 0.07$  \\
$kT_2$  &       $-$     &       $2.99_{-0.16}^{+0.10}$  &       $-$     &       $-$     &       $-$     &       $2.88 \pm 0.10$       &       $-$     &       $-$     \\
$\sigma_{\mathrm{lin, }T}$ / $\sigma_{\mathrm{log, }T}$ / $\alpha_{T}$  &       $-$     &       $-$     &       $1.11 \pm 0.07$       &       $0.194 \pm 0.011$       &       $-$     &       $-$     &       $1.13 \pm 0.07$       &       $1.3 \pm 0.3$   \\
$v_\mathrm{bulk, 1}$    &       $-183 \pm 5$    &       $-193 \pm 17$   &       $-185 \pm 5$  &       $-185 \pm 5$    &       $-191 \pm 4$    &       $-205 \pm 15$     &       $-192 \pm 5$    &       $-192 \pm 5$    \\
$v_\mathrm{rms, 1}$     &       $126 \pm 7$     &       $154 \pm 21$    &       $130 \pm 6$  &       $130 \pm 7$     &       $113 \pm 6$     &       $158 \pm 20$     &       $114 \pm 6$     &       $114 \pm 6$     \\
$v_\mathrm{bulk, 2}$    &       $-$     &       $-180 \pm 9$    &       $-$     &       $-$     &       $-$     &       $-187 \pm 6$  &       $-$     &       $-$     \\
$v_\mathrm{rms, 2}$     &       $-$     &       $121 \pm 12$    &       $-$     &       $-$     &       $-$     &       $103 \pm 8$  &       $-$     &       $-$     \\
Fe              &       $1.94 \pm 0.06$ &       $2.06 \pm 0.07$ &       $2.05 \pm 0.07$       &       $2.07 \pm 0.07$ &       $1.72 \pm 0.05$ &       $1.82 \pm 0.05$       &       $1.82 \pm 0.05$ &       $1.81 \pm 0.05$ \\
Si/Fe   &       $1.23 \pm 0.08$ &       $1.03 \pm 0.07$ &       $1.04 \pm 0.07$   &       $1.02 \pm 0.07$ &       $1.25 \pm 0.08$ &       $1.03 \pm 0.07$   &       $1.03 \pm 0.07$ &       $1.05 \pm 0.07$ \\
S/Fe    &       $1.14 \pm 0.06$ &       $1.04 \pm 0.06$ &       $1.04 \pm 0.05$   &       $1.02 \pm 0.05$ &       $1.17 \pm 0.05$ &       $1.07 \pm 0.05$   &       $1.05 \pm 0.05$ &       $1.08 \pm 0.05$ \\
Ar/Fe   &       $0.87 \pm 0.05$ &       $0.92 \pm 0.06$ &       $0.92 \pm 0.06$   &       $0.89 \pm 0.05$ &       $0.86 \pm 0.05$ &       $0.93 \pm 0.06$   &       $0.90 \pm 0.05$ &       $0.92 \pm 0.05$ \\
Ca/Fe   &       $1.04 \pm 0.06$ &       $1.13 \pm 0.07$ &       $1.13 \pm 0.07$   &       $1.11 \pm 0.06$ &       $0.96 \pm 0.05$ &       $1.04 \pm 0.06$   &       $1.03 \pm 0.06$ &       $1.04 \pm 0.06$ \\
Cr/Fe   &       $1.27 \pm 0.20$ &       $1.30 \pm 0.20$ &       $1.31 \pm 0.20$   &       $1.32 \pm 0.20$ &       $1.18 \pm 0.17$ &       $1.18 \pm 0.17$   &       $1.21 \pm 0.17$ &       $1.19 \pm 0.17$ \\
Mn/Fe   &       $1.2 \pm 0.3    $       &       $1.2 \pm 0.3$   &       $1.2 \pm 0.3$        &       $1.2 \pm 0.3$   &       $1.0 \pm 0.3$   &       $0.9 \pm 0.3$        &       $1.0 \pm 0.3$   &       $0.9 \pm 0.3    $\\
Ni/Fe   &       $1.40 \pm 0.17$ &       $1.27 \pm 0.16$ &       $1.25 \pm 0.15$   &       $1.28 \pm 0.16$ &       $1.28 \pm 0.16$ &       $1.15 \pm 0.15$   &       $1.16 \pm 0.15$ &       $1.14 \pm 0.14$ \\ \hline
C-stat/d.o.f.   &       $2967/2602$     &       $2822/2598$     &       $2820/2601$     &       $2824/2601      $       &       $22108/26123$   &       $21948/26119$   &       $21956/26122$   &       $21953/26122$   \\

\hline
\end{tabular}
\tablefoot{Measurements are derived from a broad-band fit, for different atomic codes (SPEXACT and AtomDB, respectively available in SPEX and Xspec) and (single-/multi-) temperature models. The SPEX emission measures are given in units of $10^{71}/\mathrm{m}^3$ and the Xspec normalisations are given in units of $10^{-2}/\mathrm{cm}^5$. The temperatures $kT_1$, $kT_2$, $kT_\mathrm{mean}$, and $kT_W$ are given in keV. The velocities $v_\mathrm{bulk, 1}$, $v_\mathrm{rms, 1}$, $v_\mathrm{bulk, 2}$, and $v_\mathrm{rms, 2}$ are given in km/s. The Fe abundance and X/Fe ratios are given in solar units \citep{lodders2009}. $^{(*)}$ The 2T model using SPEXACT constitutes our baseline Resolve model.
}
\end{table*}

\begin{table}[h!]
\caption{Best-fit X/Fe ratios obtained for the SPEX \texttt{clus} model.}
\label{table:results_clus}
\centering
\begin{tabular}{l c}
\hline\hline
\multicolumn{2}{c}{SPEXACT v3.08.01} \\
\multicolumn{2}{c}{\texttt{clus}} \\
\hline
Si/Fe   &       $0.89 \pm 0.04$ \\
S/Fe    &       $0.96 \pm 0.03$ \\
Ar/Fe   &       $0.96 \pm 0.04$ \\
Ca/Fe   &       $1.24 \pm 0.05$ \\
Cr/Fe   &       $1.44 \pm 0.20$ \\
Mn/Fe   &       $1.0 \pm 0.3$   \\
Ni/Fe   &       $1.24 \pm 0.15$ \\
\hline
C-stat/d.o.f.   &       3004/2708       \\

\hline
\end{tabular}
\end{table}

\begin{figure*}[h!]
\sidecaption
\centering
\includegraphics[width=12cm, trim={3.0cm 0.7cm 3.7cm 0.9cm},clip]{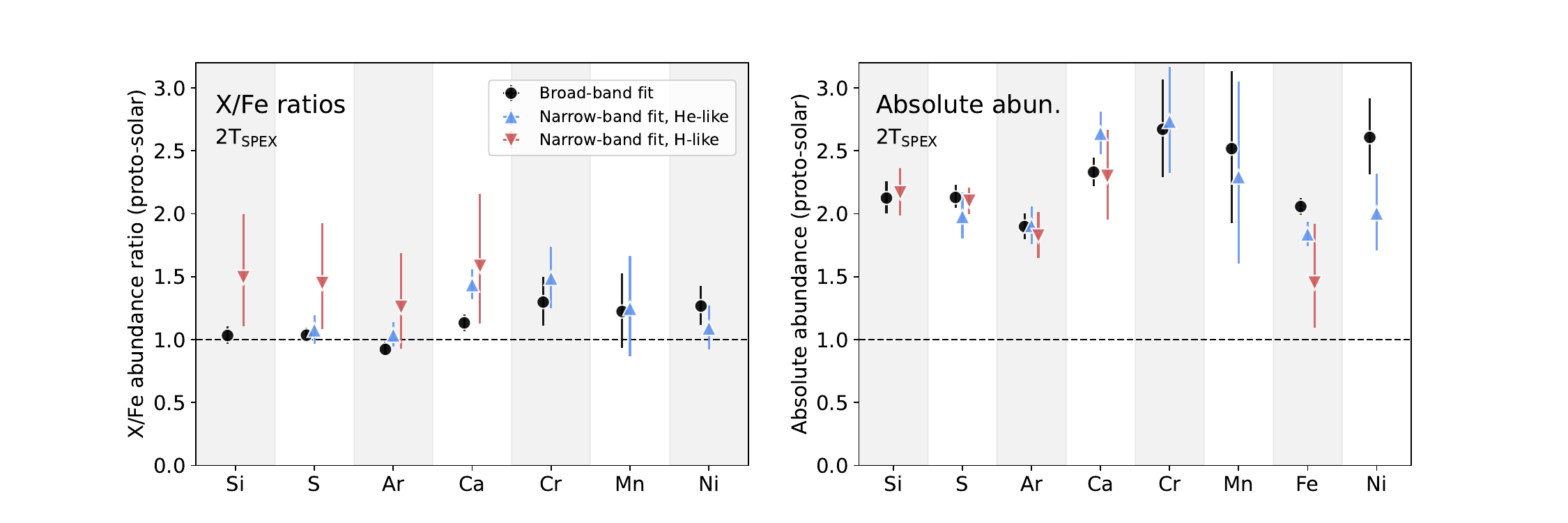} 
      \caption{Comparison between the abundances estimated from our baseline (2T$_\mathrm{SPEX}$) broad-band fit and from their narrow-band fit equivalents (i.e. fitted locally around their H-like and He-like lines; see Sect.~\ref{sec:methods:spectra:resolve} and Fig.~\ref{fig:narrowfits_app}). The left panel shows the relative X/Fe ratios while the right panel focuses on the absolute abundances.
      }
         \label{fig:broadvsnarrowfits}
\end{figure*}

As next step, Fig.~\ref{fig:broadvsnarrowfits} compares our baseline (2T$_\mathrm{SPEX}$) broad-band fit with its narrow-band fit equivalent around H-like and He-like ions. The X/Fe ratios (left panel) show overall good consistency between the three methods, however with a trend for higher estimates in the H-like based ratios. As shown in the right panel, this trend is caused by the absolute Fe abundance estimate itself with a slight (<2$\sigma$) tension appearing between broad-band and narrow-band measurements. The absolute abundances of the other elements show excellent consistency regardless of the energy band considered for the fit. One possible exception is Ca, where the He-like estimate slightly overpredicts the broad-band estimate. In turn, this increases the tension between the Ca/Fe ratio measurements (left panel), which we take into account in final ratios estimation as detailed in Sect.~\ref{sec:results:final}.

Finally, we also attempt to provide constraints on the absolute P, Cl, K, and Ti abundances. These elements have not been seen yet in the ICM, although their emission lines might in principle be detected at sufficiently high spectral resolution and statistics (Fig.~\ref{fig:spectra_app}). Using a 1T$_\mathrm{SPEX}$ for the sake of simplicity, we find $>$1$\sigma$ detections for P ($2.2\pm 2.1$~solar) and K ($1.3 \pm 1.0$~solar) and upper limits for Cl ($<1.6$~solar) and Ti ($<0.8$~solar). Future deeper observations of Centaurus could in principle help to refine further these constraints and possibly confirm a few detections.

\subsection{RGS abundances}\label{sec:results:rgs}

\begin{figure}[h!]
\centering
\includegraphics[width=0.49\textwidth, trim={1.5cm 0cm 1.5cm 0cm},clip]{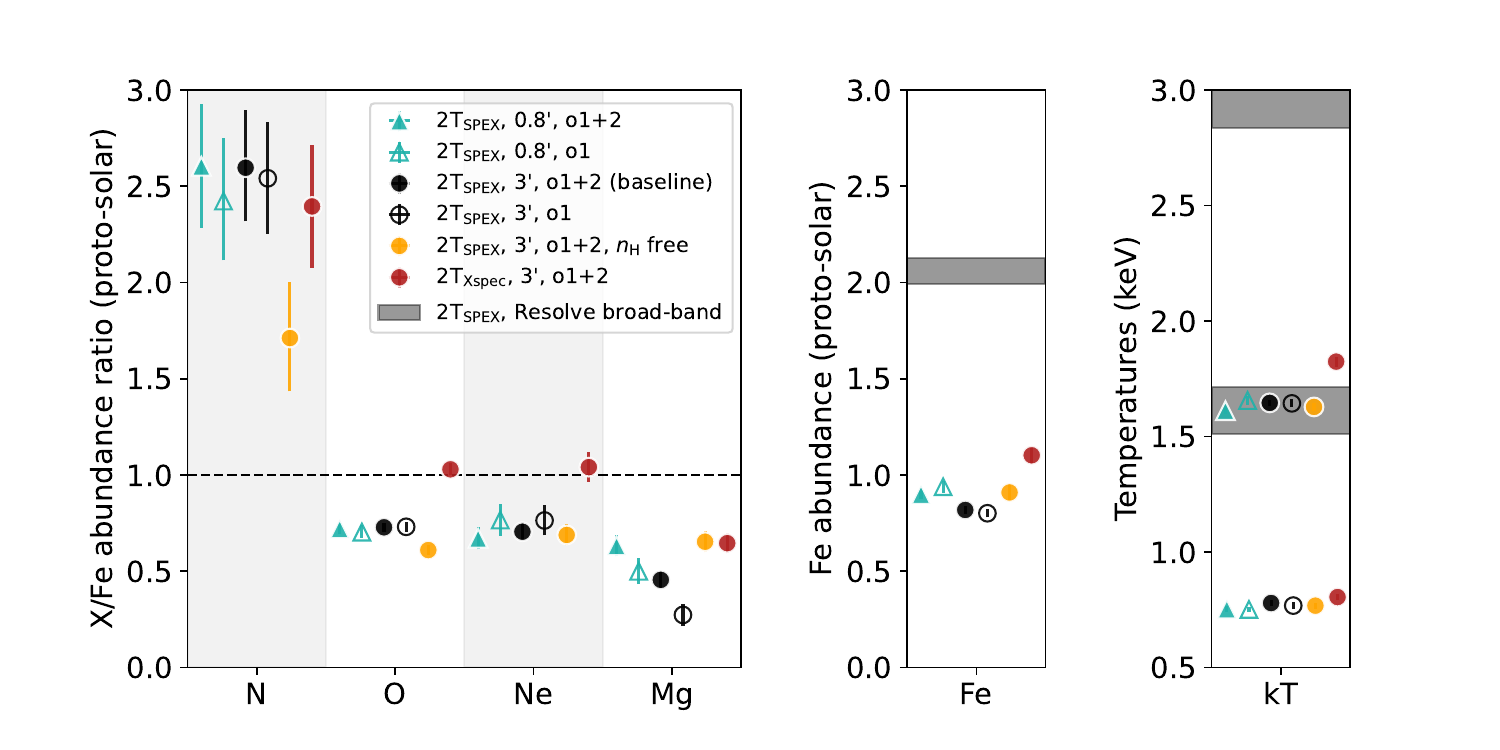} 

      \caption{Best-fit parameters for X/Fe ratios (left panel), the absolute Fe abundance (middle panel), and gas temperatures (right panel) from our RGS analysis. To account for a number of systematics, our fits are done following four different approaches (see Sect.~\ref{sec:methods:data:rgs}). For comparison, the Resolve best-fit parameters of Fe and gas temperatures are shown.
      }
         \label{fig:RGSratios}
\end{figure}

The best-fit results estimated from our RGS analysis (2T$_\mathrm{SPEX}$ model) are displayed in Fig.~\ref{fig:RGSratios} with numbers listed in Table~\ref{table:results_RGS}. As detailed in Sect.~\ref{sec:methods:spectra:rgs}, the figure compares four experiments, namely fitting spectra over 0.8$'$ or 3.0$'$ aperture on one hand and with the first order or both orders on the other hand. To maximise statistics and spatial coverage with the Resolve fits, we choose our RGS baseline model as fitting both orders over 3.0$'$ (black circles on the figure).

One immediate result to notice is on the derived temperatures (right panel): the lower temperature found in Resolve is in remarkable agreement with the higher temperature found in RGS. All in all, this strongly suggests that any combined RGS+Resolve analysis is well described by a combination of three independent temperatures. This further motivates our choice of a 2T fit as baseline model for this work. Because fitting RGS and Resolve spectra independently maintains this excellent consistency of the intermediate temperature, we choose to keep these data separated to avoid possible complications resulting from instrumental cross-calibration imperfections. A combined RGS+Resolve fit is performed and discussed thoroughly in \textcolor{blue}{Kondo et al. (to be subm)}. 

A more puzzling result, however, is that the absolute Fe abundance measured with RGS is systematically found to be a factor $\sim$2 lower than its Resolve counterpart (middle panel). While this tension is discussed further in Sect.~\ref{sec:discussion:absFe}, we note that its direct relevance to the scope of this paper (which focuses primarily on the abundance ratios) is rather limited. In our RGS analysis, ratios can be safely isolated by defining the reference element to Fe instead of H (through the \texttt{ref} parameter in SPEX), hence making the continuum dependency of our model negligible for these relative quantities. As shown in the left panel of Fig.~\ref{fig:RGSratios}, our six experiments essentially point towards a significantly super-solar N/Fe ratio. Depending on the atomic code, O/Fe and Ne/Fe are found to be either  slightly sub-solar or solar. The Mg/Fe ratio, on the other hand, is always found to be sub-solar and shows a more severe dependency on the choice of the order(s) and/or the region considered. Due to the relative lack of variation in the absolute Fe abundance with the selected region and spectral orders, this effect is to be attributed to the \ion{Mg}{xii} line itself (and the modelling of its instrumental broadening) rather than to Fe lines in the spectra. All these sources of scatter are incorporated in the estimate of our final ratios (see next section).

\begin{table*}[h!]
\caption{Best-fit values and 1$\sigma$ errors of the RGS fits on the Centaurus core for our four different experiments.} 
\label{table:results_RGS}    
\centering                        
\begin{tabular}{l @{\hspace{15pt}}| @{\hspace{15pt}}c  @{\hspace{15pt}}c @{\hspace{15pt}}| @{\hspace{15pt}}c @{\hspace{15pt}}c @{\hspace{15pt}}c @{\hspace{15pt}}c}      
\hline\hline               
\multicolumn{7}{c}{2T}\\ \hline
  & \multicolumn{2}{c|@{\hspace{15pt}}}{0.8 arcmin}     & \multicolumn{4}{c}{3.0 arcmin} \\ 
 & orders 1+2   &       order 1                 &       orders 1+2$^{(*)}$      &       order 1       &       $n_\mathrm{H}$ free     (o1+2)  &       Xspec (o1+2)\\         
\hline

EM$_1$ / norm$_1$               &       $0.156 \pm 0.005$       &       $0.157 \pm 0.005$      &       $0.271 \pm 0.012$       &       $0.277 \pm 0.013$       &       $0.16 \pm 0.005$      &               $0.133 \pm 0.004$\\
EM$_2$ / norm$_2$               &       $2.92 \pm 0.04$ &       $2.93 \pm 0.04$   &       $6.93 \pm 0.07$ &       $6.92 \pm 0.07$ &       $2.94 \pm 0.08$   &               $3.10 \pm 0.04$\\
$kT_1$  &       $0.753 \pm 0.012$       &       $0.752 \pm 0.012$       &       $0.779 \pm 0.014$      &       $0.768 \pm 0.016$       &       $0.768 \pm 0.013$       &               $0.805 \pm 0.014$\\
$kT_2$  &       $1.612 \pm 0.015$       &       $1.656 \pm 0.019$       &       $1.646 \pm 0.014$      &       $1.643 \pm 0.017$       &       $1.628 \pm 0.019$       &               $1.824 \pm 0.021$\\
Fe              &       $0.90 \pm 0.03$         &       $0.94 \pm 0.03$ &       $0.818 \pm 0.019$      &       $0.801 \pm 0.021$       &       $0.90 \pm 0.04$ &               $1.10 \pm 0.04$\\
N/Fe            &       $2.6 \pm 0.3$   &       $2.4 \pm 0.3$   &       $2.6 \pm 0.3$        &       $2.5 \pm 0.3$   &       $1.7 \pm 0.3$   &               $2.4 \pm 0.3$\\
O/Fe            &       $0.72 \pm 0.03$ &       $0.70 \pm 0.04$ &       $0.73 \pm 0.03$       &       $0.73 \pm 0.03$ &       $0.61 \pm 0.03$ &               $1.03 \pm 0.03$\\
Ne/Fe           &       $0.67 \pm 0.06$ &       $0.77 \pm 0.08$ &       $0.71 \pm 0.05$       &       $0.76 \pm 0.08$ &       $0.69 \pm 0.06$ &               $1.04 \pm 0.08$\\
Mg/Fe           &       $0.63 \pm 0.05$ &       $0.50 \pm 0.07$ &       $0.46 \pm 0.05$       &       $0.27 \pm 0.06$ &       $0.65 \pm 0.06$ &               $0.65 \pm 0.05$\\ \hline
C-stat/d.o.f.   &       $1826/1302$     &       $883/570$       &       $2012/1302$     &       $930/570$       &       $1821/1301$     &               $7216/6551$\\

\hline
\end{tabular}
\tablefoot{The SPEX emission measures are given in units of $10^{71}/\mathrm{m}^3$ and the Xspec normalisations are given in units of $10^{-2}/\mathrm{cm}^5$. The temperatures $kT_1$ and $kT_2$ are given in keV. The Fe abundance and X/Fe ratios are given in solar units \citep{lodders2009}. $^{(*)}$ The fit extracted over 3$'$ cross-dispersion aperture and including the two orders constitutes our baseline RGS model.
}
\end{table*}

\subsection{Final ratios}\label{sec:results:final}

The encouraging convergence of the X/Fe abundance ratios derived with RGS and, especially, Resolve through various approaches and assumptions constitutes a solid motivation for estimating a set of `final' ratios, with the aim of capturing robustly and conservatively the abundance pattern of the Centaurus core. Admittedly, such an estimation is somewhat arbitrary as it can be done in various ways; our results described above, however, motivate us to proceed as follows. First, we start by adopting the best-fit values $\mu^\mathrm{bl}_\mathrm{X/Fe}$ of our baseline (RGS and Resolve) models as mean values $\mu^\mathrm{final}_\mathrm{X/Fe}$ of our final ratio estimates. Then, we define the final errors $\sigma^\mathrm{final}_\mathrm{X/Fe}$ on each ratio such that, added in quadrature to the values of any experiment $i$ and $j$, their enlarged measurements $\mu^i_\mathrm{X/Fe} \pm \sqrt{ (\sigma^i_\mathrm{X/Fe})^2 + (\sigma^\mathrm{final}_\mathrm{X/Fe})^2 }$ and $\mu^j_\mathrm{X/Fe} \pm \sqrt{ (\sigma^j_\mathrm{X/Fe})^2 + (\sigma^\mathrm{final}_\mathrm{X/Fe})^2 }$ are $\le$1$\sigma$ consistent with each other.

\begin{figure*}[h!]
\sidecaption
\centering
\includegraphics[width=12cm, trim={2.1cm 0.7cm 3.0cm 1.5cm},clip]{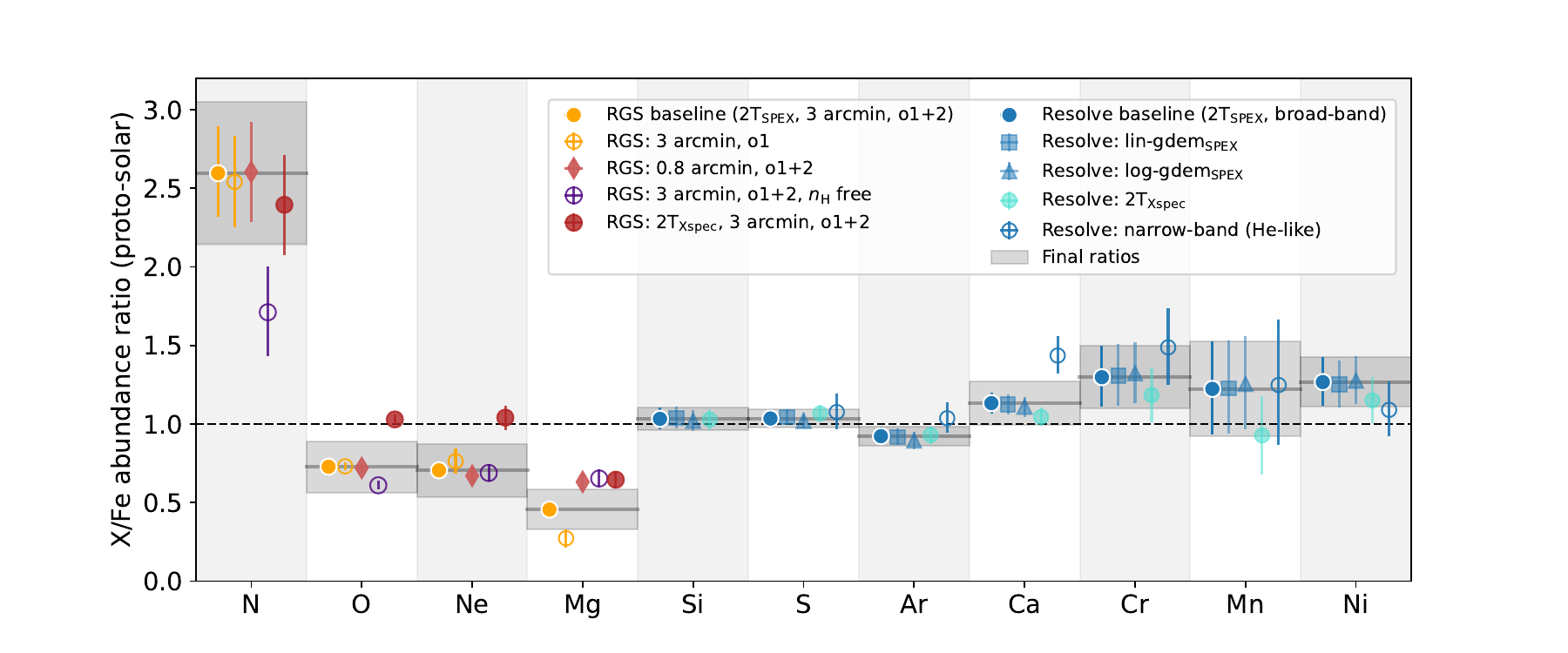} 
      \caption{Our final RGS+Resolve X/Fe abundance ratios (grey areas), shown along with individual measurements of various fitting experiments described in this paper. The procedure we used to estimate them are described in Sect.~\ref{sec:results:final}.
      }
         \label{fig:bestratios}
\end{figure*}

These final ratios are listed in Table~\ref{table:final_ratios} and are shown in Fig.~\ref{fig:bestratios} along with the most relevant individual measurements that were used in our estimate. This conservative estimate of the chemical composition of the Centaurus core is discussed in light of previous work in Sect.~\ref{sec:discussion:abunpattern}.

\begin{table}[h!]
\caption{Our final RGS+Resolve X/Fe abundance ratios with their total uncertainties.}
\label{table:final_ratios}
\centering
\begin{tabular}{l r c}
\hline\hline
Instrument      &       Ratio   &       Final value     \\
\hline
RGS     &       N/Fe    &       $2.6 \pm 0.5$   \\
        &       O/Fe    &       $0.73 \pm 0.16$ \\
        &       Ne/Fe   &       $0.71 \pm 0.17$ \\
        &       Mg/Fe   &       $0.46 \pm 0.13$ \\ \hline
Resolve &       Si/Fe   &       $1.03 \pm 0.07$ \\
        &       S/Fe            &       $1.04 \pm 0.06$ \\
        &       Ar/Fe   &       $0.92 \pm 0.06$ \\
        &       Ca/Fe   &       $1.13 \pm 0.14$ \\
        &       Cr/Fe   &       $1.30 \pm 0.20$ \\
        &       Mn/Fe   &       $1.2 \pm 0.3$   \\
        &       Ni/Fe   &       $1.27 \pm 0.16$ \\

\hline
\end{tabular}
\tablefoot{Values are in units of \citet{lodders2009} (see also Fig.~\ref{fig:bestratios}).
}
\end{table}

\section{Discussion}\label{sec:discussion}

\subsection{The absolute Fe abundance in the core of Centaurus}\label{sec:discussion:absFe}

Our RGS and Resolve spectral analyses of the Centaurus core indicate a coherent picture of the gas temperature structure; RGS tracking its cooler component, Resolve tracking its hotter component, overlapped with an intermediate component seen by both instruments at remarkably consistent temperature (Fig.~\ref{fig:RGSratios}). The Fe abundance, on the contrary, is largely at tension between RGS and Resolve: a near-solar estimate is found with RGS while Resolve suggests the gas to be twice more Fe-rich.

This peculiarity is not new as near-solar values of Fe measured with RGS in the very core of Centaurus have already been reported in the literature \citep[][]{sanders2008,deplaa2017,mao2019,fukushima2022}. Values among these studies may differ -- e.g. $1.22 \pm 0.05$ solar in \citet{deplaa2017} and $1.02 \pm 0.03$ in \citet{mao2019} -- as they are sensitive to the atomic code, the cross-dispersion aperture and other possible methodologies (e.g. the modelling of the instrumental broadening and of the background). 

\begin{figure}[h!]
\centering
\includegraphics[width=0.49\textwidth, trim={1.5cm 0cm 1.5cm 0cm},clip]{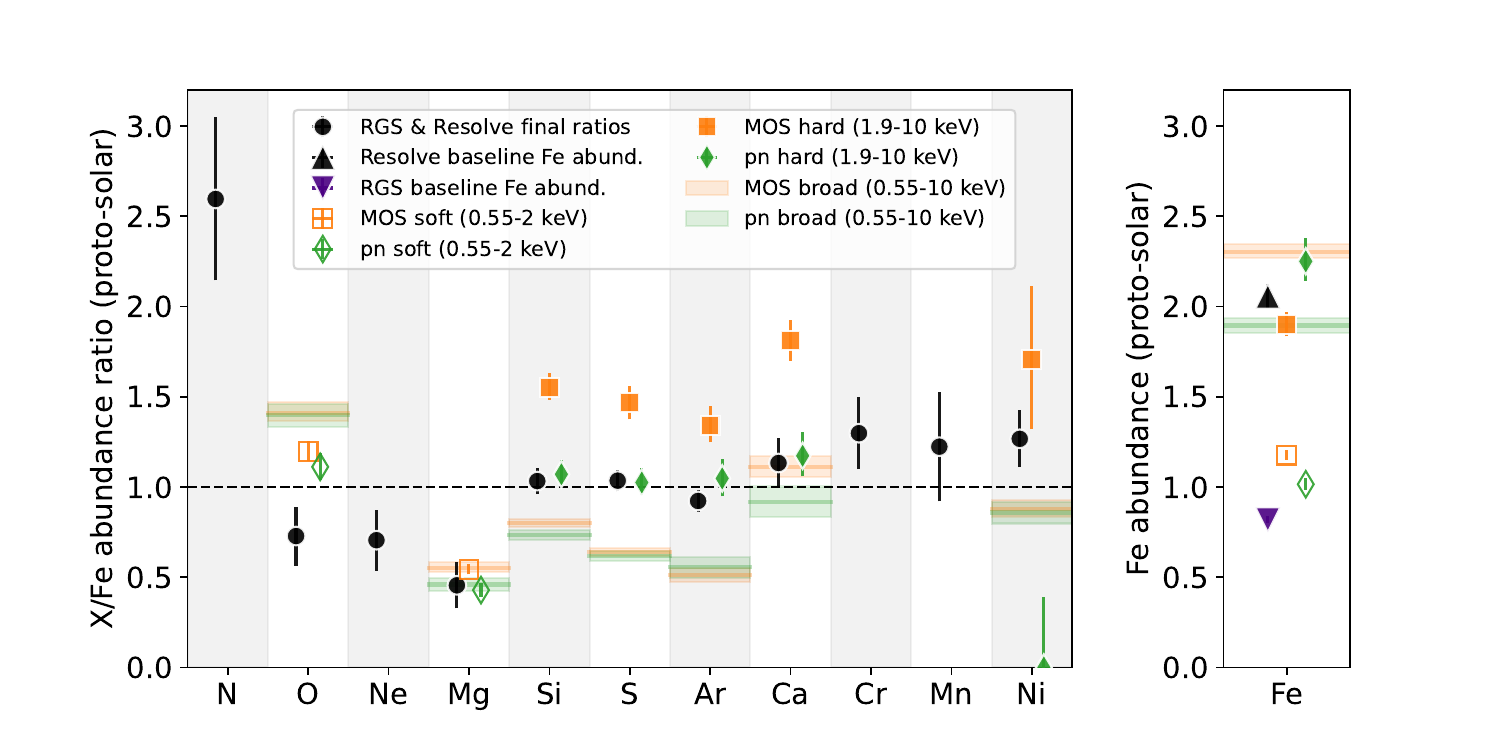}
      \caption{Final RGS+Resolve abundance measurements compared with \XMM/EPIC abundances fitted from the same Resolve extraction region. Three energy bands are considered: soft (0.55-2~keV, close to the RGS band), hard (1.9-10~keV, close to the Resolve band), and broad (0.55-10 keV). \textit{Left:} abundance ratios. \textit{Right:} absolute Fe abundance. 
      }
         \label{fig:comparCCD}
\end{figure}

On the other hand, the high Fe abundance derived with Resolve is comparable with previous values found at CCD-like spectral resolution when fitted over a broad energy band \citep{sanders2016,lakhchaura2019,fukushima2022}. To quantify this further, we compare our final high-resolution abundance pattern with estimates obtained at lower spectral resolution through a parallel \XMM/EPIC MOS\footnote{The MOS\,1 and MOS\,2 spectra are fitted jointly and are referred to as `MOS'.} and pn analysis of the same observation (extracted over the Resolve full-array sky region). For consistency with the energy bands offered by RGS and Resolve, we fitted our EPIC spectra with a 2T$_\mathrm{SPEX}$ model over (i) the full EPIC band (0.55-10~keV), (ii) a `hard', Resolve-like band (0.9-10~keV), and (iii) a `soft', RGS-like band (0.55-2~keV)\footnote{To minimise the effects due to imperfect calibration of the EPIC instrument, we choose not to fit our MOS and pn spectra below 0.55~keV.}. The results are shown in Fig.~\ref{fig:comparCCD}. Whereas the ratios (left panel) are discussed in detail in the next section, the right panel confirms that the broad- and hard-band fits find an Fe abundance of $\sim 2$--2.5~solar, very close to that of Resolve. Interestingly, the soft fits find considerably lower Fe abundances, comparable to those of our RGS fits. 

Admittedly, and as already reported in the literature \citep[e.g.][]{werner2006,deplaa2017}, measuring accurately absolute abundances with RGS is challenging. First, because by design RGS integrates the emission over the entire dispersion direction. Second, because even at high S/N the limited energy band combined with residual uncertainties on the instrumental line broadening makes it difficult to disentangle emission lines from the continuum. The dependencies of RGS abundances with $n_\mathrm{H}$ and the atomic codes add a layer of complexity, making our Fe measurement heavily sensitive to the fitting strategy (see also the measurements of  \textcolor{blue}{Kondo et al. to be subm}). We note, for instance, that \citet{fukushima2022} reported very similar fits of the Fe-L complex of NGC\,4696 (both in RGS and at lower resolution) for Fe abundances as different as 0.5 and 1.7~solar. The same authors also report diverging best-fit Fe values when various atomic codes are used to fit the spectra in those soft energies. Although disentangling and quantifying these sources of uncertainties are well beyond the scope of this work, it is not unlikely that one (or several) of the reasons above heavily biases our Fe abundance derived with RGS.

Nevertheless, the above-reported consistency with EPIC soft-band spectra rather suggests a RGS-Resolve discrepancy coming primarily from modelling biases between Fe-K and Fe-L transitions. While one possibility points at remaining atomic code uncertainties in the Fe-L complex, another possibility is that the lower Fe abundance measured in the soft band is genuine. If true, this would imply the coexistence of a cooler gas component onto NGC\,4696 (traced by RGS) being in fact less Fe-rich than the surrounding hotter ICM of the cluster itself (traced by Resolve). This possibility of a \emph{multi-metallicity} gas, with components of different temperatures associated to different gas metallicities, is investigated in detail in a separate paper by \textcolor{blue}{Pl\v{s}ek et al. (in prep)}. As projection effects along the central line-of-sight of such a bright cool core are limited, this multi-metallicity phase is rather expected to be local and somehow associated to the \textit{in situ} multiphasedness of the central gas (e.g. due to central cooling and/or gas mixing through AGN uplift). Interestingly, such a scenario may be a candidate to explain the presence of the deep central Fe abundance drop seen in Centaurus \citep[e.g.][]{sanders2016,lakhchaura2019} as well as in other systems \citep[e.g.][]{panagoulia2015,mernier2017}, since any single-metallicity fit would tend to average absolute abundances of separate gas phases. A thorough discussion on how Resolve improves our understanding of the central abundance drop in Centaurus is addressed in \textcolor{blue}{Fukushima et al. (to be subm.)}. Last but not least, it is interesting to compare the large tension of Fe abundances derived here with the more moderate discrepancies reported in soft and hard energy bands by \citet{riva2022} among a handful of relaxed systems of similar temperature. This suggests that if they exist, multi-metallicity gas phases are seen rather \textit{in situ} than in projection as they may not be ubiquitous to all cool-core clusters.

\subsection{The chemical composition of the Centaurus ICM}\label{sec:discussion:abunpattern}

Taking advantage of high spectral resolution available out to 10~keV, our measured X/Fe ratios of the Centaurus core benefit from exquisite accuracy and are of importance in our quest of constraining the chemical composition of the ICM. Nevertheless, measurements of this precision are not unprecedented and can be compared with other estimates from the literature:
\begin{enumerate}
        \item A deep study of X/Fe ratios in the Perseus core \citep{simionescu2019}, using a combination of the high-resolution spectroscopy instruments RGS+SXS, and which came as an extension of the first abundances measured by SXS \citep{hitomi2017};
        \item The measurements of central X/Fe ratios at both high and moderate resolution (using RGS and EPIC) from a sample of 44 cool-core clusters and groups (the CHEERS sample; \citealt{mernier2018}), themselves updated to more recent atomic models from an earlier study \citep{mernier2016a}.
\end{enumerate}
In addition, RGS measurements of the N/Fe ratio were taken from the CHEERS sample \citep{mao2019} and can be compared to our Centaurus estimate.

\begin{figure}[h!]
\centering
\includegraphics[width=0.49\textwidth, trim={1.5cm 0cm 1.5cm 0cm},clip]{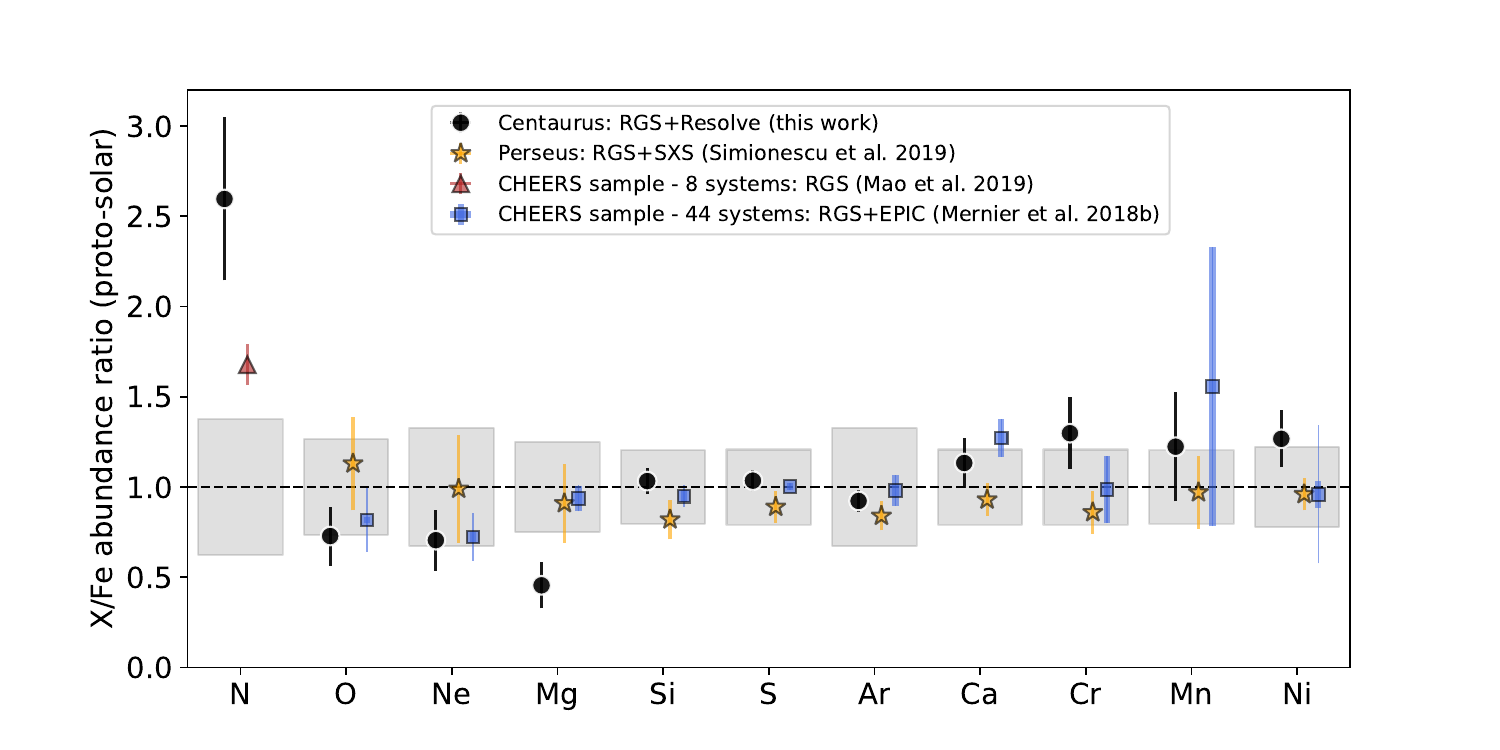}
      \caption{Our final X/Fe abundance ratios in Centaurus compared with measurements on other clusters from the literature at moderate and/or high spectral resolution \citep{mernier2018,simionescu2019,mao2019}. For comparison, solar uncertainties are shown in grey \citep{lodders2009}.
      }
         \label{fig:comparlit}
\end{figure}

This overall comparison is illustrated in Fig.~\ref{fig:comparlit}. Three trends are clearly visible. 

First, and noting the uncertainties of the solar ratios themselves (grey boxes in Fig.~\ref{fig:comparlit}), the heavier ratios measured in Resolve (i.e. from Si/Fe onwards) are formally consistent with a solar composition as well as with ratios reported in Perseus and in the CHEERS sample. This similarity is not surprising as these elements originate partly or entirely from SNIa. At face value, this can be interpreted as an isotropy of SNIa enrichment across the nearby Universe over large scales: the same SNIa explosion mechanism enriches various clusters similarly (or, if several mechanisms, in similar proportions). We note, however, that many models predict a significant fraction of Si and S to be produced by SNcc (see also next section); the lack of scatter in the Si/Fe and S/Fe ratios is thus important to mention. Another remark is that the Resolve ratios are in good agreement with values measured within the same energy band with EPIC pn while, on the contrary, they differ significantly from the EPIC MOS measurements as well as from almost all EPIC measurements obtained from a broad-band fit  (Fig.~\ref{fig:comparCCD} left).

Second, the comparison with previous measurements is more complex for O/Fe, Ne/Fe, and especially Mg/Fe. These ratios are direct tracers of the SNcc/SNIa contribution (since O, Ne, and Mg largely originate from SNcc) and, as of today, RGS is the only instrument capable of constraining these ratios at high spectral resolution. Whereas there is excellent agreement between our O/Fe and Ne/Fe ratios and those of the CHEERS sample (both estimated with SPEX), their values favour somewhat sub-solar abundances. Although they are formally consistent with the ratios measured in Perseus, the SPEX-Xspec tension reported in Fig.~\ref{fig:RGSratios} reveals non-negligible atomic uncertainties for these ratios. On the other hand, we do find a significant Mg/Fe tension between Centaurus and other systems. In fact, the Mg/Fe ratio of Centaurus reaches barely half the average estimate of the CHEERS sample, even though they were all measured with consistent instruments and atomic codes. Even more surprising is the remarkable Mg/Fe agreement between RGS and EPIC measurements both within soft and broad energy bands (Fig.~\ref{fig:comparCCD} left), arguing against the possibility of an instrumental issue specific to our RGS observation. It is interesting to note that, to our knowledge, such a low Mg/Fe ratio is not seen in Milky Way stars \citep{weinberg2019} nor in stellar populations of elliptical galaxies \citep{conroy2014,gountanis2025} -- both populations having not aged enough to be sufficiently enriched with SNIa products. Such a low Mg/Fe ratio in the ICM is thus potentially puzzling and exciting. Nevertheless, we cannot fully exclude a systematic bias due to incorrect fitting since, as detailed in Sect.~\ref{sec:results:rgs}, the Mg/Fe ratio shows a significant scatter across our fitting experiments described above (Fig.~\ref{fig:RGSratios}). These results are also comparable with the previous RGS analysis of the Centaurus core by \citet{fukushima2022}, who reported O/Fe, Ne/Fe and Mg/Fe measurements of typically 0.6--0.8~solar and 0.9--1.3~solar in $<18''$ and $18''$--$120''$ cross-dispersion apertures, respectively (see their figure 10). Because their methodology and extraction regions differ from ours, however, this comparison remains difficult to quantify. Overall, if the trend observed for these three ratios is genuine and inherent to Centaurus, the easiest explanation is that of a smaller SNcc contribution (or, conversely, a more important relative SNIa contribution) as compared to Perseus and other systems. This scenario, however, would also need to explain the solar Si/Fe ratios; hence it would be valid only providing that Si and S were predominantly produced by a nucleosynthesis channel different from that of Mg.

Third, we note that the N/Fe ratio observed in Centaurus is highly super-solar, potentially higher than the average CHEERS value \citep{mao2019}. Interestingly, a very high N/Fe value ($\sim$3 solar) has already been reported in Centaurus \citep{sanders2008}, but also in NGC\,1404 \citep[$2.7 \pm 0.5$ solar;][]{mernier2022} as well as in a number of relatively cool systems \citep{fukushima2023}, indicating that the core of Centaurus may not be exceptional in this respect. These results must be mitigated by the strong dependency of N/Fe with the assumed hydrogen absorption. As seen in Fig.~\ref{fig:RGSratios}, leaving $n_\mathrm{H}$ as a free parameter lowers N/Fe to the CHEERS estimates and to the 1--2~solar value previously reported in Centaurus by \citet{fukushima2022}. The same ratio even drops to solar when one assumes an absorption by atomic hydrogen only ($n_\mathrm{H} = 8.56 \times 10^{20}$~atoms/cm$^2$). In every case, however, these estimates are interesting. Whereas our current picture of the ICM enrichment argues in favour of an early epoch of SNIa and SNcc products injection into the intergalactic gas before cluster assembly \citep[e.g.][]{simionescu2015,urban2017,mernier2017,biffi2017,biffi2018}, less is clear about the history of the enrichment from AGB stars. The origin of the latter, rather found in low-mass stellar populations -- together with the (scattered) super-solar N/Fe values found here and in the core of other systems (see however \citealt{kara2025}) -- make it plausible that this enrichment channel is rather recent, directly connected to the red-and-dead stellar population of clusters and ellipticals, and \textit{in fine} largely decoupled from the early SNIa and SNcc components discussed above. However, and interestingly, very high N/O values have been reported in high-redshift galaxies with \textit{JWST}, potentially opening the possibility of N production and enrichment through early globular clusters or even supermassive stars \citep{marques2024}. Better measurements of N/Fe in clusters and groups may clarify the picture. We also note that, besides AGB stars, other mechanisms could be responsible for N enrichment. For instance, \citet{prantzos2018} showed that high N abundances in the thin disk of our Milky Way might have been produced by rapidly rotating massive stars at low metallicities. Whether such an (earlier) enrichment channel dominates over (later) AGB stars to explain super-solar N/Fe found at Mpc scales is an open question, which may be (at least partly) answered by determining the scatter of N/Fe from system to system, and its dependency with cluster stellar masses.

In summary, the chemical composition of the Centaurus ICM core shares a lot of similarities with that of Perseus in being remarkably close to the chemical composition of the Solar System. At first glance, and given the apparent lack of variation of chemical composition with cluster radius seen at lower spectral resolution \citep{simionescu2015,mernier2017}, it is tempting to interpret this finding as additional evidence supporting the early enrichment scenario -- in which all clusters must share an universal abundance pattern (established early on everywhere in the intergalactic medium) despite central metallicities differing from system to system. Pushing this argument further, one could even speculate that (i) the stellar population of the BCG has (always had) negligible impact on the central ICM enrichment; and (ii) the composition of our Solar System might in fact be causally connected to the universally pre-enriched gas, for example through the early condensation of the Milky Way hot circumgalactic medium onto its thin disk during its formation -- albeit the diversity of stellar $\alpha$/Fe ratios found in the Galactic disk somewhat nuances this connection \citep{bensby2014,weinberg2019}.

The above considerations, however, should be taken with a (large) grain of salt. First, as stressed earlier, the tension in the Mg/Fe ratio refrains us from calling the compositions of Perseus and Centaurus identical. We cannot exclude this difference to be genuine, as it could be explained by a non-negligible contribution from the stellar populations of NGC\,1275 (i.e. Perseus' BCG) and NGC\,4696. In such a scenario, both BCGs may have potentially enriched their surrounding ICM lately, yet somewhat differently. 
Second, if the bulk of the ICM enrichment had completed as early as the epoch of formation of massive early-type galaxies found in clusters, one would naturally expect the composition of the ICM to reflect the $\alpha$-enhanced stellar composition of such galaxies \citep{conroy2014}. Instead, the near-solar ICM composition implies either a connection with less massive ellipticals (with the note that the latter are substantially younger on average; see fig.~18 in \citealt{conroy2014}), or a non-negligible SNIa enrichment since the downsizing epoch. 
Third, and more generally speaking, finding a central abundance pattern common to many systems does not necessarily mean that their cores have not been enriched recently. In fact, recent episodes of central enrichment (from e.g. the BCG) may have occurred, yet their chemical signature may remain partly `hidden' into a continuous accumulation of metals trapped in cluster's potential well across their history. In this sense, any large-scale system with sufficiently matured chemical history may converge towards a `universal' abundance pattern, regardless of the number and epoch(s) of its bulk metal injection episode(s). The era of exquisite abundance accuracy delivered by \XRISM and promised by \textit{NewAthena} will help quantify the scatter of central abundance patterns in many nearby relaxed systems, hence shed more light on these open questions.

\section{Comparison with stellar nucleosynthesis yields}\label{sec:yields}

The abundance pattern found in the ICM is the result of billions of SNcc and SNIa having exploded, presumably (but perhaps not exclusively) in the first half of age of the Universe. Therefore, it constitutes a remarkable integrated record of elemental yields of these stellar populations and ICM abundances can be used to constrain supernova explosion and enrichment models (Sect.~\ref{sec:intro}). Given the excellent accuracy of our measured X/Fe ratios, in this section we compare them to a series of nucleosynthesis yields with the aim of determining which combination of models fits our data best. Such an exercise has been made in a number of previous studies, to which we refer the reader for a detailed formalism \citep[e.g.][]{deplaa2007,sato2007,degrandi2009,mernier2016b,hitomi2017,simionescu2019,mernier2020,fukushima2022,kara2025}.

In short, a given linear combination of, for example, SNcc+SNIa yield models will provide the total number of atoms $N_\mathrm{X,tot}$ of a given element X as
\begin{equation}
        N_\mathrm{X,tot} = a\,N_\mathrm{X,cc} + b\,N_\mathrm{X,Ia},
\end{equation}
where $a$ and $b$ are constants to optimise over the whole abundance pattern. These numbers of atoms are related to the yields as
\begin{equation}
        N_{\mathrm{X,}k} = \frac{M_{\mathrm{X,}k}}{\mu_\mathrm{X}},
\end{equation}
where $k$ stands for either `cc' or `Ia', $M_{\mathrm{X},k}$ is the mass of element X produced by either SNcc or SNIa and $\mu_\mathrm{X}$ is the atomic weight of element X. In the case of SNIa, the $M_{\mathrm{X,Ia}}$ yields are tabulated as such since they depend directly on the considered explosion model. In the case of SNcc, however, the $M_{\mathrm{X,cc}}$ yields depend on the mass of the progenitor and one needs to integrate the yields over a given stellar mass distribution -- i.e. an IMF:
\begin{equation}
        M_{\mathrm{X},cc} = \frac{\int^{M_\mathrm{cut}}_{M_\mathrm{low}} M^\mathrm{tab}_{\mathrm{X},cc}(m)\, m^\alpha \, dm}{\int^{M_\mathrm{cut}}_{M_\mathrm{low}} \, m^\alpha dm},
\end{equation}
where $M^\mathrm{tab}_{\mathrm{X},cc}(m)$ is the (initial) tabulated yield as a fonction of progenitor mass $m$, $M_\mathrm{low}$ and $M_\mathrm{cut}$ are respectively the lower and cut masses (usually adopted from the available yield models), and $\alpha$ is the slope of the IMF (which we keep at $\alpha = -2.35$; \citealt{salpeter1955}). We use the publicly available code \texttt{abunfit}\footnote{\href{https://github.com/mernier/abunfit}{https://github.com/mernier/abunfit}} to fit a a large number of yield models and combinations to our final ratios. For convenience, these fits are labelled from Fit~A to Fit~H.

\begin{table*}[h!]
\caption{Results of various combinations of yield models fitted on our final abundance pattern.}
\label{table:SNyields}
\centering
\begin{tabular}{l c l c c c r}
\hline\hline
Fit     &       $Z_\mathrm{SNcc/AGB}$   &       Combination     &       Ratios  &       Favoured models  &       $f_\mathrm{Ia}$ &       $\chi^2$/d.o.f. \\
\hline
A       &               &       SNcc    &       [O-Ni]/Fe       &       No13\_SNcc\_0.004         &       $0.31$  &       18.1/9\\
        &               &       + SNIa  &               &        + Le18\_300-0-c3       &               &       \\
A       &       $\odot$ &       SNcc    &       [O-Ni]/Fe       &       No13\_SNcc\_0.02         &       $0.28$  &       20.5/9\\
        &               &       + SNIa  &               &        + Ba06\_DDTc   &               &       \\
        \hline
B       &               &       SNcc    &       [O-Ni]/Fe       &       Su16\_N20       &       $0.54$  &       6.1/8\\
        &               &       + SNIa(near-$M_\mathrm{Ch}$)    &               &        + Se13\_N40     &               &       \\
        &               &       + SNIa(sub-$M_\mathrm{Ch}$)     &               &         + Sh18\_M085\_5050\_Z0\_01     &               &       \\      
B       &       $\odot$ &       SNcc    &       [O-Ni]/Fe       &       No13\_SNcc\_0.02        &       $0.90$  &       6.4/8\\
        &               &       + SNIa(near-$M_\mathrm{Ch}$)    &               &        + Fi14\_N1def   &               &       \\
        &               &        + SNIa(sub-$M_\mathrm{Ch}$)    &               &        + Sh18\_M085\_5050\_Z0\_01      &               &       \\
        \hline
C       &       $\odot$ &       SNcc    &       [O-Ni]/Fe       &       No13\_0.02         &       $0.55$  &       6.2/8\\
        &               &       + SNIa(def)     &               &        + Fi14\_N100Ldef        &               &       \\
        &               &       + SNIa(del-det) &               &        + Ba06\_DDTf    &               &       \\
        \hline
D       &       $\odot$ &       SNcc    &       [Si-Ni]/Fe      &       Le25\_A22S03         &       $0.23$  &       7.4/6\\
        &               &       + SNIa  &               &        + Fi14\_N100Hdef       &               &       \\
        \hline
E       &       $\odot$ &       SNcc    &       [Si-Ni]/Fe      &       Le25\_A22S03    &       $0.54$  &       4.3/5\\ 
        &               &       + SNIa(near-$M_\mathrm{Ch}$)    &               &        + Se13\_N150    &               &       \\
        &               &       + SNIa(sub-$M_\mathrm{Ch}$)     &               &        + Sh18\_M085\_5050\_Z0\_01      &               &       \\
        \hline
F       &       $\odot$ &       SNcc    &       [Si-Ni]/Fe      &       Le25\_A22S03         &       $0.30$  &       5.0/5\\
        &               &       + SNIa(def)     &               &        + Fi14\_N100Hdef        &               &       \\
        &               &       + SNIa(del-det) &               &        + Ba06\_DDTe    &               &       \\
        \hline
G       &               &       AGB     &       all     &       No13\_AGB\_0.004        &       $0.58$  &       4.9/8\\ 
        &               &       + SNcc  &               &       + Su16\_N20     &               &       \\
        &               &       + SNIa(near-$M_\mathrm{Ch}$)    &               &        + Se13\_N40*    &               &       \\
        &               &       + SNIa(sub-$M_\mathrm{Ch}$)     &               &        + Sh18\_M085\_5050\_Z0\_01*     &               &       \\
G       &       $\odot$ &       AGB     &       all     &       No13\_AGB\_0.02 &       $0.50$  &       10.2/8\\ 
        &               &       + SNcc  &               &       + No13\_SNcc\_0.02      &               &       \\
        &               &       + SNIa(near-$M_\mathrm{Ch}$)    &               &        + Se13\_N40*    &               &       \\
        &               &       + SNIa(sub-$M_\mathrm{Ch}$)     &               &        + Sh18\_M085\_5050\_Z0\_01*     &               &       \\
        \hline
H       &               &       AGB     &       all     &       No13\_AGB\_0.004         &       $0.69$  &       6.1/8\\
        &               &       + SNcc  &               &       + No13\_HNe\_0.008         &               &       \\
        &               &       + SNIa(def)     &               &        + Fi14\_N100Ldef*               &               &       \\
        &               &       + SNIa(del-det) &               &        + Ba06\_DDTf*           &               &       \\
H       &       $\odot$         &       AGB     &       all     &       No13\_AGB\_0.02         &       $0.69$  &       6.4/8\\
        &               &       + SNcc  &               &       + No13\_HNe\_0.02         &               &       \\
        &               &       + SNIa(def)     &               &        + Fi14\_N100Ldef*               &               &       \\
        &               &       + SNIa(del-det) &               &        + Ba06\_DDTf*           &               &       \\
        \hline
\hline
\end{tabular}
\tablefoot{In each fit, if the favoured combination is a SNcc (and AGB) model with sub-solar initial metallicity, we also display the most favoured combination where these models are solar (marked by $\odot$). Models that are pre-selected are indicated with an asterisk (see text).
}
\end{table*}

\begin{figure*}[h!]
\centering
\includegraphics[width=0.32\textwidth, trim={0cm 0cm 0cm 0cm},clip]{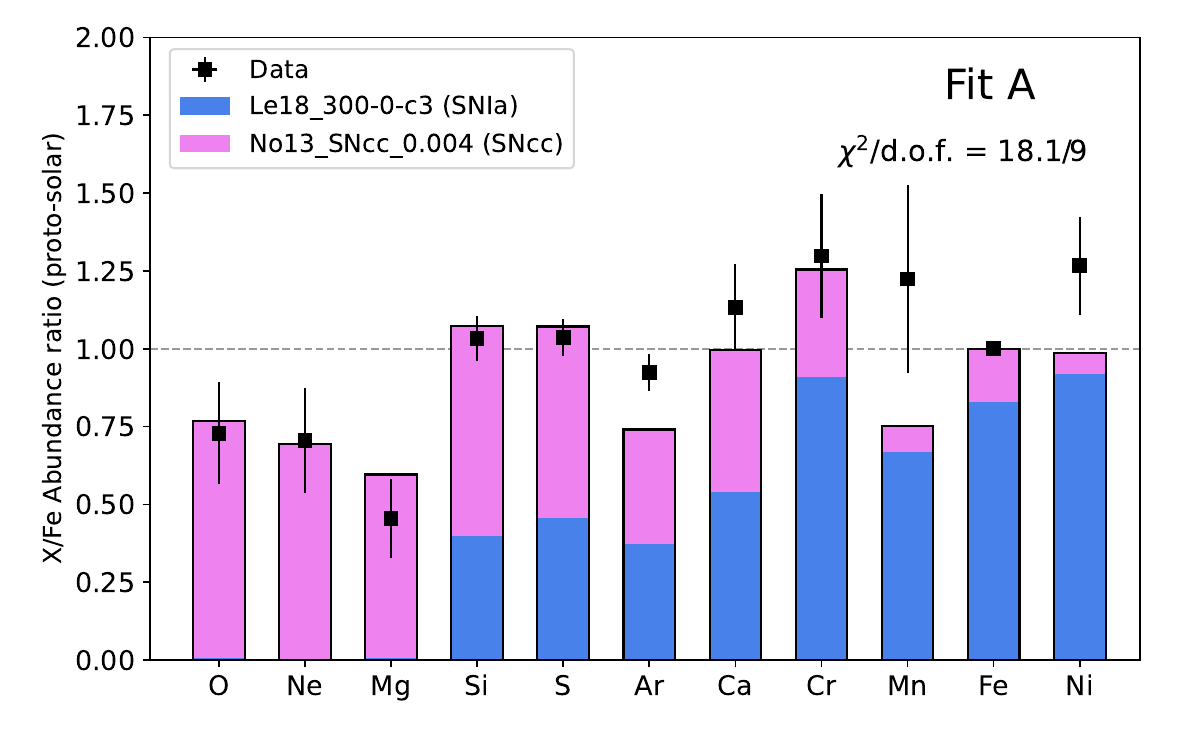} 
\includegraphics[width=0.32\textwidth, trim={0cm 0cm 0cm 0cm},clip]{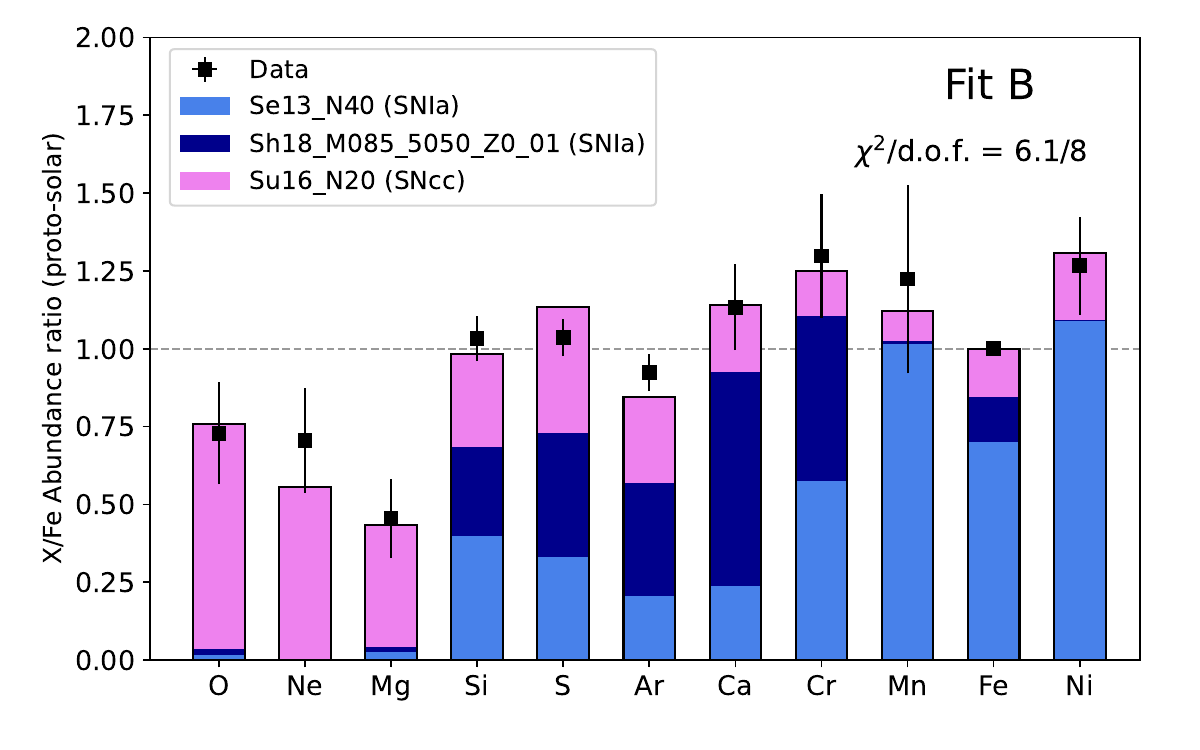} 
\includegraphics[width=0.32\textwidth, trim={0cm 0cm 0cm 0cm},clip]{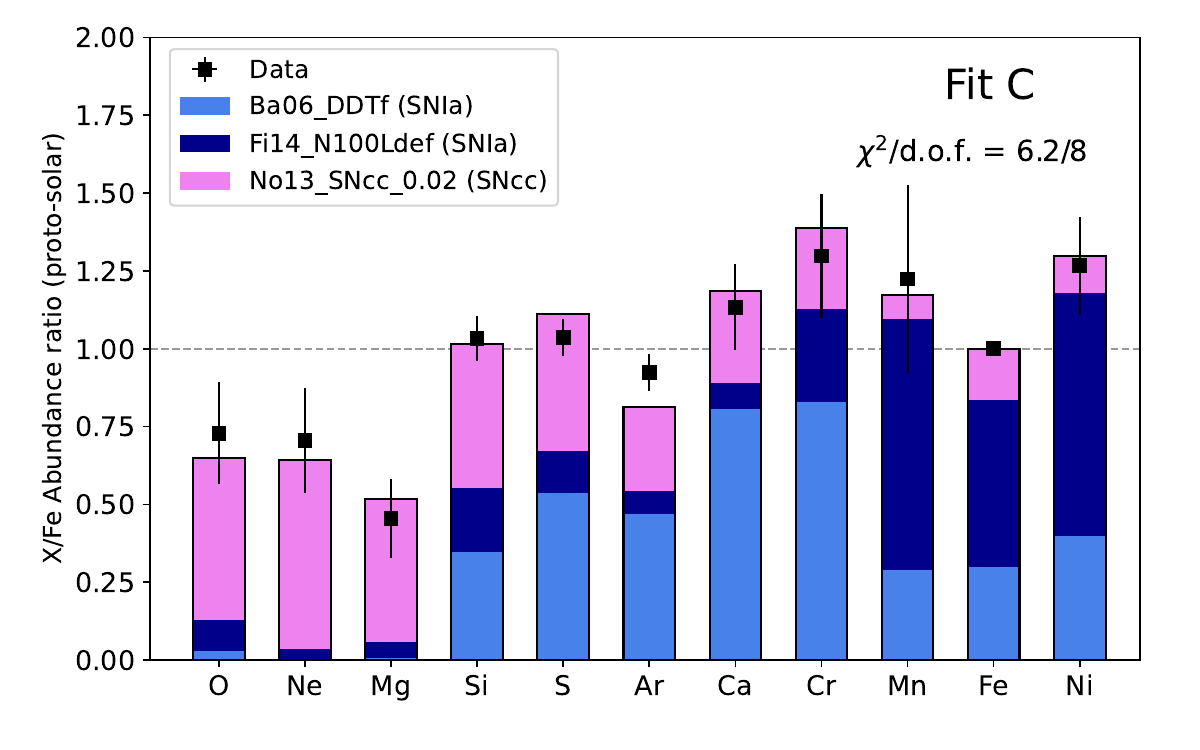} \\
\includegraphics[width=0.32\textwidth, trim={0cm 0cm 0cm 0cm},clip]{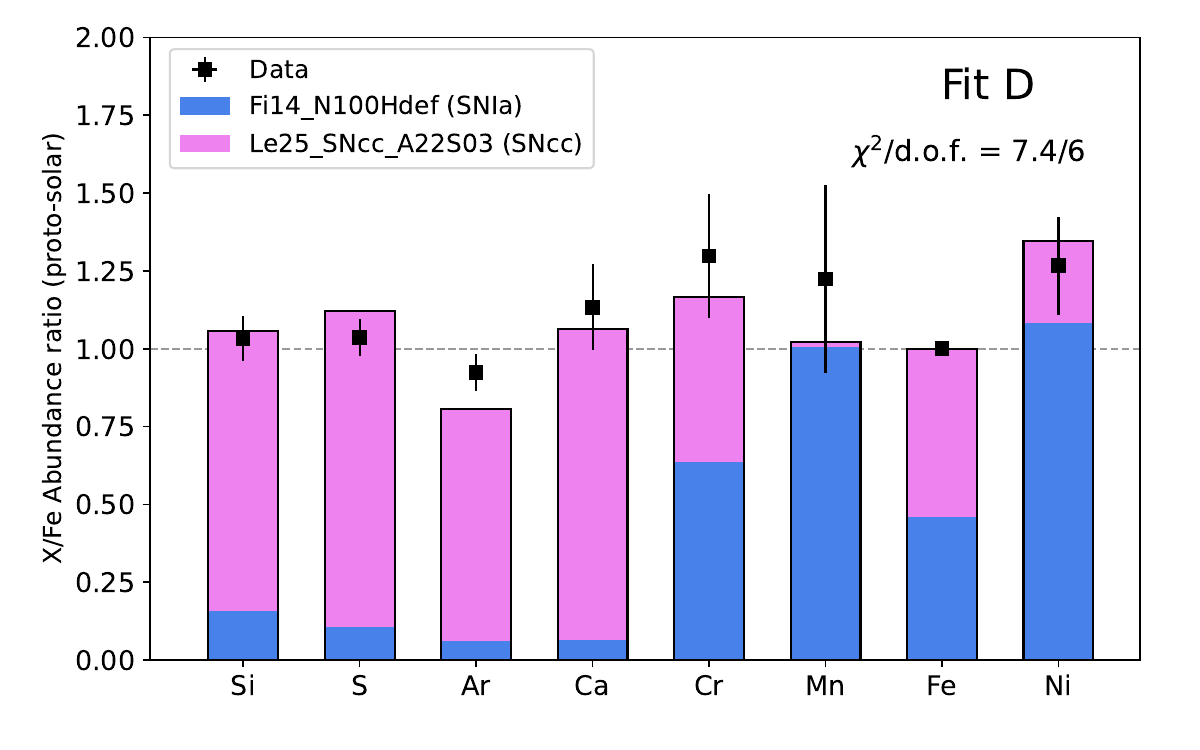} 
\includegraphics[width=0.32\textwidth, trim={0cm 0cm 0cm 0cm},clip]{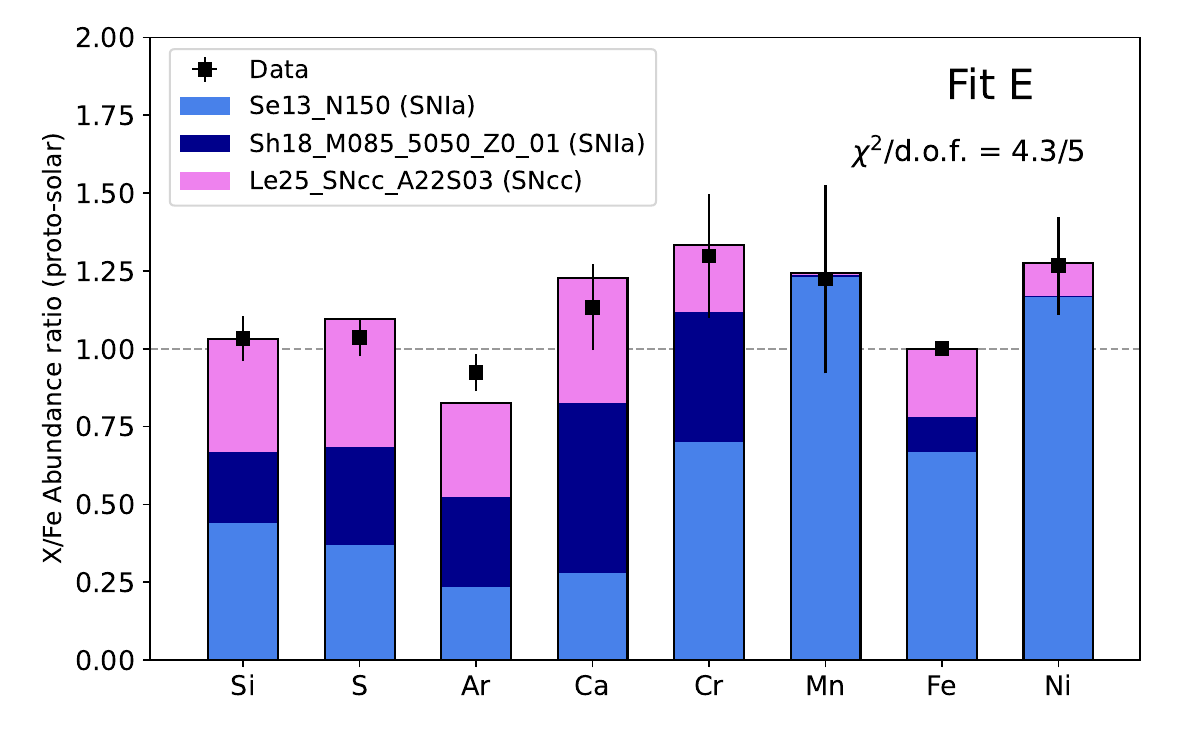} 
\includegraphics[width=0.32\textwidth, trim={0cm 0cm 0cm 0cm},clip]{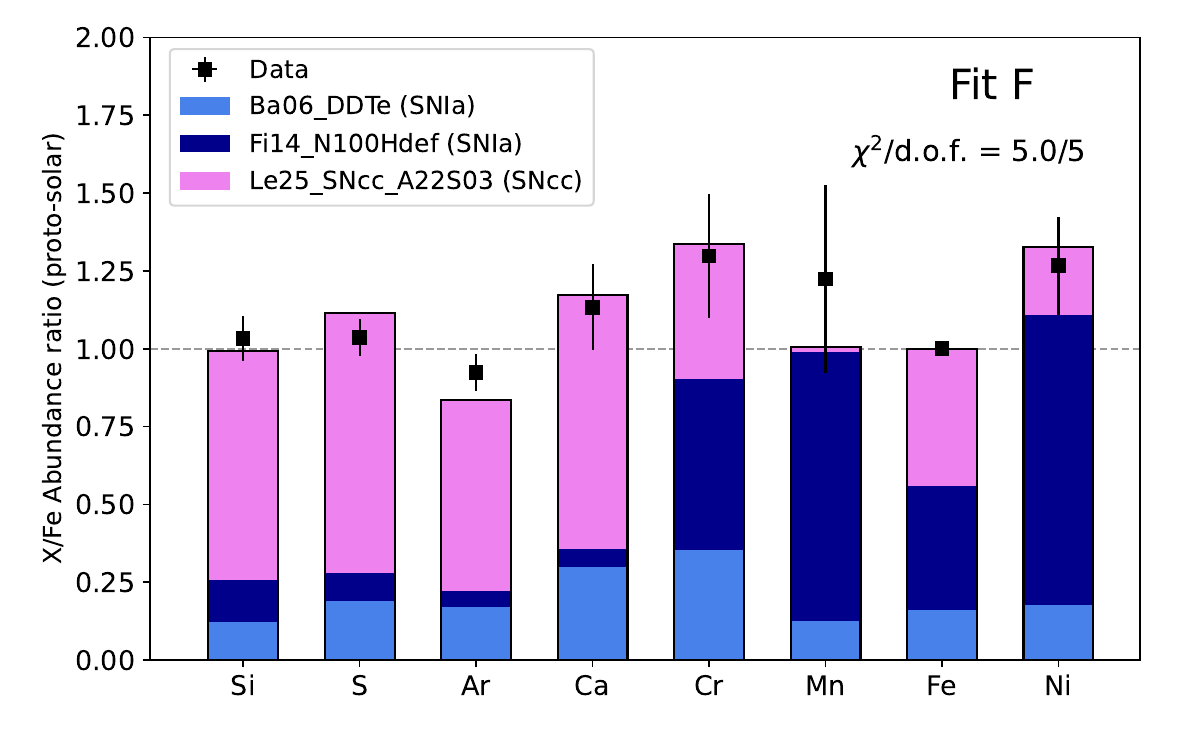} \\
\hspace*{5.9cm}
\includegraphics[width=0.32\textwidth, trim={0cm 0cm 0cm 0cm},clip]{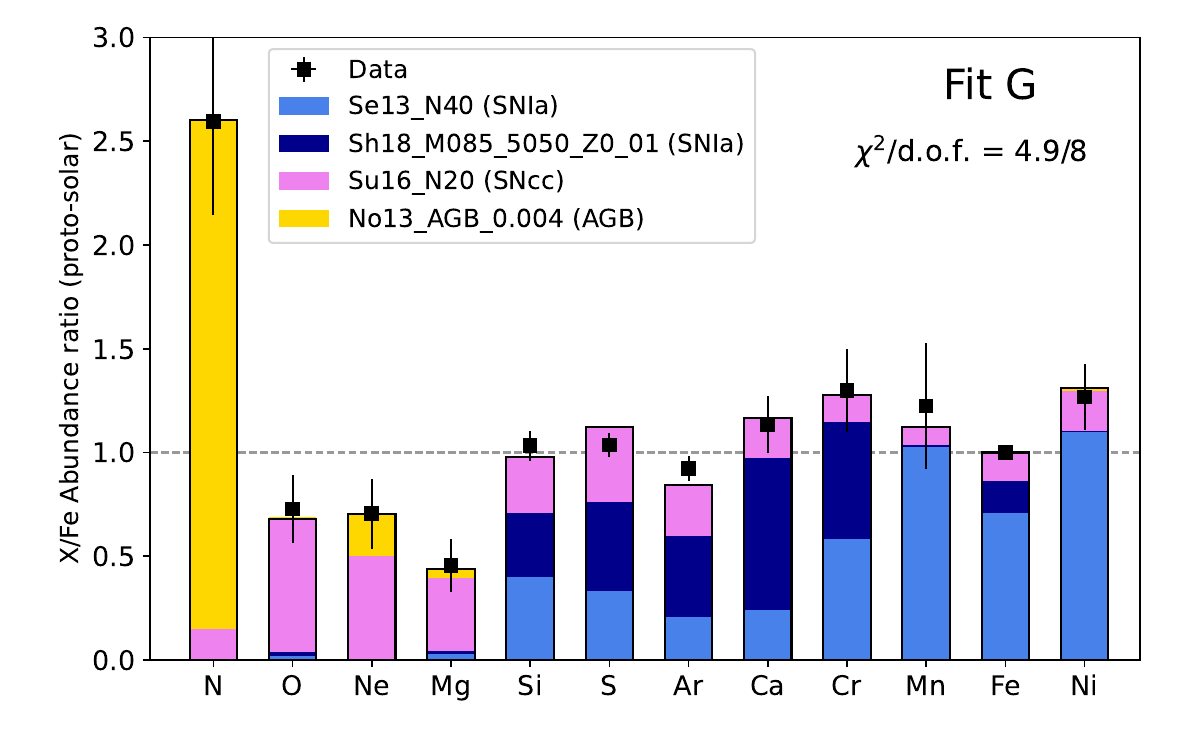} 
\includegraphics[width=0.32\textwidth, trim={0cm 0cm 0cm 0cm},clip]{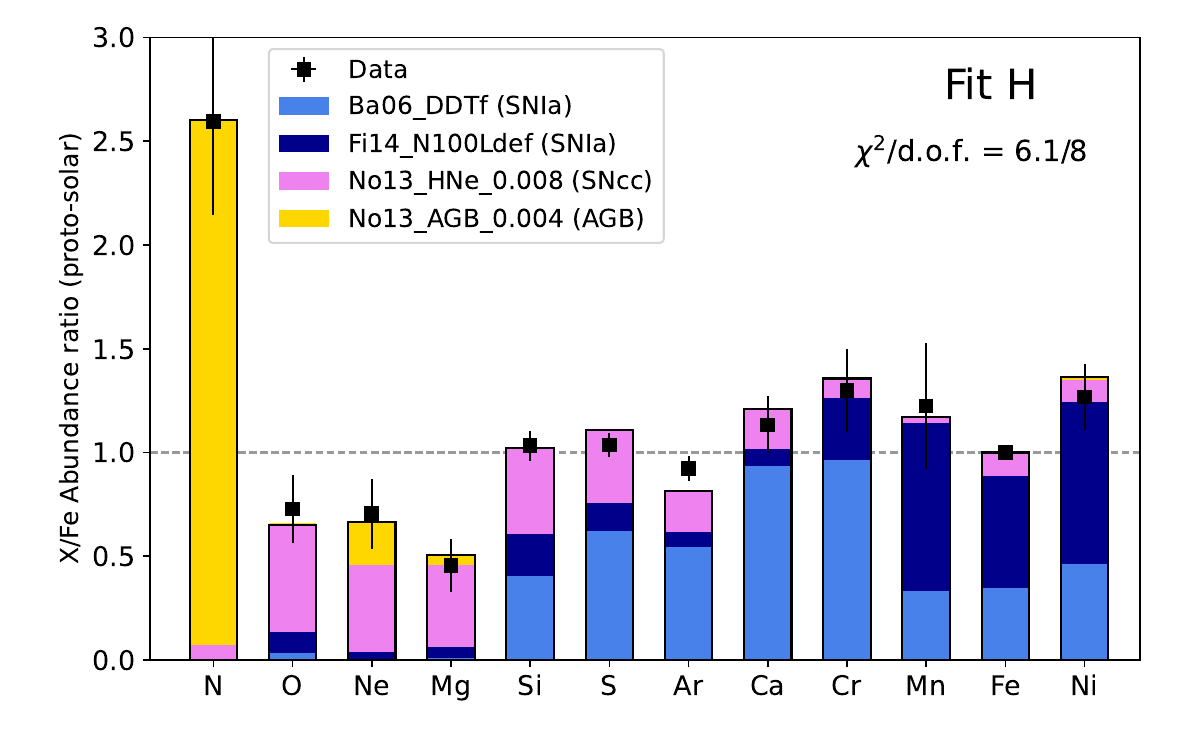}

      \caption{Best-fit combination of AGB, SNcc, and/or SNIa yield models (coloured histograms) on our Centaurus final abundance pattern (black data points). AGB and SNcc contributions are yellow and pink, respectively. SNIa contributions are (light and dark) blue. Top panels (Fits~A--C) are for all ratios except N/Fe and ignore the AGB contribution. Middle panels (Fits~D--F) are for ratios measured with Resolve only. Bottom panels (Fits~G, H) are for all ratios and include an AGB contribution. These fits are described further in the text (see also Table~\ref{table:SNyields}.
      }
         \label{fig:SNyields}
\end{figure*}

A full list of yield models and references is given in Appendix~\ref{sec:app:yields}. Although the details of each and every model fall naturally beyond the scope of this paper, it is important to emphasise the main categories of models as well as the physical parameters of interest on which they depend.
\begin{enumerate}
        \item AGB yield models depend on the metallicity of the initial progenitor $Z_\mathrm{init}$, which varies between 0 and 0.02 (i.e. solar) and are adopted from \citet{karakas2010,nomoto2006}.
        \item SNcc yield models also depend on $Z_\mathrm{init}$ within the same range as for AGB models. Besides standard yield models \citep{nomoto2006,kobayashi2006,nomoto2013}, we use two models including neutrino transport and calibrated to reproduce key properties of SN1987A \citep{sukhbold2016} as well as a model including convective processes \citep{leung2025} and specifically tailored to reproduce the ICM abundance pattern of Perseus measured with \hitomi \citep{simionescu2019}. Additionally, we also include hypernova (HNe) models from \citep{nomoto2013}.
        \item SNIa models depend primarily on their explosion mechanisms and, though less directly, on the mass of their progenitors:
        \begin{itemize}
                \item Near-$M_\mathrm{Ch}$ models assume either a deflagration \citep{iwamoto1999,maeda2010,fink2014,leung2018} or a delayed-detonation \citep{iwamoto1999,badenes2006,maeda2010,seitenzahl2013,ohlmann2014,leung2018}. We also consider the intriguing possibility of hybrid CONe WDs \citep{kromer2015} as well as gravitationally confined detonations \citep{seitenzahl2016};
                \item Sub-$M_\mathrm{Ch}$ models typically assume a (simple or double) detonation \citep{sim2010,sim2012, leung2020} which, in some cases, is triggered under the double-degenerate scenario \citep{pakmor2010,kromer2013,shen2018}.
        \end{itemize}
In turn, constraining these models may thus lead to a better understanding of SNIa progenitors. Our yield model fitting results are listed in Table~\ref{table:SNyields}, shown in Fig.~\ref{fig:SNyields} and discussed below.
\end{enumerate}

\subsection{Fit A: SNcc+SNIa models}\label{sec:yields:1SNIa}

For the sake of simplicity, we start with SNcc and SNIa models only, thus we consider all ratios except N/Fe. The first, most straightforward attempt is that of a simple SNcc+SNIa combination (Fit~A). The best-fit combination obtained with this setup (Fig.~\ref{fig:SNyields}, top left) is a SNcc model \citep[No13\_SNcc\_0.004;][]{nomoto2013} with $Z_\mathrm{init} = 0.004$ and a near-$M_\mathrm{Ch}$ delayed-detonation model \citep[Le18\_300-0-c3;][]{leung2018}. The rather low $Z_\mathrm{init}$ of SNcc suggested by the fit, however, is surprising given the near-solar stellar metallicities observed in BCGs \cite[e.g.][]{oliva2015,edwards2024}. Indeed, if one assumes that the bulk of SNcc having enriched cluster cores had exploded during (or after) the BCG formation, it is natural to expect a high initial metallicity of their progenitors. In this context, we also report in Table~\ref{table:SNyields} the most favoured combination that includes a solar $Z_\mathrm{init}$. This alternative combination has a slightly worse $\chi^2$/d.o.f than our best-fit combination. However, we note that both fits suffer from a large $\chi^2$/d.o.f. which indicates that the observed ratios are not well reproduced. This is especially true for Ar/Fe and Mn/Fe which are clearly underestimated by the models. 

\subsection{Fits B \& C: SNcc+SNIa+SNIa models}\label{sec:yields:2SNIa}

Since the quality of Fit~A is rather poor, we explore the possibility of adding an extra SNIa component in our fit. The physical motivation of this improvement, based on the diversity of SNIa as supported by observations, is twofold: (i) near-$M_\mathrm{Ch}$ and sub-$M_\mathrm{Ch}$ SNIa may coexist in nature; or (ii) the near-$M_\mathrm{Ch}$ channel may dominate but with a diversity of explosion scenarios (i.e. deflagration and delayed-detonation models may coexist). We include these two additional scenarios in Fits~B and C, respectively. As shown in the upper middle and right panels of Fig.~\ref{fig:SNyields}, both experiments provide a clearly better fit to the data. In Fit~C, we note that the favoured delayed-detonation model (DDTf; \citealt{badenes2006}) is among the least recent calculations from our list. Interestingly, \citet{deplaa2007} found a model from the same family to fit best their abundance pattern over a sample of clusters (observed with \XMM/EPIC). 

\subsection{Fits D--F: Resolve ratios only}\label{sec:yields:Resolve}

Even though our RGS ratios have been measured carefully and account for various systematic uncertainties, our peculiar Mg/Fe measurement obligates us to take these values with caution. More generally, if multi-phase \emph{and} multi-metallicity gas is at play in the Centaurus core (Sect.~\ref{sec:discussion:absFe}), the chemical composition of different phases ought to be treated separately. To be conservative in this respect, we reproduce our first three experiments accounting this time for the Resolve ratios only (from Si/Fe to Ni/Fe) as safe tracers of the hot phase. This provides Fits~D, E, and F which are shown in the middle three panels of Fig.~\ref{fig:SNyields}.

Unsurprisingly, fitting our restricted abundance pattern with two SNIa models (Fits~E \& F) provides acceptable results like in their extended Fits~B \& C counterparts. More surprising is that a simple SNcc+SNIa combination is now able to reproduce our Resolve ratios satisfactorily (Fit~D). This is mostly achieved through a better reproduction of Mn/Fe (via a deflagration SNIa model) and of Ar/Fe, though the latter remains somewhat underestimated. In fact, the fit now uses the updated SNcc model from \citet{leung2025} which tends to lower the S/Ar ratio towards unity, as observations suggest \citep[e.g.][]{simionescu2019}. Interestingly, this close-to-acceptable fit now postulates that the wide majority of Si, S, Ar, and Ca in the ICM may have been produced by SNcc. This scenario disagrees with the common conception (seen in our other fits and in earlier attempts; e.g. \citealt{mernier2016b,simionescu2019}) that these intermediate-mass elements would be produced by a mix of SNcc and SNIa in comparable proportions. We note, however, that if we extend this particular combination to the RGS ratios, this SNcc model largely overproduces O/Fe, Ne/Fe, and Mg/Fe. It is thus valid only under the prior condition that those three ratios are not relevant to the hot ICM phase.

\subsection{Fits G \& H: Including AGB contribution}\label{sec:yields:AGB}

Contrasting with our previous attempts (Fits D--F), another interesting exercise is to obtain a `full picture' of the Centaurus enrichment channels by considering now all ratios at hand, thus including N/Fe as well. Investigating this ratio inevitably introduces an additional AGB component, which we add in our fits (i.e. AGB+SNcc+SNIa+SNIa). To avoid our data from being overfitted, we decide to keep the SNIa models to those favoured in Fits~B \& C. In compensation, we explore all AGB and SNcc models from our list. We obtain Fits~G and H, accounting for respectively the coexistence of (i) near- and sub-$M_\mathrm{Ch}$ progenitors and (ii) deflagration and delayed-detonation explosions (Fig.~\ref{fig:SNyields}, bottom panels).

We find that both experiments provide a largely acceptable fit to our data, with no clear way to favour one over the other. While they both favour sub-solar $Z_\mathrm{init}$ for their SNcc and/or AGB progenitors, we note that forcing $Z_\mathrm{init} = 0.02$ provides equally acceptable results (Table~\ref{table:SNyields}). To better constrain these initial metallicities will necessarily imply a more accurate determination of N--Mg/Fe ratios, which are to be obtained from non-dispersive high-resolution spectroscopy and improvement of atomic models (Sect.~\ref{sec:results:rgs}). Another item of notice suggested by the models is the non-negligible contribution of AGB stars in the total Ne enrichment. If true, this may affect the consistency and interpretation of our Fits A--C (excluding the AGB contribution in Ne/Fe) but also a number of similar fits attempted previously in the literature.

\subsection{General remarks and considerations}\label{sec:yields:summary}

We end this section by summarising our nucleosynthesis yield comparison made above, as well as by adding a few points of discussion on its strengths and limitations.

\begin{itemize}
        \item The question on whether a unique SNIa explosion model can explain the Centaurus ICM enrichment depends largely on the interpretation of the ratios measured by RGS (O--Mg/Fe). If these lighter ratios are genuine and relevant to the hot phase ICM, then at least two distinct SNIa channels are required to explain the chemical composition of the latter. If, on the contrary, these ratios are affected by unknown systematics and/or trace a cooler component with a different enrichment history (as possibly hinted by the lower Fe abundance measured in RGS), then only one (near-$M_\mathrm{Ch}$, deflagration) SNIa explosion model is sufficient to explain our data.
        \item If two SNIa explosion channels co-exist in nature, whether this witnesses a diversity of SNIa progenitors (near- and sub-$M_\mathrm{Ch}$ channels) or of SNIa explosion channels (deflagration and delayed-detonation) cannot be clearly established by our data. 
        \item As already noted by \citet{simionescu2019}, the major challenge in fitting supernova yield models on observational data resides primarily on the S/Ar ratio, which is systematically overproduced by the models. Recent improvement in SNcc yield calculations include now an updated treatment of convective processes \citep{leung2025}. Like for Perseus, this new model is preferred by our Resolve ratios in Centaurus as well; however it does not reproduce best our RGS ratios. 
        \item A key follow-up question is thus whether the O--Mg/Fe ratios observed here are unique to Centaurus or whether, to some extent, they often differ from cluster to cluster. Although RGS is currently crucial in this respect, future non-dispersive high-resolution spectroscopy will certainly be transformative in light of this question.
        \item Despite their excellent accuracies, our current measurements of lighter and intermediate ratios (O--Ca/Fe) do not allow us to constrain the initial metallicity of SNcc progenitors. A better understanding of the systematics at play in these measurements is necessary to address this question.
        \item Under the assumption that the N enrichment originates from AGB stars (see also Sect. \ref{sec:discussion:abunpattern}), the AGB contribution is important not only for N enrichment, but also potentially for Ne enrichment. 
        \item One quantity of interest in such a comparison is $f_\mathrm{Ia}$, namely the fraction of SNIa over the total number of supernovae (i.e. SNcc+SNIa) effectively responsible for the ICM enrichment, which can be directly inferred from the best-fit parameters $a$ and $b$. As seen in Table~\ref{table:SNyields}, our series of fits above generate $f_\mathrm{Ia}$ values typically between $\sim 23$--90\% (with individual uncertainties of $\sim$5--6\%). This large, mostly unconstrained range is explained by the wide diversity of SNcc and SNIa yields as predicted by these competitive models \citep[see also][]{degrandi2009}.
        \item Another important aspect to consider is the relative uncertainty of nucleosynthesis models themselves to predict the yield of a given element. One well known example is that of the $^{12}$C($\alpha$,$\gamma$)$^{16}$O reaction rate \citep[e.g.][]{kunz2002}, whose uncertainties are still large despite its crucial importance in nucleosynthesis calculations. In particular, an increase in the $\alpha$-capture efficiency would considerably suppress C burning, hence result in a modification of the Ne and Mg yields \citep{kobayashi2020,xin2025}. Generally speaking, such inaccuracies are of course very challenging to quantify since they may vary from model to model and from element to element, given their respective physical assumptions and limitations. Taking this effect into account may ultimately require more complex methods than a simple linear fit as performed in the paper and in previous literature. For instance, a future trajectory may consist of comparing ICM abundances with full chemical evolution models that account for the metallicity evolution of supernovae progenitors.
\end{itemize}

\section{Conclusions}\label{sec:conclusions}

In this study, we performed a thorough analysis of the \XRISM/Resolve PV data of the core of the Centaurus cluster. With 287~ks of total net exposure obtained at $\sim$5~eV spectral resolution in the 2--10~keV band for the first time for this system, this work provides accurate constraints on the absolute Fe abundance as well as the Si/Fe, S/Fe, Ar/Fe, Ca/Fe, Cr/Fe, Mn/Fe, and Ni/Fe abundance ratios in the integrated vicinity of the central BCG, NGC\,4696. In addition, we used the most relevant \XMM/RGS data at hand to obtain accurate measurements of the N/Fe, O/Fe, Ne/Fe, and Mg/Fe ratios, hence completing our high spectral resolution view on the chemical composition in the core ICM of Centaurus. Our results are summarised as follows.

\begin{itemize}
        \item Beyond their remarkable precision, our Resolve abundance measurements are also robust towards a number of systematic effects, such as (i) the choice of the (multi-T) modelling, (ii) independent estimates from two up-to-date atomic codes (SPEXACT and AtomDB), and (iii) fitting effects such as locally imperfect calibration of the ARF. Abundance ratios measured with RGS are also relatively stable, though they do show a degree of dependency on the atomic code that was used (O/Fe and Ne/Fe) or on the extraction region and spectral order(s) considered (Mg/Fe). We accounted for these additional systematic uncertainties in our final estimates.
        \item Despite the encouraging convergence of the X/Fe ratio measurements, a factor $\sim$2 discrepancy of absolute Fe abundance was measured between Resolve and RGS. This discrepancy, which has been reported a number of times in previous work, is also seen at CCD resolution with \XMM/EPIC when separating soft- and hard-band spectral analyses. This apparent mismatch between Fe-L and Fe-K measurements may reveal atomic code issues and/or the intriguing possibility of the central multi-phase gas of NGC\,4696 of being (locally) multi-metallicity. At second order, this Resolve-RGS discrepancy in Fe may also witness extraction regions that are too different or large systematic uncertainties in absolute abundance estimates with RGS (due to the limited energy band and instrumental line broadening).
        \item In line with high-resolution spectroscopy (RGS and \hitomi/SXS) measurements obtained in the core of Perseus \citep{simionescu2019}, the chemical composition of Centaurus is formally consistent with that of our Solar System for the majority of the ratios probed in this work. However, the super-solar N/Fe ratio which (as also reported and discussed in previous studies) suggests an enrichment channel from AGB stars that is decoupled from (and more recent than) the early SNcc and SNIa channels. In addition, and perhaps even more surprisingly, the Mg/Fe ratio is found to be significantly sub-solar, in tension with similar measurements in Perseus. Therefore, while our current measurements are not sufficient to prove that the Centaurus chemical composition differs from solar, they should {not} be interpreted as decisive evidence of a uniform solar composition in clusters either. More clusters need to be observed at high-resolution spectroscopy to assess and interpret the (non-)universality of the central ICM composition and connect it with the chemical history of cluster cores. 
        \item We also compared our derived abundance ratios with various combinations of SNIa, SNcc, and AGB nucleosynthesis yield models, aiming to constrain the physics and/or progenitors of the stellar population responsible for the ICM enrichment. Unless the lighter ratios (measured with RGS) are biased and/or decoupled from the hotter gas phase, our data suggest that at least two SNIa explosion and/or progenitor channels are necessary to enrich the ICM in the observed proportions. Additionally, we note an improvement of recent SNcc model updates to fit the observed ratios. This is particularly relevant for S/Ar, although this ratio remains imperfectly reproduced by the models.
\end{itemize}

Future \XRISM and \XMM/RGS observations of other nearby relaxed clusters will help answer the open questions addressed in this work and further improve our understanding of metal enrichment in the ICM.

\begin{acknowledgements}
      This paper is dedicated to the memory of our colleague Katja Pottschmidt who passed away on June 17, 2025. Besides her invaluable expertise on the \XRISM and \nustar missions and on the astrophysics of compact objects, her kindness and generosity will be dearly missed among the X-ray astronomical community. We thank the referee for their useful comments which contributed to increase the overall quality and readiness of this manuscript. We thank Ralf Ballhausen, Frederick S. Porter, Anna Ogorza\l{}ek, and Ming Sun for insightful comments and discussion. This work was supported by the JSPS Core-to-Core Program (JPJSCCA20220002). The material is based upon work supported by NASA under award number 80GSFC21M0002. NW and TP were supported by the GACR EXPRO grant 21-13491X.
\end{acknowledgements}

\bibliographystyle{aa}
\bibliography{aa57442-25}

\begin{appendix}

\section{Parameters of the \texttt{clus} model} \label{sec:app:clus}

We report the parameters of the \texttt{clus} model\footnote{The data reproduction package of the \texttt{clus} model for these data is available at \href{https://zenodo.org/records/17902019}{https://zenodo.org/records/17902019}.} in Table~\ref{tab:clus_params}. The density, temperature and abundance profile with which these parameters were obtained are reported in \cite{majumder2025}.

\begin{table}[h!]
    \centering
\caption{\texttt{clus} model parameters for the Centaurus core.} 
\label{tab:clus_params}
    \begin{tabular}{cc|cc}
    \hline
    \hline
    Parameter & Value & Parameter & Value\\
    \hline
        $n_{0,1}$ & 65521 m$^{-3}$ & $c$ & 0.33\\
        $r_{tc1}$ & $2.48 \times 10^{-3}r_{500}$ & $A$ & 98\\
        $\beta_1$ & 0.599 & $B$ & $6.4 \times 10^{-4}r_{500}$\\
        $n_{0,2}$ & 22685 m$^{-3}$ & $C$ & 0.9\\
        $r_{tc2}$ & $2.63 \times 10^{-2}r_{500}$ & $D$ & $0.88$ \\
        $\beta_2$ & 0.778 & $E$ & $1.97 \times 10^{-5}r_{500}$ \\
        $T_c$ & 0.74 keV & $F$ & $14r_{500}$\\
        $T_h$ & 6.75 keV & $G$ & $5.2 \times 10^{-2}$\\
        $r_{\textrm{to}}$ & $0.15r_{500}$ & $r_{500}$ & $1.88$ ($10^{22}$ cm)\tablefootmark{(a)}\\
        $\mu$ & 2.36 & $r_{out}$ & 1\\
        $r_{\textrm{tc}}$ & $4.5 \times 10^{-2}r_{500}$ & $r_{min}$ & 1\\
        $a$ & $0.12$ & $r_{max}$ & 0.06\tablefootmark{(b)}\\
        $b$ & $0.54$ & $a_v$ & 119$\sqrt{2}$ km s$^{-1}$ \tablefootmark{(c)}\\
    \hline
    \hline
    \end{tabular}
    \tablefoot{\ \tablefootmark{(a)} $r_{500} = 610$ kpc ($3000$ arcsec; \citealt{walker13}). \ \tablefootmark{(b)} For a Resolve field of view radius of $\sim 90$ arcsec divided by $r_{500} = 3000$ arcsec and $r_\mathrm{out}=1$. \ \tablefootmark{(c)} In the \texttt{clus} model, the velocity parameter describes the magnitude of micro-turbulence, which is equal to $\sqrt{2}$ times the 1D velocity dispersion.
    }
\end{table}

\FloatBarrier

\section{Narrow-band fits and their model dependencies}

\begin{figure}[h!]
\centering
\includegraphics[width=0.98\hsize, trim={1.0cm 0cm 1.5cm 0cm},clip]{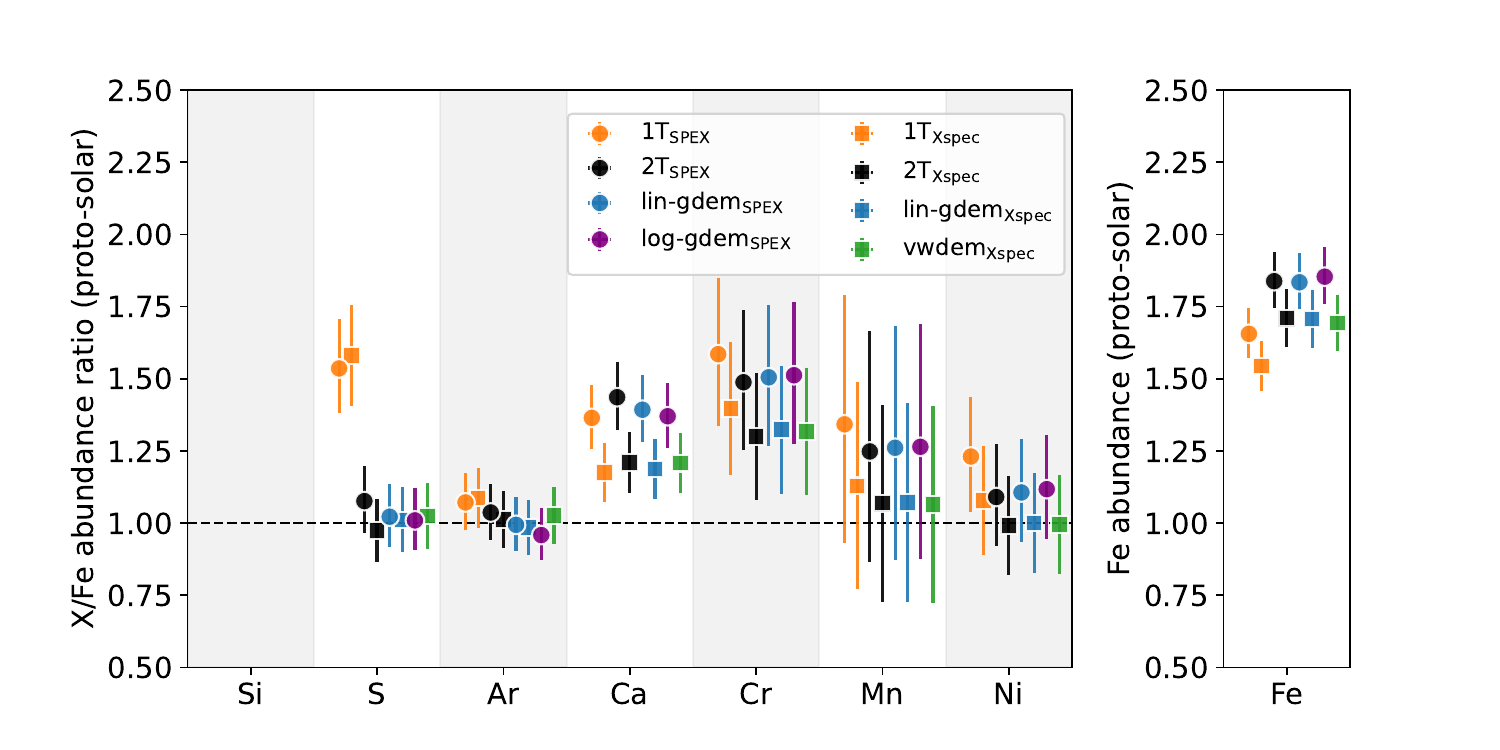} \\
\includegraphics[width=0.98\hsize, trim={1.0cm 0cm 1.5cm 0cm},clip]{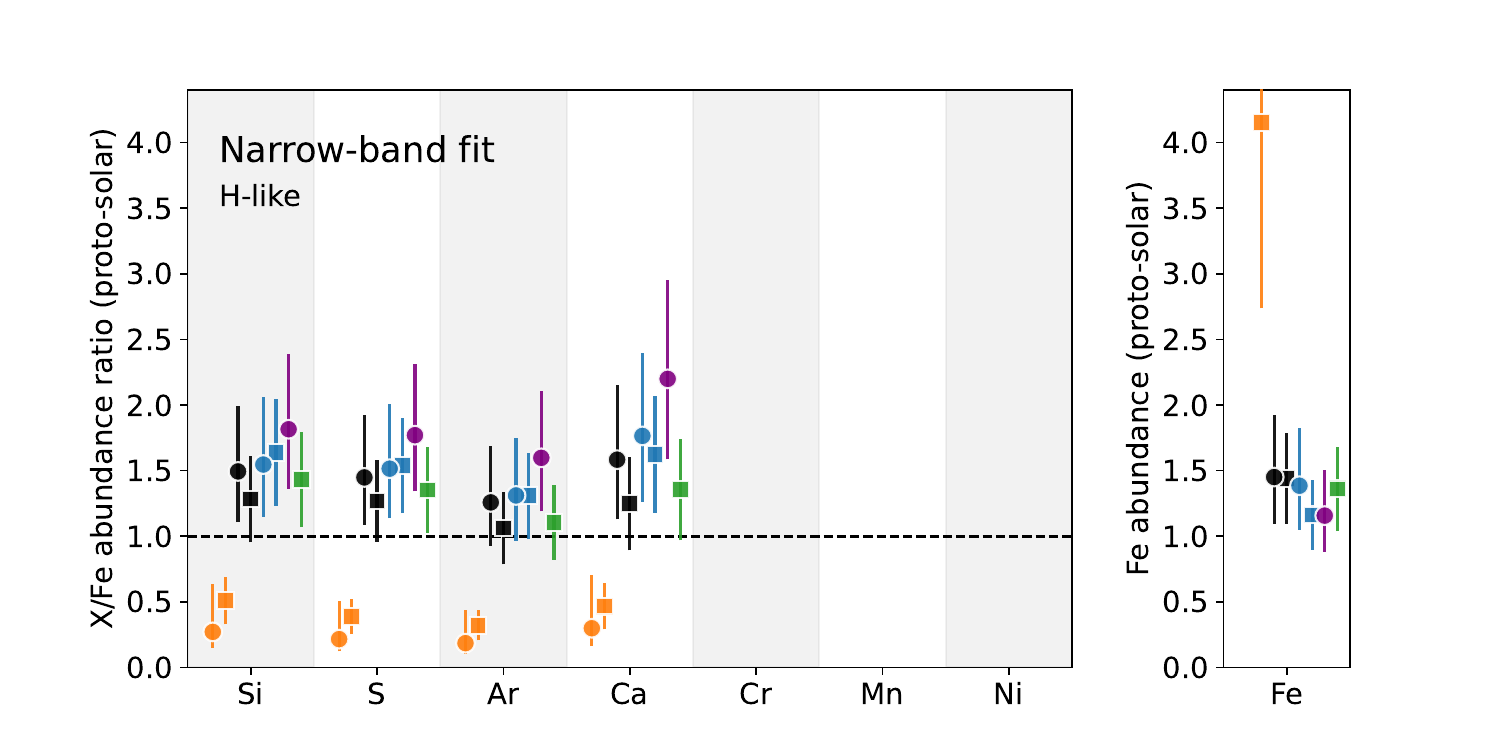}

      \caption{Same as Fig.~\ref{fig:broadfits}, now for narrow-band fits. \textit{Top:} He-like ions. \textit{Bottom:} H-like ions. Lines and fitting ranges used in this exercise are displayed in Fig.~\ref{fig:spectra_app}.
      }
         \label{fig:narrowfits_app}
\end{figure}

To complete Figs.~\ref{fig:broadfits} and \ref{fig:broadvsnarrowfits} discussed in the main text, we show in this Appendix a comparison of our different fitting model strategies for narrow-band fits (Fig.~\ref{fig:narrowfits_app}). As for broad-band fits, we note excellent consistency between our different experiments, with the notable exception of 1T models which tend to bias Fe and S/Fe as derived from He-like bands and to severely overestimate Fe as derived from H-like bands (implying in turn a strong underestimation of all X/Fe ratios). Together with the poorer fitting quality of the 1T models (Table~\ref{table:results_Resolve}), this indicates that the gas needs to be modelled with at least two temperature components. Such bias is comparable to the inverse-Fe bias reported at moderate spectral resolution in prevous works \citep[e.g.][]{rasia2008,simionescu2009,gastaldello2021}.

\FloatBarrier

\section{Simultaneous vs. combined RGS\,1 \& RGS\,2 fits}\label{sec:app:rgscombine}

\begin{figure}[h!]
\centering
\includegraphics[width=0.49\textwidth, trim={1.5cm 0cm 1.5cm 0cm},clip]{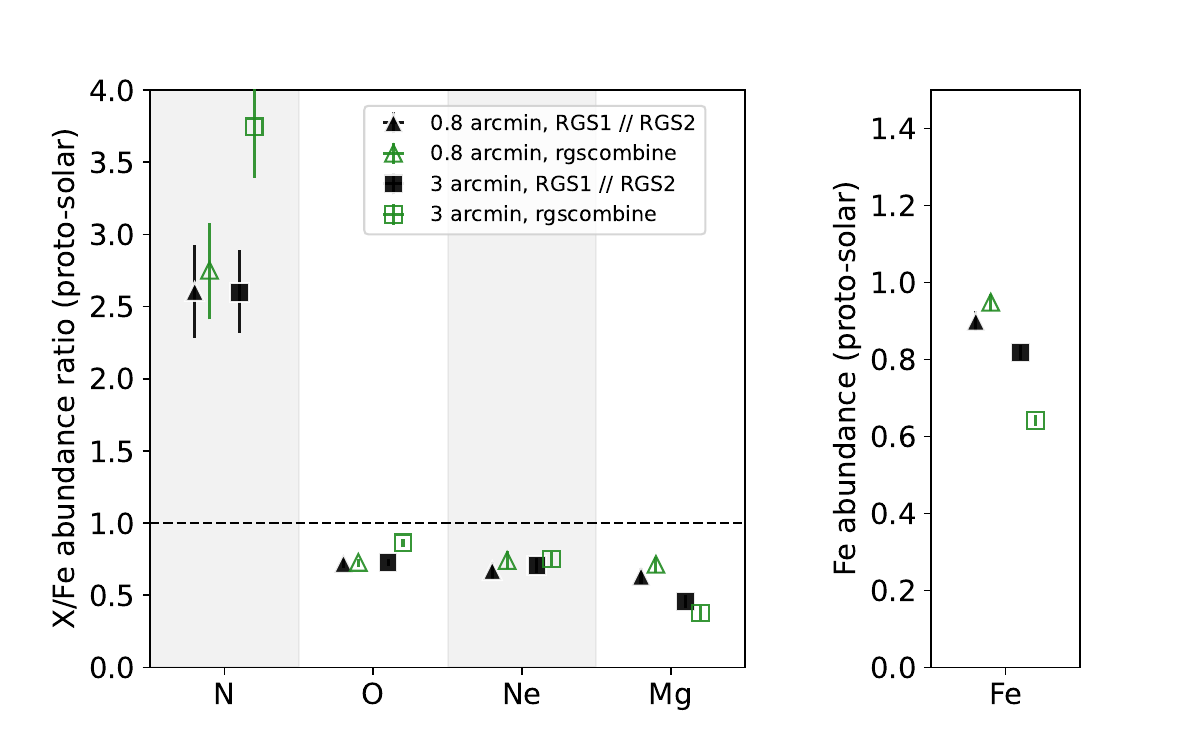} 

      \caption{Comparison of absolute Fe and X/Fe ratios derived with RGS over 0.8$'$ and 3.0$'$ cross-direction apertures using either a simultaneous RGS\,1 + RGS\,2 fit or the SAS task \texttt{rgscombine}.
      }
         \label{fig:rgscombine_app}
\end{figure}

In this Appendix section we present a comparison of Fe and the X/Fe ratios derived from RGS either fitting RGS\,1 and RGS\,2 simultaneously or combining the two instruments usin the SAS task \texttt{rgscombine} (as often seen in the literature). The results are shown in Fig.~\ref{fig:rgscombine_app}. For consistency with the rest of the paper, the results are shown in both the 0.8$'$ and the 3.0$'$ cross-dispersion apertures. Whereas the O/Fe, Ne/Fe and Mg/Fe ratios show marginal differences between the two approaches, the N/Fe ratio derived within 3.0$'$ using the combined method significantly exceeds the predictions of the three other estimates. Perhaps even more puzzling are the fairly large discrepancies obtained for the absolute Fe abundance, especially at 3.0$'$ cross-dispersion aperture. The precise reasons of such discrepancies are unknown and fall beyond the scope of this paper. While future work may address this issue more specifically, we find this issue worth reporting to the RGS cluster community.

\FloatBarrier

\section{Zoomed-in Resolve spectrum}\label{sec:app:zoomed}

\begin{figure*}[h!]
\centering
\includegraphics[width=0.99\hsize, trim={0cm 4.0cm 0cm 11.7cm},clip]{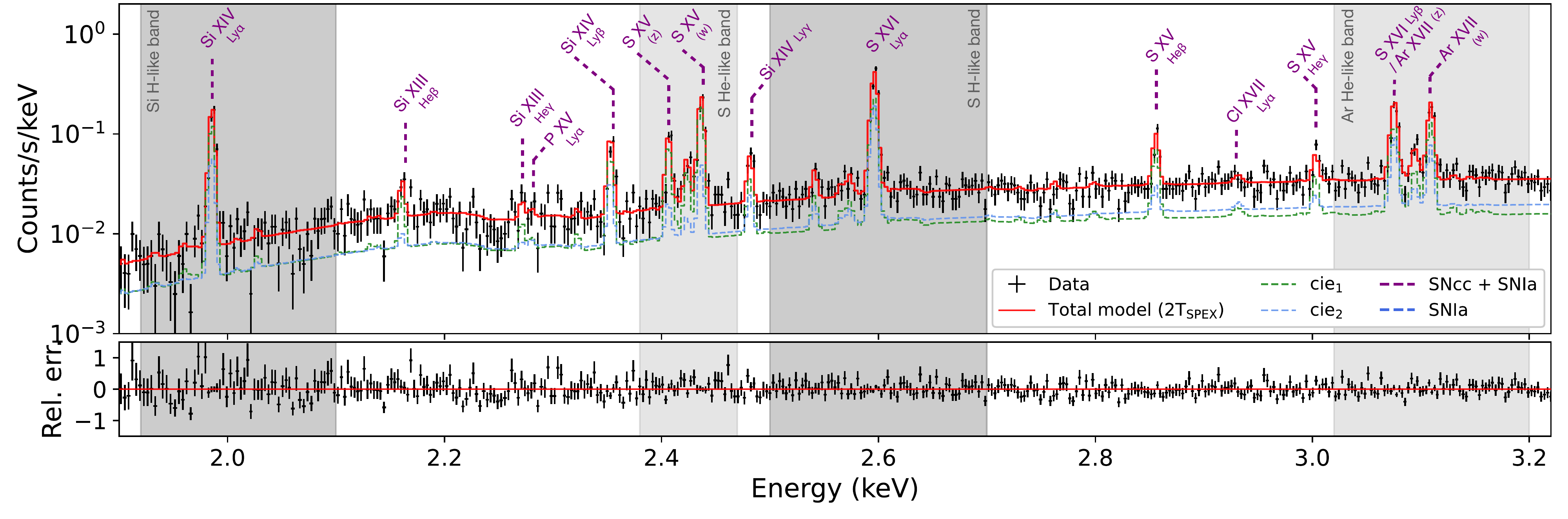}\\
\includegraphics[width=0.99\hsize, trim={0cm 4.0cm 0cm 11.7cm},clip]{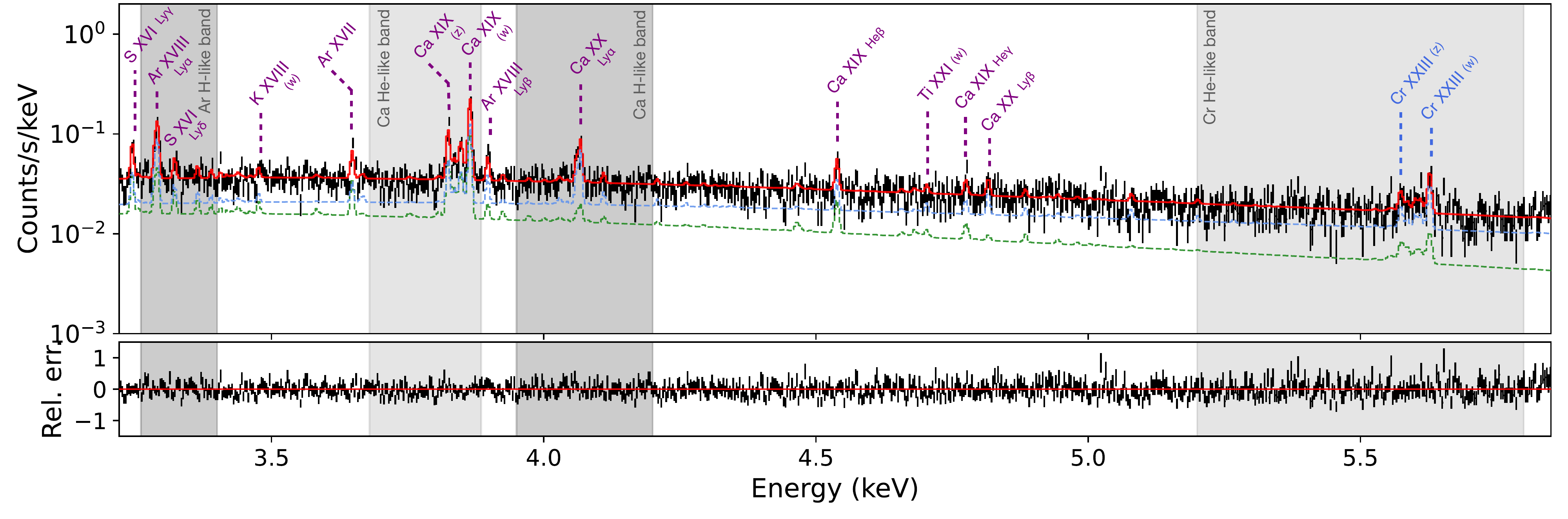} \\
\includegraphics[width=0.99\hsize, trim={0cm 4.0cm 0cm 11.7cm},clip]{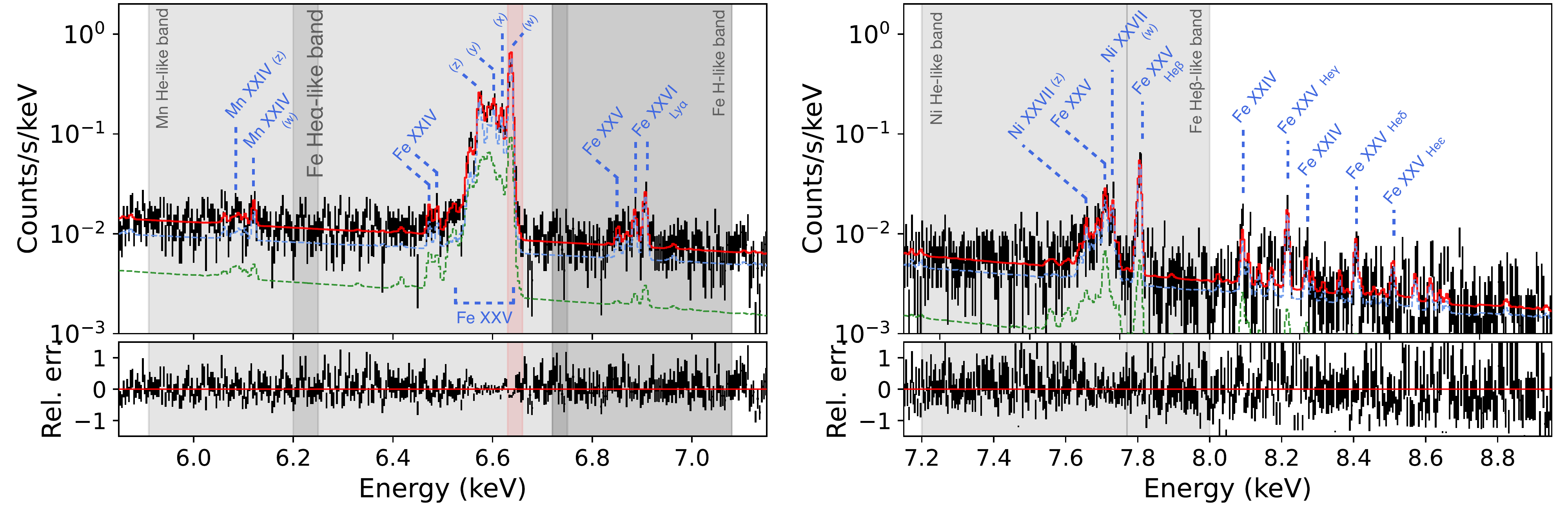}
      \caption{Zoomed-in Resolve spectrum as shown in Fig.~\ref{fig:spectra} right. Restricted energy bands used for our narrow-band fitting approach are indicated (dark and light grey for H-like and He-like elements, respectively). The red area shows the energy band that is excluded in all our fits to avoid biases induced by resonant scattering effects. The scaling of the y-axis (counts/s/keV) is kept constant between all panels for consistency.
      }
         \label{fig:spectra_app}
\end{figure*}

In support of Fig.~\ref{fig:spectra}, in this Appendix we present a detailed, zoomed-in version of the Centaurus full-array Resolve spectrum (Fig.~\ref{fig:spectra_app}). Lines of interest are annotated, as well as the energy bands used to perform our narrow-band fits.

\FloatBarrier

\section{List of SN yield models used in this work}\label{sec:app:yields}

This section lists all individual SNIa, SNcc, and AGB yield models used to fit pour final Centaurus abundance pattern. The SNIa models (Table~\ref{table:SNIa_models}) are classified either as near-$M_\mathrm{Ch}$ (deflagration, delayed-detonation) or sub-$M_\mathrm{Ch}$ (violent mergers, detonation, double-detonation). The SNcc and AGB models (Tables~\ref{table:SNcc_models} and \ref{table:AGB_models}, respectively) differ mostly through the initial metallicity of their progenitor $Z_\text{init}$. Note that we also consider hypernovae (HNe) as a possible candidate to reproduce the ratios of lighter elements.

\onecolumn

\begin{longtable}{lccl}
\caption{\label{table:SNIa_models} SNIa and SNcc yield models, taken from literature and used in this work.} \\
   \hline
   \hline
   Category & Name & Reference & Description \\
  \hline
   \endfirsthead
\caption{continued.}\\
   \hline
   \hline
   Category & Name & Reference & Description \\
  \hline
  \endhead
  \hline
   \endfoot 
\multicolumn{4}{c}{SNIa, near-$M_\text{Ch}$}  \\
\hline
Iw99 & W7 & 1 & 1D deflagration, $\rho_9 = 2.12$ \\
Iw99 & W70 & 1 & 1D deflagration, $\rho_9 = 2.12$, $Z_\text{init} = 0$ \\
Ma10 & C-DEF & 2 & 2D deflagration, $\rho_9 = 2.9$ \\
Fi14 & N1def & 3 & 3D deflagration, $\rho_9 = 2.9$, 1 ignition spot \\
Fi14 & N3def & 3 & 3D deflagration, $\rho_9 = 2.9$, 3 ignition spots \\
Fi14 & N5def & 3 & 3D deflagration, $\rho_9 = 2.9$, 5 ignition spots \\
Fi14 & N10def & 3 & 3D deflagration, $\rho_9 = 2.9$, 10 ignition spots \\
Fi14 & N20def & 3 & 3D deflagration, $\rho_9 = 2.9$, 20 ignition spots \\
Fi14 & N40def & 3 & 3D deflagration, $\rho_9 = 2.9$, 40 ignition spots \\
Fi14 & N100Ldef & 3 & 3D deflagration, $\rho_9 = 1.0$, 100 ignition spots \\
Fi14 & N100def & 3 & 3D deflagration, $\rho_9 = 2.9$, 100 ignition spots \\
Fi14 & N100Hdef & 3 & 3D deflagration, $\rho_9 = 5.5$, 100 ignition spots \\
Fi14 & N150def & 3 & 3D deflagration, $\rho_9 = 2.9$, 150 ignition spots \\
Fi14 & N200def & 3 & 3D deflagration, $\rho_9 = 2.9$, 200 ignition spots \\
Fi14 & N300Cdef & 3 & 3D deflagration, $\rho_9 = 2.9$, 300 centred ignition spots \\
Fi14 & N1600def & 3 & 3D deflagration, $\rho_9 = 2.9$, 1600 ignition spots \\
Fi14 & N1600Cdef & 3 & 3D deflagration, $\rho_9 = 2.9$, 1600 centred ignition spots \\
Le18 & 050-1-c3-P & 4 &  2D deflagration, $\rho_9 = 0.5$, $Z_\text{init} = 0.02$ \\
Le18 & 100-1-c3-P & 4 &  2D deflagration, $\rho_9 = 1.0$, $Z_\text{init} = 0.02$ \\
Le18 & 300-1-c3-P & 4 &  2D deflagration, $\rho_9 = 3.0$, $Z_\text{init} = 0.02$ \\
Le18 & 500-1-c3-P & 4 &  2D deflagration, $\rho_9 = 5.0$, $Z_\text{init} = 0.02$ \\
Le18 & W7new & 4 & 1D deflagration, W7 model (1) with updated electron capture and rates \\
\hline
Iw99 & WDD1 & 1 & 1D delayed-detonation, $\rho_9 = 2.12$, $\rho_\text{T,7} = 1.7$ \\
Iw99 & WDD2 & 1 & 1D delayed-detonation, $\rho_9 = 2.12$, $\rho_\text{T,7} = 2.2$ \\
Iw99 & WDD3 & 1 & 1D delayed-detonation, $\rho_9 = 2.12$, $\rho_\text{T,7} = 3.0$ \\
Iw99 & CDD1 & 1 & 1D delayed-detonation, $\rho_9 = 1.37$, $\rho_\text{T,7} = 1.7$ \\
Iw99 & CDD2 & 1 & 1D delayed-detonation, $\rho_9 = 1.37$, $\rho_\text{T,7} = 2.2$ \\
Ba06 & DDTa & 5,6 & 1D delayed-detonation, fits the Tycho SNR, $\rho_\text{T,7} = 3.9$ \\
Ba06 & DDTc & 5,6 & 1D delayed-detonation, fits the Tycho SNR, $\rho_\text{T,7} = 2.2$ \\
Ba06 & DDTe & 5,6 & 1D delayed-detonation, fits the Tycho SNR, $\rho_\text{T,7} = 1.3$ \\
Ma10 & C-DDT & 2 & 2D delayed-detonation, $\rho_9 = 2.9$, $\rho_\text{T,7} = 1.0$ \\
Ma10 & O-DDT & 2 & 2D delayed-detonation, $\rho_9 = 2.9$, $\rho_\text{T,7} = 1.0$, off-centre ignition \\
Se13 & N1 & 7 &  3D delayed-detonation, $\rho_9 = 2.9$, 1 ignition spot \\
Se13 & N3 & 7 &  3D delayed-detonation, $\rho_9 = 2.9$, 3 ignition spots \\
Se13 & N5 & 7 &  3D delayed-detonation, $\rho_9 = 2.9$, 5 ignition spots \\
Se13 & N10 & 7 &  3D delayed-detonation, $\rho_9 = 2.9$, 10 ignition spots \\
Se13 & N20 & 7 &  3D delayed-detonation, $\rho_9 = 2.9$, 20 ignition spots \\
Se13 & N40 & 7 &  3D delayed-detonation, $\rho_9 = 2.9$, 40 ignition spots \\
Se13 & N100L & 7 &  3D delayed-detonation, $\rho_9 = 1.0$, 100 ignition spots \\
Se13 & N100 & 7 &  3D delayed-detonation, $\rho_9 = 2.9$, 100 ignition spots \\
Se13 & N100H & 7 &  3D delayed-detonation, $\rho_9 = 5.5$, 100 ignition spots \\
Se13 & N150 & 7 &  3D delayed-detonation, $\rho_9 = 2.9$, 150 ignition spots \\
Se13 & N200 & 7 &  3D delayed-detonation, $\rho_9 = 2.9$, 200 ignition spots \\
Se13 & N300C & 7 &  3D delayed-detonation, $\rho_9 = 2.9$, 300 centred ignition spots \\
Se13 & N1600 & 7 &  3D delayed-detonation, $\rho_9 = 2.9$, 1600 ignition spots \\
Se13 & N1600C & 7 &  3D delayed-detonation, $\rho_9 = 2.9$, 1600 centred ignition spots \\
Oh14 & N100\_c50 & 8 &  3D delayed-detonation, N100 model (7), homogeneous with 50\% C  \\
Oh14 & N100\_rpc20 & 8 &  3D delayed-detonation, N100 model (7), depleted core with 20\% C  \\
Oh14 & N100\_rpc32 & 8 &  3D delayed-detonation, N100 model (7), depleted core with 32\% C  \\
Oh14 & N100\_rpc40 & 8 &  3D delayed-detonation, N100 model (7), depleted core with 40\% C  \\
Le18 & 100-0-c3 & 4 &  2D delayed-detonation, $\rho_9 = 1.0$, $Z_\text{init} = 0$ \\
Le18 & 100-0.1-c3 & 4 &  2D delayed-detonation, $\rho_9 = 1.0$, $Z_\text{init} = 0.002$ \\
Le18 & 100-0.5-c3 & 4 &  2D delayed-detonation, $\rho_9 = 1.0$, $Z_\text{init} = 0.01$ \\
Le18 & 100-1-c3 & 4 &  2D delayed-detonation, $\rho_9 = 1.0$, $Z_\text{init} = 0.02$ \\
Le18 & 100-2-c3 & 4 &  2D delayed-detonation, $\rho_9 = 1.0$, $Z_\text{init} = 0.04$ \\
Le18 & 100-3-c3 & 4 &  2D delayed-detonation, $\rho_9 = 1.0$, $Z_\text{init} = 0.06$ \\
Le18 & 100-5-c3 & 4 &  2D delayed-detonation, $\rho_9 = 1.0$, $Z_\text{init} = 0.1$ \\
Le18 & 100-1-c3 & 4 &  2D delayed-detonation, $\rho_9 = 1.0$, $Z_\text{init} = 0.02$ \\
Le18 & 300-0-c3 & 4 &  2D delayed-detonation, $\rho_9 = 3.0$, $Z_\text{init} = 0$ \\
Le18 & 300-0.1-c3 & 4 &  2D delayed-detonation, $\rho_9 = 3.0$, $Z_\text{init} = 0.002$ \\
Le18 & 300-0.5-c3 & 4 &  2D delayed-detonation, $\rho_9 = 3.0$, $Z_\text{init} = 0.01$ \\
Le18 & 300-1-c3 & 4 &  2D delayed-detonation, $\rho_9 = 3.0$, $Z_\text{init} = 0.02$ \\
Le18 & 300-2-c3 & 4 &  2D delayed-detonation, $\rho_9 = 3.0$, $Z_\text{init} = 0.04$ \\
Le18 & 300-3-c3 & 4 &  2D delayed-detonation, $\rho_9 = 3.0$, $Z_\text{init} = 0.06$ \\
Le18 & 300-5-c3 & 4 &  2D delayed-detonation, $\rho_9 = 3.0$, $Z_\text{init} = 0.1$ \\
Le18 & 300-1-c3 & 4 &  2D delayed-detonation, $\rho_9 = 3.0$, $Z_\text{init} = 0.02$ \\
Le18 & 500-0-c3 & 4 &  2D delayed-detonation, $\rho_9 = 5.0$, $Z_\text{init} = 0$ \\
Le18 & 500-0.1-c3 & 4 &  2D delayed-detonation, $\rho_9 = 5.0$, $Z_\text{init} = 0.002$ \\
Le18 & 500-0.5-c3 & 4 &  2D delayed-detonation, $\rho_9 = 5.0$, $Z_\text{init} = 0.01$ \\
Le18 & 500-1-c3 & 4 &  2D delayed-detonation, $\rho_9 = 5.0$, $Z_\text{init} = 0.02$ \\
Le18 & 500-2-c3 & 4 &  2D delayed-detonation, $\rho_9 = 5.0$, $Z_\text{init} = 0.04$ \\
Le18 & 500-3-c3 & 4 &  2D delayed-detonation, $\rho_9 = 5.0$, $Z_\text{init} = 0.06$ \\
Le18 & 500-5-c3 & 4 &  2D delayed-detonation, $\rho_9 = 5.0$, $Z_\text{init} = 0.1$ \\
Le18 & 500-1-c3 & 4 &  2D delayed-detonation, $\rho_9 = 5.0$, $Z_\text{init} = 0.02$ \\
 \hline
Kr15 & hyCONe & 9 &  3D deflagration, hybrid CONe based on N5def (3) \\
Se16 & GCD200 & 10 &  3D gravitationally confined detonation, $\rho_9 = 2.9$, ignition at $R=200$~km \\
\hline
\multicolumn{4}{c}{SNIa, sub-$M_\text{Ch}$}  \\
\hline
Pa10 & 0.9\_0.9 & 11 & WD-WD violent merger, $M_\text{WD} \simeq 0.9 + 0.9$, $\rho_9 = 1.4\times 10^{-2}$ \\
Kr13 & 0.9\_0.76 & 12 & WD-WD violent merger, $M_\text{WD} = 0.9 + 0.76$, $\rho_9 = 1.91\times 10^{-3}$ \\
Si10 & det\_0.81 & 13 &  1D pure detonation, $\rho_9 = 1 \times 10^{-2}$, $M_\mathrm{WD} = 0.81~M_\odot$ \\
Si10 & det\_0.88 & 13 &  1D pure detonation, $\rho_9 = 1.45 \times 10^{-2}$, $M_\mathrm{WD} = 0.88~M_\odot$ \\
Si10 & det\_0.97 & 13 &  1D pure detonation, $\rho_9 = 2.4 \times 10^{-2}$, $M_\mathrm{WD} = 0.97~M_\odot$ \\
Si10 & det\_1.06 & 13 &  1D pure detonation, $\rho_9 = 4.15 \times 10^{-2}$, $M_\mathrm{WD} = 1.06~M_\odot$ \\
Si10 & det\_1.15 & 13 &  1D pure detonation, $\rho_9 = 7.9 \times 10^{-2}$, $M_\mathrm{WD} = 1.15~M_\odot$ \\
Si12 & CSDD-S & 14 &  2D double-detonation, $\rho_9 = 0.85$, $M_\mathrm{WD} = 0.79~M_\odot$, converging shocks \\
Si12 & CSDD-L & 14 &  2D double-detonation, $\rho_9 = 0.381$, $M_\mathrm{WD} = 0.66~M_\odot$, converging shocks \\
Si12 & ELDD-S & 14 &  2D double-detonation, $\rho_9 = 0.85$, $M_\mathrm{WD} = 0.79~M_\odot$, edge-lit \\
Si12 & ELDD-L & 14 &  2D double-detonation, $\rho_9 = 0.381$, $M_\mathrm{WD} = 0.66~M_\odot$, edge-lit \\
Si12 & HeD-S & 14 &  2D double-detonation, $\rho_9 = 0.85$, $M_\mathrm{WD} = 0.79~M_\odot$, He detonation only \\
Si12 & HeD-L & 14 &  2D double-detonation, $\rho_9 = 0.381$, $M_\mathrm{WD} = 0.66~M_\odot$, He detonation only \\
Ma15 & CO15e7 & 15 &  2D pure detonation, $\rho_9 = 15 \times 10^{-2}$, $M_\mathrm{WD} = 1.23~M_\odot$ \\
Ma15 & ONe10e7 & 15 &  2D pure detonation, ONe WD, $\rho_9 = 10 \times 10^{-2}$, $M_\mathrm{WD} = 1.18~M_\odot$ \\
Ma15 & ONe13e7 & 15 &  2D pure detonation, ONe WD, $\rho_9 = 13 \times 10^{-2}$, $M_\mathrm{WD} = 1.21~M_\odot$ \\
Ma15 & ONe15e7 & 15 &  2D pure detonation, ONe WD, $\rho_9 = 15 \times 10^{-2}$, $M_\mathrm{WD} = 1.23~M_\odot$ \\
Ma15 & ONe17e7 & 15 &  2D pure detonation, ONe WD, $\rho_9 = 17 \times 10^{-2}$, $M_\mathrm{WD} = 1.24~M_\odot$ \\
Ma15 & ONe20e7 & 15 &  2D pure detonation, ONe WD, $\rho_9 = 20 \times 10^{-2}$, $M_\mathrm{WD} = 1.25~M_\odot$ \\
Sh18 & M08\_3070\_Z0\_1 & 16 & 3D D$^6$, $M_\text{WD} = 0.8$, C/O = 30/70, $Z_\text{init} = 0$, $^{12}$C+$^{16}$O rate = 1 \\
Sh18 & M08\_3070\_Z0\_01 & 16 & 3D D$^6$, $M_\text{WD} = 0.8$, C/O = 30/70, $Z_\text{init} = 0$, $^{12}$C+$^{16}$O rate = 0.1 \\
Sh18 & M08\_3070\_Z0005\_1 & 16 & 3D D$^6$, $M_\text{WD} = 0.8$, C/O = 30/70, $Z_\text{init} = 0.005$, $^{12}$C+$^{16}$O rate = 1 \\
Sh18 & M08\_3070\_Z0005\_01 & 16 & 3D D$^6$, $M_\text{WD} = 0.8$, C/O = 30/70, $Z_\text{init} = 0.005$, $^{12}$C+$^{16}$O rate = 0.1 \\
Sh18 & M08\_3070\_Z001\_1 & 16 & 3D D$^6$, $M_\text{WD} = 0.8$, C/O = 30/70, $Z_\text{init} = 0.01$, $^{12}$C+$^{16}$O rate = 1 \\
Sh18 & M08\_3070\_Z001\_01 & 16 & 3D D$^6$, $M_\text{WD} = 0.8$, C/O = 30/70, $Z_\text{init} = 0.01$, $^{12}$C+$^{16}$O rate = 0.1 \\
Sh18 & M08\_3070\_Z002\_1 & 16 & 3D D$^6$, $M_\text{WD} = 0.8$, C/O = 30/70, $Z_\text{init} = 0.02$, $^{12}$C+$^{16}$O rate = 1 \\
Sh18 & M08\_3070\_Z002\_01 & 16 & 3D D$^6$, $M_\text{WD} = 0.8$, C/O = 30/70, $Z_\text{init} = 0.02$, $^{12}$C+$^{16}$O rate = 0.1 \\
Sh18 & M08\_5050\_Z0\_1 & 16 & 3D D$^6$, $M_\text{WD} = 0.8$, C/O = 50/50, $Z_\text{init} = 0$, $^{12}$C+$^{16}$O rate = 1 \\
Sh18 & M08\_5050\_Z0\_01 & 16 & 3D D$^6$, $M_\text{WD} = 0.8$, C/O = 50/50, $Z_\text{init} = 0$, $^{12}$C+$^{16}$O rate = 0.1 \\
Sh18 & M08\_5050\_Z0005\_1 & 16 & 3D D$^6$, $M_\text{WD} = 0.8$, C/O = 50/50, $Z_\text{init} = 0.005$, $^{12}$C+$^{16}$O rate = 1 \\
Sh18 & M08\_5050\_Z0005\_01 & 16 & 3D D$^6$, $M_\text{WD} = 0.8$, C/O = 50/50, $Z_\text{init} = 0.005$, $^{12}$C+$^{16}$O rate = 0.1 \\
Sh18 & M08\_5050\_Z001\_1 & 16 & 3D D$^6$, $M_\text{WD} = 0.8$, C/O = 50/50, $Z_\text{init} = 0.01$, $^{12}$C+$^{16}$O rate = 1 \\
Sh18 & M08\_5050\_Z001\_01 & 16 & 3D D$^6$, $M_\text{WD} = 0.8$, C/O = 50/50, $Z_\text{init} = 0.01$, $^{12}$C+$^{16}$O rate = 0.1 \\
Sh18 & M08\_5050\_Z002\_1 & 16 & 3D D$^6$, $M_\text{WD} = 0.8$, C/O = 50/50, $Z_\text{init} = 0.02$, $^{12}$C+$^{16}$O rate = 1 \\
Sh18 & M08\_5050\_Z002\_01 & 16 & 3D D$^6$, $M_\text{WD} = 0.8$, C/O = 50/50, $Z_\text{init} = 0.02$, $^{12}$C+$^{16}$O rate = 0.1 \\
Sh18 & M085\_3070\_Z0\_1 & 16 & 3D D$^6$, $M_\text{WD} = 0.85$, C/O = 30/70, $Z_\text{init} = 0$, $^{12}$C+$^{16}$O rate = 1 \\
Sh18 & M085\_3070\_Z0\_01 & 16 & 3D D$^6$, $M_\text{WD} = 0.85$, C/O = 30/70, $Z_\text{init} = 0$, $^{12}$C+$^{16}$O rate = 0.1 \\
Sh18 & M085\_3070\_Z0005\_1 & 16 & 3D D$^6$, $M_\text{WD} = 0.85$, C/O = 30/70, $Z_\text{init} = 0.005$, $^{12}$C+$^{16}$O rate = 1 \\
Sh18 & M085\_3070\_Z0005\_01 & 16 & 3D D$^6$, $M_\text{WD} = 0.85$, C/O = 30/70, $Z_\text{init} = 0.005$, $^{12}$C+$^{16}$O rate = 0.1 \\
Sh18 & M085\_3070\_Z001\_1 & 16 & 3D D$^6$, $M_\text{WD} = 0.85$, C/O = 30/70, $Z_\text{init} = 0.01$, $^{12}$C+$^{16}$O rate = 1 \\
Sh18 & M085\_3070\_Z001\_01 & 16 & 3D D$^6$, $M_\text{WD} = 0.85$, C/O = 30/70, $Z_\text{init} = 0.01$, $^{12}$C+$^{16}$O rate = 0.1 \\
Sh18 & M085\_3070\_Z002\_1 & 16 & 3D D$^6$, $M_\text{WD} = 0.85$, C/O = 30/70, $Z_\text{init} = 0.02$, $^{12}$C+$^{16}$O rate = 1 \\
Sh18 & M085\_3070\_Z002\_01 & 16 & 3D D$^6$, $M_\text{WD} = 0.85$, C/O = 30/70, $Z_\text{init} = 0.02$, $^{12}$C+$^{16}$O rate = 0.1 \\
Sh18 & M085\_5050\_Z0\_1 & 16 & 3D D$^6$, $M_\text{WD} = 0.85$, C/O = 50/50, $Z_\text{init} = 0$, $^{12}$C+$^{16}$O rate = 1 \\
Sh18 & M085\_5050\_Z0\_01 & 16 & 3D D$^6$, $M_\text{WD} = 0.85$, C/O = 50/50, $Z_\text{init} = 0$, $^{12}$C+$^{16}$O rate = 0.1 \\
Sh18 & M085\_5050\_Z0005\_1 & 16 & 3D D$^6$, $M_\text{WD} = 0.85$, C/O = 50/50, $Z_\text{init} = 0.005$, $^{12}$C+$^{16}$O rate = 1 \\
Sh18 & M085\_5050\_Z0005\_01 & 16 & 3D D$^6$, $M_\text{WD} = 0.85$, C/O = 50/50, $Z_\text{init} = 0.005$, $^{12}$C+$^{16}$O rate = 0.1 \\
Sh18 & M085\_5050\_Z001\_1 & 16 & 3D D$^6$, $M_\text{WD} = 0.85$, C/O = 50/50, $Z_\text{init} = 0.01$, $^{12}$C+$^{16}$O rate = 1 \\
Sh18 & M085\_5050\_Z001\_01 & 16 & 3D D$^6$, $M_\text{WD} = 0.85$, C/O = 50/50, $Z_\text{init} = 0.01$, $^{12}$C+$^{16}$O rate = 0.1 \\
Sh18 & M085\_5050\_Z002\_1 & 16 & 3D D$^6$, $M_\text{WD} = 0.85$, C/O = 50/50, $Z_\text{init} = 0.02$, $^{12}$C+$^{16}$O rate = 1 \\
Sh18 & M085\_5050\_Z002\_01 & 16 & 3D D$^6$, $M_\text{WD} = 0.85$, C/O = 50/50, $Z_\text{init} = 0.02$, $^{12}$C+$^{16}$O rate = 0.1 \\
Sh18 & M09\_3070\_Z0\_1 & 16 & 3D D$^6$, $M_\text{WD} = 0.9$, C/O = 30/70, $Z_\text{init} = 0$, $^{12}$C+$^{16}$O rate = 1 \\
Sh18 & M09\_3070\_Z0\_01 & 16 & 3D D$^6$, $M_\text{WD} = 0.9$, C/O = 30/70, $Z_\text{init} = 0$, $^{12}$C+$^{16}$O rate = 0.1 \\
Sh18 & M09\_3070\_Z0005\_1 & 16 & 3D D$^6$, $M_\text{WD} = 0.9$, C/O = 30/70, $Z_\text{init} = 0.005$, $^{12}$C+$^{16}$O rate = 1 \\
Sh18 & M09\_3070\_Z0005\_01 & 16 & 3D D$^6$, $M_\text{WD} = 0.9$, C/O = 30/70, $Z_\text{init} = 0.005$, $^{12}$C+$^{16}$O rate = 0.1 \\
Sh18 & M09\_3070\_Z001\_1 & 16 & 3D D$^6$, $M_\text{WD} = 0.9$, C/O = 30/70, $Z_\text{init} = 0.01$, $^{12}$C+$^{16}$O rate = 1 \\
Sh18 & M09\_3070\_Z001\_01 & 16 & 3D D$^6$, $M_\text{WD} = 0.9$, C/O = 30/70, $Z_\text{init} = 0.01$, $^{12}$C+$^{16}$O rate = 0.1 \\
Sh18 & M09\_3070\_Z002\_1 & 16 & 3D D$^6$, $M_\text{WD} = 0.9$, C/O = 30/70, $Z_\text{init} = 0.02$, $^{12}$C+$^{16}$O rate = 1 \\
Sh18 & M09\_3070\_Z002\_01 & 16 & 3D D$^6$, $M_\text{WD} = 0.9$, C/O = 30/70, $Z_\text{init} = 0.02$, $^{12}$C+$^{16}$O rate = 0.1 \\
Sh18 & M09\_5050\_Z0\_1 & 16 & 3D D$^6$, $M_\text{WD} = 0.9$, C/O = 50/50, $Z_\text{init} = 0$, $^{12}$C+$^{16}$O rate = 1 \\
Sh18 & M09\_5050\_Z0\_01 & 16 & 3D D$^6$, $M_\text{WD} = 0.9$, C/O = 50/50, $Z_\text{init} = 0$, $^{12}$C+$^{16}$O rate = 0.1 \\
Sh18 & M09\_5050\_Z0005\_1 & 16 & 3D D$^6$, $M_\text{WD} = 0.9$, C/O = 50/50, $Z_\text{init} = 0.005$, $^{12}$C+$^{16}$O rate = 1 \\
Sh18 & M09\_5050\_Z0005\_01 & 16 & 3D D$^6$, $M_\text{WD} = 0.9$, C/O = 50/50, $Z_\text{init} = 0.005$, $^{12}$C+$^{16}$O rate = 0.1 \\
Sh18 & M09\_5050\_Z001\_1 & 16 & 3D D$^6$, $M_\text{WD} = 0.9$, C/O = 50/50, $Z_\text{init} = 0.01$, $^{12}$C+$^{16}$O rate = 1 \\
Sh18 & M09\_5050\_Z001\_01 & 16 & 3D D$^6$, $M_\text{WD} = 0.9$, C/O = 50/50, $Z_\text{init} = 0.01$, $^{12}$C+$^{16}$O rate = 0.1 \\
Sh18 & M09\_5050\_Z002\_1 & 16 & 3D D$^6$, $M_\text{WD} = 0.9$, C/O = 50/50, $Z_\text{init} = 0.02$, $^{12}$C+$^{16}$O rate = 1 \\
Sh18 & M09\_5050\_Z002\_01 & 16 & 3D D$^6$, $M_\text{WD} = 0.9$, C/O = 50/50, $Z_\text{init} = 0.02$, $^{12}$C+$^{16}$O rate = 0.1 \\
Sh18 & M10\_3070\_Z0\_1 & 16 & 3D D$^6$, $M_\text{WD} = 1.0$, C/O = 30/70, $Z_\text{init} = 0$, $^{12}$C+$^{16}$O rate = 1 \\
Sh18 & M10\_3070\_Z0\_01 & 16 & 3D D$^6$, $M_\text{WD} = 1.0$, C/O = 30/70, $Z_\text{init} = 0$, $^{12}$C+$^{16}$O rate = 0.1 \\
Sh18 & M10\_3070\_Z0005\_1 & 16 & 3D D$^6$, $M_\text{WD} = 1.0$, C/O = 30/70, $Z_\text{init} = 0.005$, $^{12}$C+$^{16}$O rate = 1 \\
Sh18 & M10\_3070\_Z0005\_01 & 16 & 3D D$^6$, $M_\text{WD} = 1.0$, C/O = 30/70, $Z_\text{init} = 0.005$, $^{12}$C+$^{16}$O rate = 0.1 \\
Sh18 & M10\_3070\_Z001\_1 & 16 & 3D D$^6$, $M_\text{WD} = 1.0$, C/O = 30/70, $Z_\text{init} = 0.01$, $^{12}$C+$^{16}$O rate = 1 \\
Sh18 & M10\_3070\_Z001\_01 & 16 & 3D D$^6$, $M_\text{WD} = 1.0$, C/O = 30/70, $Z_\text{init} = 0.01$, $^{12}$C+$^{16}$O rate = 0.1 \\
Sh18 & M10\_3070\_Z002\_1 & 16 & 3D D$^6$, $M_\text{WD} = 1.0$, C/O = 30/70, $Z_\text{init} = 0.02$, $^{12}$C+$^{16}$O rate = 1 \\
Sh18 & M10\_3070\_Z002\_01 & 16 & 3D D$^6$, $M_\text{WD} = 1.0$, C/O = 30/70, $Z_\text{init} = 0.02$, $^{12}$C+$^{16}$O rate = 0.1 \\
Sh18 & M10\_5050\_Z0\_1 & 16 & 3D D$^6$, $M_\text{WD} = 1.0$, C/O = 50/50, $Z_\text{init} = 0$, $^{12}$C+$^{16}$O rate = 1 \\
Sh18 & M10\_5050\_Z0\_01 & 16 & 3D D$^6$, $M_\text{WD} = 1.0$, C/O = 50/50, $Z_\text{init} = 0$, $^{12}$C+$^{16}$O rate = 0.1 \\
Sh18 & M10\_5050\_Z0005\_1 & 16 & 3D D$^6$, $M_\text{WD} = 1.0$, C/O = 50/50, $Z_\text{init} = 0.005$, $^{12}$C+$^{16}$O rate = 1 \\
Sh18 & M10\_5050\_Z0005\_01 & 16 & 3D D$^6$, $M_\text{WD} = 1.0$, C/O = 50/50, $Z_\text{init} = 0.005$, $^{12}$C+$^{16}$O rate = 0.1 \\
Sh18 & M10\_5050\_Z001\_1 & 16 & 3D D$^6$, $M_\text{WD} = 1.0$, C/O = 50/50, $Z_\text{init} = 0.01$, $^{12}$C+$^{16}$O rate = 1 \\
Sh18 & M10\_5050\_Z001\_01 & 16 & 3D D$^6$, $M_\text{WD} = 1.0$, C/O = 50/50, $Z_\text{init} = 0.01$, $^{12}$C+$^{16}$O rate = 0.1 \\
Sh18 & M10\_5050\_Z002\_1 & 16 & 3D D$^6$, $M_\text{WD} = 1.0$, C/O = 50/50, $Z_\text{init} = 0.02$, $^{12}$C+$^{16}$O rate = 1 \\
Sh18 & M10\_5050\_Z002\_01 & 16 & 3D D$^6$, $M_\text{WD} = 1.0$, C/O = 50/50, $Z_\text{init} = 0.02$, $^{12}$C+$^{16}$O rate = 0.1 \\
Sh18 & M11\_3070\_Z0\_1 & 16 & 3D D$^6$, $M_\text{WD} = 1.1$, C/O = 30/70, $Z_\text{init} = 0$, $^{12}$C+$^{16}$O rate = 1 \\
Sh18 & M11\_3070\_Z0\_01 & 16 & 3D D$^6$, $M_\text{WD} = 1.1$, C/O = 30/70, $Z_\text{init} = 0$, $^{12}$C+$^{16}$O rate = 0.1 \\
Sh18 & M11\_3070\_Z0005\_1 & 16 & 3D D$^6$, $M_\text{WD} = 1.1$, C/O = 30/70, $Z_\text{init} = 0.005$, $^{12}$C+$^{16}$O rate = 1 \\
Sh18 & M11\_3070\_Z0005\_01 & 16 & 3D D$^6$, $M_\text{WD} = 1.1$, C/O = 30/70, $Z_\text{init} = 0.005$, $^{12}$C+$^{16}$O rate = 0.1 \\
Sh18 & M11\_3070\_Z001\_1 & 16 & 3D D$^6$, $M_\text{WD} = 1.1$, C/O = 30/70, $Z_\text{init} = 0.01$, $^{12}$C+$^{16}$O rate = 1 \\
Sh18 & M11\_3070\_Z001\_01 & 16 & 3D D$^6$, $M_\text{WD} = 1.1$, C/O = 30/70, $Z_\text{init} = 0.01$, $^{12}$C+$^{16}$O rate = 0.1 \\
Sh18 & M11\_3070\_Z002\_1 & 16 & 3D D$^6$, $M_\text{WD} = 1.1$, C/O = 30/70, $Z_\text{init} = 0.02$, $^{12}$C+$^{16}$O rate = 1 \\
Sh18 & M11\_3070\_Z002\_01 & 16 & 3D D$^6$, $M_\text{WD} = 1.1$, C/O = 30/70, $Z_\text{init} = 0.02$, $^{12}$C+$^{16}$O rate = 0.1 \\
Sh18 & M11\_5050\_Z0\_1 & 16 & 3D D$^6$, $M_\text{WD} = 1.1$, C/O = 50/50, $Z_\text{init} = 0$, $^{12}$C+$^{16}$O rate = 1 \\
Sh18 & M11\_5050\_Z0\_01 & 16 & 3D D$^6$, $M_\text{WD} = 1.1$, C/O = 50/50, $Z_\text{init} = 0$, $^{12}$C+$^{16}$O rate = 0.1 \\
Sh18 & M11\_5050\_Z0005\_1 & 16 & 3D D$^6$, $M_\text{WD} = 1.1$, C/O = 50/50, $Z_\text{init} = 0.005$, $^{12}$C+$^{16}$O rate = 1 \\
Sh18 & M11\_5050\_Z0005\_01 & 16 & 3D D$^6$, $M_\text{WD} = 1.1$, C/O = 50/50, $Z_\text{init} = 0.005$, $^{12}$C+$^{16}$O rate = 0.1 \\
Sh18 & M11\_5050\_Z001\_1 & 16 & 3D D$^6$, $M_\text{WD} = 1.1$, C/O = 50/50, $Z_\text{init} = 0.01$, $^{12}$C+$^{16}$O rate = 1 \\
Sh18 & M11\_5050\_Z001\_01 & 16 & 3D D$^6$, $M_\text{WD} = 1.1$, C/O = 50/50, $Z_\text{init} = 0.01$, $^{12}$C+$^{16}$O rate = 0.1 \\
Sh18 & M11\_5050\_Z002\_1 & 16 & 3D D$^6$, $M_\text{WD} = 1.1$, C/O = 50/50, $Z_\text{init} = 0.02$, $^{12}$C+$^{16}$O rate = 1 \\
Sh18 & M11\_5050\_Z002\_01 & 16 & 3D D$^6$, $M_\text{WD} = 1.1$, C/O = 50/50, $Z_\text{init} = 0.02$, $^{12}$C+$^{16}$O rate = 0.1 \\
Le25 & 100-050-2-S50 & 17,18 & 2D detonation, $M_\text{WD} = 1.0$, $M_\text{He} = 0.05$, $Z_\text{init} = 0.02$, spherical det.  \\
Le25 & 110-050-2-B50 & 17,18 & 2D detonation, $M_\text{WD} = 1.1$, $M_\text{He} = 0.05$, $Z_\text{init} = 0.02$, ring-shaped det. \\

\end{longtable}
\tablefoot{The inner core densities $\rho_9$ are given in units of $10^9$ g/cm$^3$. The transitional deflagration-to-detonation densities $\rho_{T,7}$ are given in units of $10^7$ g/cm$^3$. The mass of the WD ($M_\text{WD}$) and, if relevant, of the He layer ($M_\text{He}$) are given in units of $M_\sun$. The Sh18 models are based on dynamically driven double-degenerate double-detonation models (DDDDDD or, more simply, D$^6$).}\\

\tablebib{(1)~\citet{iwamoto1999};
(2) \citet{maeda2010}; (3) \citet{fink2014};
(4) \citet{leung2018}; (5) \citet{badenes2003}; (6) \citet{badenes2006}; (7) \citet{seitenzahl2013}
(8) \citet{ohlmann2014}; (9) \citet{kromer2015}; (10) \citet{seitenzahl2016};
(11) \citet{pakmor2010}; (12) \citet{kromer2013}; (13) \citet{sim2010};
(14) \citet{sim2012};
(15) \citet{marquardt2015};
(16) \citet{shen2018};
(16) \citet{leung2020};
(18) \citet{leung2025}.}

\begin{longtable}{lccl}
\caption{\label{table:SNcc_models} SNcc (and HNe) yield models, taken from literature and used in this work.} \\
   \hline
   \hline
   Category & Name & Reference & Description \\
  \hline
   \endfirsthead
\caption{continued.}\\
   \hline
   \hline
   Category & Name & Reference & Description \\
  \hline
  \endhead
  \hline
   \endfoot 
\multicolumn{4}{c}{SNcc}  \\ 
\hline
No13 & Z0 & 1,2,3 & $Z_\text{init} = 0$ \\
No13 & Z0.001 & 1,2,3 & $Z_\text{init} = 0.001$ \\
No13 & Z0.004 & 1,2,3 & $Z_\text{init} = 0.004$ \\
No13 & Z0.008 & 3 & $Z_\text{init} = 0.008$ \\
No13 & Z0.02 & 1,2,3 & $Z_\text{init} = 0.02$ \\
Su16 & N20 & 4 & Low $Z_\text{init}$, incl. neutrino transport, calibrated to SN1987A  \\
Su16 & W18 & 4 & $Z_\text{init} \simeq 0.007$, incl. neutrino transport, calibrated to SN1987A  \\
Le25 & A22S03 & 5 & $Z_\text{init} = 0.02$, convective process optimised to ICM abundances of Perseus \\
\hline
\multicolumn{4}{c}{HNe}  \\ 
\hline
No13 & HNe\_Z0 & 1,2,3 & Hypernova, $Z_\text{init} = 0$ \\
No13 & HNe\_Z0.001 & 1,2,3 & Hypernova, $Z_\text{init} = 0.001$ \\
No13 & HNe\_Z0.004 & 1,2,3 & Hypernova, $Z_\text{init} = 0.004$ \\
No13 & HNe\_Z0.008 & 3 & Hypernova, $Z_\text{init} = 0.008$ \\
No13 & HNe\_Z0.02 & 1,2,3 & Hypernova, $Z_\text{init} = 0.02$ \\

\end{longtable}

\tablebib{(1) \citet{nomoto2006}; (2) \citet{kobayashi2006};
(3) \citet{nomoto2013}; (4) \citet{sukhbold2016}; (5) \citet{leung2025}.}

\begin{longtable}{lccl}
\caption{\label{table:AGB_models} AGB yield models, taken from literature and used in this work.} \\
   \hline
   \hline
   Category & Name & Reference & Description \\
  \hline
   \endfirsthead
\caption{Continued.}\\
   \hline
   \hline
   Category & Name & Reference & Description \\
  \hline
  \endhead
  \hline
   \endfoot 
\multicolumn{4}{c}{SNcc}  \\ 
\hline
No13 & Z0 & 1,2 & $Z_\text{init} = 0$ \\
No13 & Z0.001 & 1,2 & $Z_\text{init} = 0.001$ \\
No13 & Z0.004 & 1,2 & $Z_\text{init} = 0.004$ \\
No13 & Z0.008 & 2 & $Z_\text{init} = 0.008$ \\
No13 & Z0.02 & 1,2 & $Z_\text{init} = 0.02$ \\

\end{longtable}

\tablebib{(1) \citet{karakas2010}; (2) \citet{nomoto2013}.}

\end{appendix}
\end{document}